\documentclass[%
 preprint,
 amsmath,amssymb,
 aps, physrev,
]{revtex4-2}

\usepackage{CJK}
\usepackage{tikz}
\usepackage{graphicx,graphbox,makecell,tablefootnote}
\usepackage{longtable}
\usepackage{multirow}
\usepackage{booktabs}
\allowdisplaybreaks[4]
\usepackage{diagbox}
\usepackage{alltt}

\usepackage{dcolumn}
\usepackage{bm}

\definecolor{darkred}{RGB}{173,34,48}
\definecolor{lightgreen}{rgb}{0.56, 0.69, 0.19}
\definecolor{lightblue}{rgb}{0.36, 0.51, 0.71}
\definecolor{lightyellow}{rgb}{0.88, 0.61, 0.14}
\definecolor{darkgreen}{rgb}{0.6, 0.6, 0.35}
\definecolor{lightred}{rgb}{0.99, 0.36, 0.02}
\definecolor{box1}{rgb}{0.46, 0.6, 0.45}
\definecolor{box2}{rgb}{0.62, 0.56, 0.43}
\definecolor{box3}{rgb}{0.72, 0.65, 0.17}

\usepackage[hang]{footmisc}
\makeatletter
\ifFN@para
\else
  \long\def\@makefntext#1{%
    \ifFN@hangfoot
      \bgroup
      \setbox\@tempboxa\hbox{%
        \ifdim\footnotemargin>0pt
          \hb@xt@\footnotemargin{\@makefnmark\hss}%
        \else
          \@makefnmark\hskip-\footnotemargin      
        \fi
      }%
      \leftmargin\wd\@tempboxa
      \rightmargin\z@
      \linewidth \columnwidth
      \advance \linewidth -\leftmargin
      \parshape \@ne \leftmargin \linewidth
      \footnotesize
      \@setpar{{\@@par}}%
      \leavevmode
      \llap{\box\@tempboxa}%
      \parskip\hangfootparskip\relax
      \parindent\hangfootparindent\relax
    \else
      \parindent1em
      \noindent
      \ifdim\footnotemargin>\z@
        \hb@xt@ \footnotemargin{\hss\@makefnmark}%
      \else
        \ifdim\footnotemargin=\z@
          \llap{\@makefnmark}%
        \else
          \llap{\hb@xt@ -\footnotemargin{\@makefnmark\hss}}%
        \fi
      \fi
    \fi
    \footnotelayout#1%
    \ifFN@hangfoot
      \par\egroup
    \fi
  }
\fi
\makeatother

\usepackage{slashed}

\usepackage{tikz}

\begin{document}

\preprint{APS/123-QED}

\begin{CJK*}{UTF8}{}
\CJKfamily{gbsn}
\title{\textbf{Notes on conformal integrals: Coulomb branch amplitudes, magic identities and bootstrap} 
}%

\author{Song He}
\email{songhe@itp.ac.cn}
\affiliation{Institute of Theoretical Physics, Chinese Academy of Sciences, Beijing 100190, China}
\affiliation{School of Fundamental Physics and Mathematical Sciences, Hangzhou Institute for Advanced Study and ICTP-AP, UCAS, Hangzhou 310024, China}
\affiliation{Peng Huanwu Center for Fundamental Theory, Hefei 230026, China}
\author{Xuhang Jiang}%
 \email{xhjiang@itp.ac.cn}
 \affiliation{Institute of Theoretical Physics, Chinese Academy of Sciences, Beijing 100190, China}

\author{Jiahao Liu}
 \email{liujiahao@itp.ac.cn}
\affiliation{Institute of Theoretical Physics, Chinese Academy of Sciences, Beijing 100190, China}
\affiliation{School of Physical Sciences, University of Chinese Academy of Sciences, No.19A Yuquan Road, Beijing 100049, China}
\author{Yao{-}Qi Zhang}
\email{zhangyaoqi@itp.ac.cn}
\affiliation{Institute of Theoretical Physics, Chinese Academy of Sciences, Beijing 100190, China}
\affiliation{School of Physical Sciences, University of Chinese Academy of Sciences, No.19A Yuquan Road, Beijing 100049, China}


\date{\today}

\begin{abstract}
We study multi-loop conformal integrals for four-point correlators of planar ${\cal N}=4$ super-Yang-Mills theory, and in particular those contributing to Coulomb branch amplitudes in the ten-dimensional lightlike limit, where linear combinations of such integrals are determined by the large R-charge octagons exactly known from integrability. Exploiting known results for integrands, we review those combinations of dual conformal invariant (DCI) integrals that must evaluate to determinants of ladders, generalizing the simplest cases of Basso-Dixon fishnet integrals; in this way, we summarize all-loop predictions for the integrands (which are extracted from $f$-graphs) contributing to components of Coulomb branch amplitudes, such as next-to-fishnet integrals. Moreover, this exercise produces new ``magic identities", {\it i.e.} certain combinations of DCI integrals equal zero, and we enumerate and simplify such identities up to six loops explicitly. 

On the other hand, most of these individual integrals have not been computed beyond three loops, and as a first step we consider a bootstrap program for DCI integrals based on their leading singularities and the space of pure functions. We bootstrap the $3$ nontrivial DCI integrals for four-loop Coulomb branch amplitudes (providing an independent verification of the four-loop magic identity), which all take remarkably simple form as weight-$8$ single-valued harmonic polylogarithms. We also compute all leading singularities and a large portion of the pure functions for the $34$ DCI integrals contributing to five-loop amplitudes, where not only some integrals evaluate to functions beyond harmonic polylogarithms, but they also contain lower-weight pieces individually. 

\end{abstract}

\maketitle
\end{CJK*}

\section{Introduction}
Recent years have seen enormous progress in the study of scattering amplitudes and other physical quantities in quantum field theory, especially in the most symmetric of all four-dimensional gauge theories, the ${\cal N}=4$ supersymmetric Yang-Mills (SYM) theory in the planar limit. The two complementary sides of the story are the development of powerful new tools for computations, and consequently the discovery of hidden simplicity and mathematical structures (which usually leads to new computational tools {\it etc.}). In this story, correlation functions of half-BPS operators (the simplest local operators in the theory) have played a key role and attracted a lot of interest because they stand at the crossroads of perturbative calculations,  integrability and conformal bootstrap.

There are two seemingly unrelated relations between half-BPS correlators and scattering amplitudes in two different theories (see the nice review~\cite{Heslop:2022xgp} and references therein). At strong coupling they are dual to type-IIB supergravity amplitudes in string theory on AdS$_5\times S^5$ background via AdS/CFT; at weak coupling, we have the famous duality relating lightlike limits of correlators to null polygonal Wilson loops~\cite{Alday:2010zy,Adamo:2011dq} and equivalently (the square of) planar scattering amplitudes in ${\cal N}=4$ SYM~\cite{Alday:2007hr,Drummond:2007aua,Brandhuber:2007yx,Bern:2008ap,Drummond:2008aq,Eden:2010zz,Arkani-Hamed:2010zjl,Mason:2010yk,Eden:2010ce,Caron-Huot:2010ryg,Eden:2011yp,Eden:2011ku}. Furthermore, this ``triality" holds at the integrand level, and has provided important insights and rich data for both correlators and amplitudes: the $\ell$-loop planar integrand of the simplest four-point correlator of the stress-tensor supermultiplet already contains the $\ell$-loop integrand of four-point amplitudes and $(\ell-1)$-loop five-point amplitudes {\it etc.}~\cite{Eden:2010zz, Ambrosio:2013pba, Heslop:2018zut}, in various lightlike limits. In fact, the most accurate perturbative data for amplitude integrands are obtained from that of the four-point correlators, and, in the other direction, the relation to squared amplitudes also imposes powerful constraints on planar correlators. Following earlier works~\cite{Gonzalez-Rey:1998wyj, Eden:1998hh, Eden:1999kh, Eden:2000mv, Bianchi:2000hn}, the planar integrand of the four-point correlator was determined for three to seven loops in~\cite{Eden:2011we, Eden:2012tu, Ambrosio:2013pba} (the integrated results are known up to three loops~\cite{Drummond:2013nda}, and for results of higher Kaluza-Klein modes and higher points, see~\cite{Chicherin:2015edu,Chicherin:2018avq,Caron-Huot:2021usw,Bargheer:2022sfd,Caron-Huot:2023wdh}). The key breakthrough was the discovery of a hidden permutation symmetry for the integrand of the correlator~\cite{Eden:2011we}, which when combined with correlator / amplitude duality, has allowed for a powerful {\it graphical} bootstrap program that determines four-point integrands (for both correlators and amplitudes) up to ten loops~\cite{Bourjaily:2011hi, Bourjaily:2015bpz, Bourjaily:2016evz}. Very recently, a new, universal property has been found for the correlators which greatly improves the graphical bootstrap: this single rule has very efficiently determined correlator (and amplitude) integrands up to eleven~\cite{He:2024cej} and even twelve loops~\cite{toappear:2}! 

Although the planar loop integrand for four-point correlators and amplitudes has been determined to impressively high orders, very little is known about integrated results in general. The one- and two-loop correlators involve conformal integrals which are dual to the well-known box and double-box integrals respectively, but already the three-loop case needs rather non-trivial integrals which were bootstrapped in~\cite{Drummond:2013nda}. These integrals contain new {\it leading singularities} and new transcendental functions beyond the simplest (single-valued) harmonic polylogarithms~\cite{Remiddi:1999ew,Brown:2004ugm,Dixon:2012yy,Schnetz:2013hqa}, which, at the {\it symbol} level~\cite{Goncharov:2010jf, Duhr:2011zq, Duhr:2012fh} means new symbol letters beyond the simplest alphabet. Our original motivation was to similarly bootstrap all ($32$) conformal integrals at four loops: while we have succeeded for the majority of these integrals, the remaining few turn out to be very difficult to determine (including two integrals that contain elliptic pieces~\cite{toappear:1}). Therefore, except for some families of special cases such as ladders~\cite{Usyukina:1993ch} and fishnet integrals~\cite{Basso:2017jwq,Basso:2021omx,Aprile:2023gnh}, it remains an important open question how to systematically compute higher-loop, four-point conformal integrals analytically. 

On the other hand, in a remarkable paper~\cite{Caron-Huot:2021usw}, the authors have discovered a hidden {\it ten-dimensional} conformal symmetry which not only allows one to package all higher-charge four-point correlators into the 10d correlator (whose integrands can be expressed in terms of $f$-graphs), but also provide exact, integrated results for 10d lightlike limit of the correlators known as large R-charge correlator or the octagon~\cite{Coronado:2018ypq}. As explained in~\cite{Caron-Huot:2021usw}, the latter has a nice physical interpretation as the four-point Coulomb branch amplitudes (with non-vanishing v.e.v. for scalar fields), thus via the duality to the octagon these amplitudes (with different R-charge components) have been solved to all orders, which are given by determinants of ladder integrals. As we will review shortly, the interplay between (10d) correlator integrands using $f$-graphs, the 10d lightlike limit as Coulomb branch amplitudes and the exact results provided by the octagon proves extremely powerful for constraining both the integrand and integrated results for {\it dual conformal invariant} (DCI) integrals contributing to Coulomb branch amplitudes. 

In the first part of this paper we will summarize some ``data" obtained from these constraints. In sec.~\ref{sec:review} we will discuss how the 10d limit can be used to determine coefficients of some $f$-graphs for the correlator integrand: already the consistency with possible 10d limit implies vanishing of a large class of such coefficients (extending previous zeros from {\it square rules}~\cite{Bourjaily:2016evz}), and we also obtain intriguing connections between some $f$-graphs and special DCI integrals such as Basso-Dixon fishnet and generalizations such as next-to fishnet~\cite{Caron-Huot:2021usw}. We have checked all these constraints using both explicit results for $f$-graphs up to eleven loops~\cite{He:2024cej} and even higher-loop checks from square rules {\it etc.}. Moreover, in sec.~\ref{sec:magic} we will continue the study of the so-called ``magic identities" initiated in~\cite{Caron-Huot:2021usw}; it is well known that many DCI integrals are related to each other: already the original two-loop ``magic identity" for double-box integral~\cite{Drummond:2006rz} implies numerous higher-loop relations, {\it e.g.} three-loop ``tennis-court" integral equals the ladder integral; by exploiting the fact that certain combinations of DCI integrals equal to various components of Coulomb branch amplitudes which are given by determinants of ladders, the authors of~\cite{Caron-Huot:2021usw} have derived more non-trivial ``magic identities" at four and five loops. Note that these are highly nontrivial identities among DCI integrals which cannot be explained from other methods such as integration-by-parts identities {\it etc.}. We simplify such identities by subtracting known ``zeros" which can be understood from lower-loop identities, {\it e.g.} they can be these ladder-type identities (originated from double-box case) or those generated by ``inverse boxing" of lower-loop ones. In this way, we enumerate such non-trivial, simplified magic identities through six loops, which shows how these seemingly different DCI integrals can be closely related to each other. We will also comment on periods of conformal integrals~\cite{Broadhurst:1995km, Schnetz:2008mp, Brown:2009ta} (or the underlying $f$-graphs), which can serve as consistency checks of such magic identities.

All these studies have shown that, although it seems too difficult to compute all these conformal integrals for four loops and above, those DCI integrals contributing to Coulomb branch amplitudes may be relatively tamed. Indeed in sec.~\ref{sec:bootstrap} we will attempt to bootstrap such individual DCI integrals at four and five loops. Recall that for four loops, out of the $32$ conformal integrals, only $8$ contribute to the Coulomb branch amplitudes where $5$ of them are all equal to the ladder integral (by the original ladder-type magic identity)~\cite{Drummond:2006rz}. As shown in~\cite{Caron-Huot:2021usw}, the remaining $3$ nontrivial DCI integrals (dubbed $d2$, $f2$, and $f$) form a new magic identity, and our main goal is to bootstrap them individually, which then provides an independent check for the identity. Note that the $d_2$ integral and $f_2$ integral have been obtained before: the former was bootstrapped in~\cite{Drummond:2013nda}~\footnote{In fact $d_2$ and $f_2$ integral can be directly computed by the package \texttt{HyperlogProcedures}~\cite{Schnetz:2013hqa,Borinsky:2022lds} (but not for $f$ integral), which proves to be extremely useful for these computations.} and the latter is the simplest $2\times 2$ fishnet integral~\cite{Basso:2017jwq}. We obtain all these integrals analytically in terms of single-valued harmonic polylogarithms (SVHPL)~\cite{Brown:2004ugm,Drummond:2012bg,Schnetz:2013hqa,Brown:2015ztw}: we first analyze their leading singularities and the accompanying function space consists of (parity odd or even) SVHPL at weight-$8$; By imposing boundary conditions such as asymptotic expansion of these integrals~\cite{Chicherin:2018avq}, it becomes very straightforward to localize these functions, which all take remarkably simple form in terms of SVHPL functions. 

We then move to five loops: there are $34$ DCI integrals but many of them are either known ({\it e.g.} equal to ladder integral) or can be computed using \texttt{HyperlogProcedures}, but we find that some of these integrals are more complicated than expected. They can evaluate to functions that are no longer SVHPL but (single-valued) multiple polylogarithms~\cite{Chavez:2012kn,Drummond:2013nda,Schnetz:2021ebf} (in analogy with ``hard" integral at three loops or general conformal integrals at four loops). More importantly, some of these integrals contain lower-weight pieces, {\it e.g.} one can easily see that some of them actually give weight-$9$ functions. We compute all leading singularities of these DCI integrals, which then in principle allows us to bootstrap their maximal-transcendental part, and we also present some of the accompanying pure functions (both for maximal, weight-$10$ cases and for some weight-$9$ parts). We end with conclusions and some open equations in sec.~\ref{sec:discussions}.  


\section{From $f$-graphs to Coulomb-branch amplitudes/octagons}\label{sec:review}

We consider $\mathcal{N} = 4$ super Yang-Mills with the gauge group $\mathrm{SU}(N_c)$. There are $6$ real scalars, $\phi^I(x)$ where $I=1,2,\ldots,6$ in the fundamental representation of the internal symmetry group. By contracting with the coordinates $y_I$ satisfying $y_I y^I=0$, 
\begin{equation}
\phi^I(x)\rightarrow \phi(x,y)=y_I \phi^I(x),
\end{equation}
the half BPS operators fall into symmetric traceless representations of $\mathrm{SO}(6)$ which are then represented as the products of $\phi(x, y)$. In this paper, we focus on the single trace operator with general charge $k_i$,
\begin{equation}
\mathcal{O}_{k_i}(x_i)=\frac{1}{k_i}\left(\frac{2}{4\pi^2 N_c}\right)^{\frac{k_i}{2}}\mathrm{Tr}\left[\phi(x,y)^{k_i}\right],
\end{equation}
where $k_i\geq 2$ since we are considering the $\mathrm{SU}(N_c)$ group. When $k_i=2$, we come back to the lowest charge half-BPS operator which is also called the stress-tensor multiplet since it contains the stress-tensor and the on-shell Lagrangian.

We are interested in the connected correlator of four such operators in the planar limit. Perturbatively, we have
\begin{equation}
N_c^2\langle  \mathcal{O}_{k_{1}}\mathcal{O}_{k_{2}} \mathcal{O}_{k_{3}} \mathcal{O}_{k_{4}} \rangle_{\rm c} =G^{\rm free}_{k_1 k_2 k_3 k_4} + \sum_{\ell=1}^{\infty} G^{(\ell)}_{k_1 k_2 k_3 k_4}  +O(1/N_c^2),
\end{equation}
where the loop corrections $G^{(\ell)}_{k_1 k_2 k_3 k_4}$ can be computed using the Lagrangian insertion method, {\it i.e.} the $(4+\ell)$-point correlator with four external operators and $\ell$  chiral Lagrangians at Born level \cite{Intriligator:1998ig,Eden:2011we}:
\begin{equation}\label{eq:integrand}
\begin{aligned}
    G^{(\ell)}_{k_{1}k_{2}k_{3}k_{4}}
&= \frac{(-g^2)^\ell}{\ell!}\int \frac{d^4x_{5}}{\pi^2} \cdots \frac{d^4x_{4+n}}{\pi^2} \mathcal{G}^{(\ell)}_{k_{1}k_{2}k_{3}k_{4}} \\
&=\frac{(-g^2)^\ell}{\ell!}\int \frac{d^4x_{5}}{\pi^2} \cdots \frac{d^4x_{4+n}}{\pi^2} \langle \mathcal{O}_{k_{1}}\mathcal{O}_{k_{2}} \mathcal{O}_{k_{3}} \mathcal{O}_{k_{4}}\mathcal{L}(x_{5})\cdots \mathcal{L}(x_{4+\ell})\rangle^{(0)}.
\end{aligned}
\end{equation}
For the four-point correlator, the non-renormalization theorem \cite{Eden:2000bk} states that the loop integrand $\mathcal{G}^{(\ell)}_{k_{1}k_{2}k_{3}k_{4}}$ can factor out a universal prefactor,
\begin{equation}\label{eq:non}
\mathcal{G}_{k_{1}k_{2}k_{3}k_{4}}= R_{1234}\left(2x_{12}^2x_{13}^2x_{14}^2x_{23}^2x_{24}^2x_{34}^2\right) \mathcal{H}^{(\ell)}_{k_{1} k_{2} k_{3} k_{4}},
\end{equation}
where $\mathcal{H}^{(\ell)}_{k_{1} k_{2} k_{3} k_{4}}$ is the integrand for the reduced correlator and 
\begin{equation}
R_{1234}= d_{13}^2 d_{24}^2 x_{13}^2x_{24}^2 + d_{12}d_{23}d_{34}d_{14}\left(x_{13}^2x_{24}^2-x_{12}^2x_{34}^2-x_{14}^2 x_{23}^2\right) +(1\leftrightarrow 2)  +  (1\leftrightarrow 4),
\end{equation}
where $d_{ij}=-\frac{y_{ij}^2}{x_{ij}^2}$. Notice that the operator $\mathcal{O}_{k_i}(x_i,y_i)$ carries $k_i$ charges and the mass dimension $k_i$ while the prefactor is degree $2$ of both $y_i$ and $x_i$, therefore, the reduced integrand carries $k_i-2$ charges and dimension $k_i+2$.

\subsection{Review: four-point correlators and Coulomb branch amplitudes}
\paragraph{$f$-graphs for correlator integrands} Especially, for the lowest charge ($k_i=2$) half-BPS operator, the reduced integrand $\mathcal{H}^{(\ell)}_{2222}$ is just a rational function of $x_{ij}^2$ which has at most simple poles as a consequence of OPE~\cite{Green:2020eyj} and has conformal weight 4 with respect to each $x_i$. More importantly, it is invariant under an arbitrary permutation of $x_1,\cdots,x_{4+\ell}$ {\it i,e,} the full permutation $S_{4+\ell}$ instead of only $S_4\times S_{\ell}$~\cite{Eden:2011we}. The full permutation symmetry indicates that it is useful to package the $\ell$-loop integrands into $(4+\ell)$-point $f$-graphs where the solid line between vertex $i$ and $j$ represents a pole $1/x_{ij}^2$ and the dashed line denotes a numerator $x_{ij}^2$. Each $f$-graph (after performing \texttt{Expand[]} in Mathematica) represents a permutation invariant sum of terms, divided by the order of the automorphism group of the $f$-graph $\left|\mathrm{aut}(f)\right|$, so that the coefficient of any term is equal to 1. For example, when $\ell\leq 3$~\cite{Eden:2012tu},
\begin{equation}\label{eq:f123}
\begin{aligned}
&\mathcal{H}^{(1)}_{2222}=\includegraphics[align=c,scale=0.3]{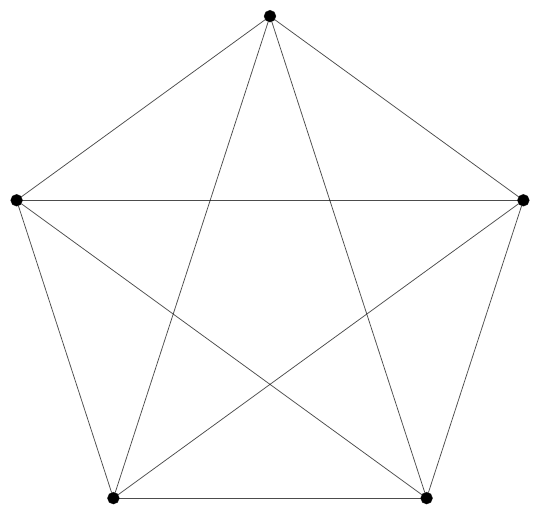}=\frac{1}{\prod_{1\leq i<j\leq5}x_{ij}^2},\\
&\mathcal{H}^{(2)}_{2222}=\includegraphics[align=c,scale=0.3]{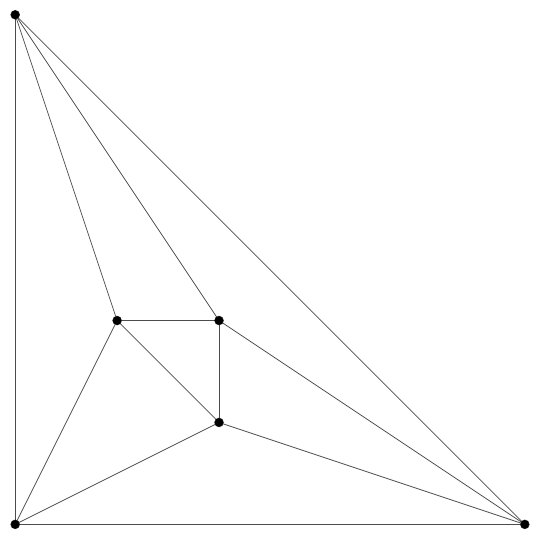}=\frac{1}{48} \frac{x_{12}^2x_{34}^2x_{56}^2 + S_{6}\,\text{permutations}}{\prod_{1\leq i < j \leq 6}x_{ij}^2},\\
&\mathcal{H}^{(3)}_{2222}=\includegraphics[align=c,scale=0.3]{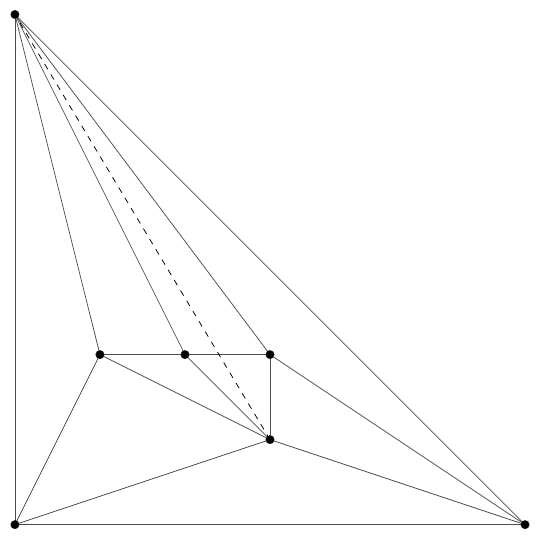}=\frac{1}{20}\frac{(x_{12}^2)^2\left(x_{34}^2x_{45}^2 x_{56}^2 x_{67}^2 x_{37}^2\right) + S_{7}\,\text{permutations}}{\prod_{1\leq i < j \leq 7}x_{ij}^2}.\\
\end{aligned}
\end{equation}
Generally, $\mathcal{H}^{(\ell)}_{2222}$ can be written as a linear combination of all $f$-graphs with $(4+\ell)$-vertecies $\mathcal{H}^{(\ell)}_{2222}=\sum_{i=1}^{\mathcal N_{\ell}}c_{i}^{(\ell)}f_{i}^{(\ell)}$ where $\mathcal{N}_{\ell}$ denotes the number of $(4+\ell)$-point $f$-graphs and $c_i^{(\ell)}$ are some rational coefficients. Considering the cusp limit where two consecutive separations $x_{12}^2=x_{23}^2=0$, these coefficients are fixed up to $\ell=11$ \cite{He:2024cej} and even $\ell=12$~\cite{toappear:2}. 

According to \eqref{eq:integrand}-\eqref{eq:f123}, we obtain the explicit expression of the correlator, written as the so-called {\it conformal integrals},
\begin{equation}
\begin{aligned}
G^{(1)}_{2222} &= -2g^2 R_{1234} \times g_{1234}, \\
G^{(2)}_{2222} &= 2g^4R_{1234}\bigg(h_{12;34}+h_{34;12}+h_{14;23}+h_{23;14}+h_{13;24}+h_{24;13}\nonumber\\
 &\qquad\qquad \left. \qquad+ \frac{1}{2} \left(x_{12}^2x_{34}^2 + x_{13}^2 x_{24}^2 + x_{14}^2 x_{23}^2\right)[g_{1234}]^2\right),
 \end{aligned}
\end{equation}
where the one-loop box and two-loop ladder are defined as 
\begin{equation}
    \begin{aligned}
        g_{1234} &=   \int\frac{d^{4}x_{5}}{x_{15}^2 x_{25}^2 x_{35}^2 x_{45}^2} ,\nonumber\\
h_{13;24} & = x_{24}^2 \int\frac{d^{4}x_{5}d^{4}x_{6}}{(x_{15}^2x_{25}^2x_{45}^2)x_{56}^2(x_{26}^2x_{36}^2x_{46}^2)}.
    \end{aligned}
\end{equation}
We can also identify the integral to a Feynman diagram in either momentum or position space. For example, the two-loop ladder integral can be drawn as
\begin{equation}
h_{13;24}\Leftrightarrow\includegraphics[align=c]{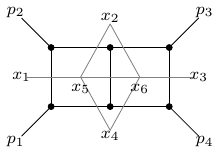},
\end{equation}
where the position space diagram (gray line) is dual to the momentum space diagram (black line). And the conformal integral $h_{13;24}$ is equivalent to the double-box integral with 4 external legs.

At $\ell=3$, there are $5$ conformal integrals from \eqref{eq:f123}.
\begin{figure}[htbp]
\hspace{-12cm}
    \includegraphics[scale=1]{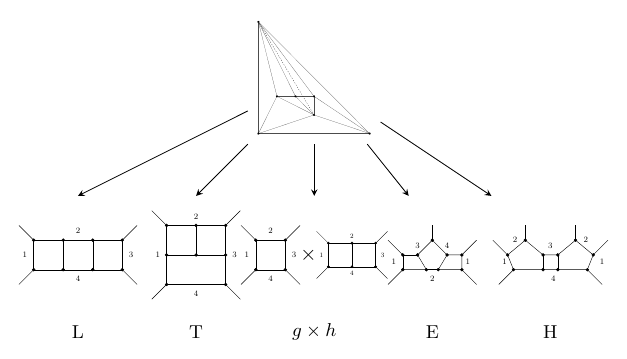}
    \caption{\label{fig:conf3} 5 conformal integrals at $\ell=3$.}
\end{figure}

\begin{equation}
\begin{aligned}
L_{13 ; 24} & =\int \prod_{a=5}^7 d^4 x_a\frac{\left(x_{24}^2\right)^2}{\left(\pi^2\right)^3} \int \frac{d^4 x_5 d^4 x_6 d^4 x_7}{\left(x_{15}^2 x_{25}^2 x_{45}^2\right) x_{56}^2\left(x_{26}^2 x_{46}^2\right) x_{67}^2\left(x_{27}^2 x_{37}^2 x_{47}^2\right)} \\
T_{13 ; 24} & =\int \prod_{a=5}^7 d^4 x_a\frac{x_{24}^2}{\left(\pi^2\right)^3} \int \frac{d^4 x_5 d^4 x_6 d^4 x_7 x_{17}^2}{\left(x_{15}^2 x_{25}^2\right)\left(x_{16}^2 x_{46}^2\right)\left(x_{27}^2 x_{37}^2 x_{47}^2\right) x_{56}^2 x_{57}^2 x_{67}^2}  \\
E_{12 ; 34} & =\int \prod_{a=5}^7 d^4 x_a\frac{x_{23}^2 x_{24}^2}{\left(\pi^2\right)^3} \int \frac{d^4 x_5 d^4 x_6 d^4 x_7 x_{16}^2}{\left(x_{15}^2 x_{25}^2 x_{35}^2\right) x_{56}^2\left(x_{26}^2 x_{36}^2 x_{46}^2\right) x_{67}^2\left(x_{17}^2 x_{27}^2 x_{47}^2\right)} \\
H_{12 ; 34} & =\int \prod_{a=5}^7 d^4 x_a\frac{x_{14}^2 x_{23}^2 x_{34}^2}{\left(\pi^2\right)^3} \int \frac{d^4 x_5^4 d_6 d^4 x_7 x_{57}^2}{\left(x_{15}^2 x_{25}^2 x_{35}^2 x_{45}^2\right) x_{56}^2\left(x_{36}^2 x_{46}^2\right) x_{67}^2\left(x_{17}^2 x_{27}^2 x_{37}^2 x_{47}^2\right)} \\
(gh)_{13 ; 24} & =\int \prod_{a=5}^7 d^4 x_a\frac{x_{13}^2\left(x_{24}^2\right)^2}{\left(\pi^2\right)^3} \int \frac{d^4 x_5}{x_{15}^2 x_{25}^2 x_{35}^2 x_{45}^2} \int \frac{d^4 x_6 d^4 x_7}{\left(x_{16}^2 x_{26}^2 x_{46}^2\right) x_{67}^2\left(x_{27}^2 x_{37}^2 x_{47}^2\right)}
\end{aligned}
\end{equation}

We remark that at $\ell=1,2,\cdots, 5$, there are $1,2,5, 32, 255$ conformal integrals, but the number of DCI integrals contributing to Coloumb branch amplitudes are smaller: already at two and three loops only $1$ (double box) and $2$ (ladder and tennis court) contribute, while at four and five loops only $8$ and $34$ DCI integrals contribute respectively. These integrals are defined in Appendix~\ref{app:definition}. There are $229, 1873, \cdots 66928343$ such DCI integrals (with nonzero coefficients) that contribute to Coulomb branch amplitudes at $\ell=6, 7, \cdots, 11$~\cite{Bourjaily:2016evz, He:2024cej}. 


For higher charge correlators, the reduced integrand can be written as a sum of different $R$-charge structures
\begin{equation}
\mathcal{H}^{(\ell)}_{k_{1}k_{2}k_{3}k_{4}} =  \sum_{\substack{b_{ij}\\ k_{i} =2+\sum_{j} b_{ij}} }  \mathcal{F}^{(\ell)}_{\{b_{ij}\}}
\times \prod_{1\leq i<j \leq 4} \left(d_{ij}\right)^{b_{ij}}.
\end{equation}
For example, $\mathcal{H}_{2222}^{(\ell)}=F_{\{0,0,0,0,0,0\}}^{(\ell)}$. The higher-charge integrand was bootstraped up to $\ell=5$ in \cite{Chicherin:2015edu, Chicherin:2018avq}. Surprisingly, the integrals that appear for higher-charge correlators are exactly those of the stress-tensor multiplet, only different by their coefficients. In fact, \cite{Caron-Huot:2021usw} indicates that all correlators with different R-charges come from a single master four-point correlator $\mathsf{G}(x_{ij}^2,y_{ij}^2)$
\begin{equation}\label{eq:co10}
\mathsf{G}(x_{ij}^2,y_{ij}^2)=\langle  \mathsf{O}(x_1,y_1)\mathsf{O}(x_2,y_2) \mathsf{O}(x_3,y_3) \mathsf{O}(x_4,y_4) \rangle, \quad \text{with}\;\mathsf{O}(x,y)\equiv\sum_{k=2}^{\infty}\mathrm{O}_{k}(x,y).
\end{equation}
Similar to \eqref{eq:non}, the loop integrand of the master correlator can pull out a universal factor
\begin{equation}
\mathsf{G}^{(\ell)}
= \frac{(-g^2)^\ell}{\ell!} R_{1234}\left(2x_{12}^2x_{13}^2 x_{14}^2x_{23}^2x_{24}^2x_{34}^2\right)\int \frac{dx_{5}^4}{\pi^2}\cdots \frac{dx_{4+n}^4}{\pi^2}\mathsf{H}^{(\ell)}(x_{ij}^2,y_{ij}^2),
\end{equation}
where $y_i=0$ for $i\geq 5$. As indicated in \cite{Caron-Huot:2021usw},   $\mathsf{H}^{(\ell)}(x_{ij}^2,y_{ij}^2)$ is a ten-dimensional conformal invariant object, {\it i.e.} a function of the distance in $D=10$ $X_{ij}^2\equiv x_{ij}^2 + y_{ij}^2=(1-d_{ij})x_{ij}^2$, and also possesses the full permutation symmetry $S_{4+\ell}$. A given correlator can be obtained by extracting the coefficient of the correct power of $y_i$,
\begin{equation}\label{eq:10d}
\mathcal{H}^{(\ell)}_{k_{1} k_{2} k_{3} k_{4}}(x_{ij}^2,y_{ij}^2)=\mbox{coefficient of $\left(\prod_{i=1}^4 \beta_i^{k_i-2}\right)$ in } \mathsf{H}^{(\ell)}(X_{ij}^2)
\big{|}_{y_{ij}^2 \to \beta_{i}\beta_{j}y^2_{ij}},
\end{equation}
For example, at $\ell=1$
\begin{equation}
 \mathsf{H}^{(1)}=\frac{1}{\prod_{1\leq i<j\leq5}x_{ij}^2}=\frac{1}{\prod_{1\leq i<j\leq5}x_{ij}^2(1-d_{ij})}=\frac{1}{\xi_4}g_{1234} \sum_{\{b_{ij}\}}\prod_{1\leq i<j \leq 4} \left(d_{ij}\right)^{b_{ij}},
\end{equation}
where $\xi_4=x_{12}^2x_{23}^2x_{34}^2x_{14}^2x_{13}^4x_{24}^4$. Therefore, all component correlators are equal and given by the one-loop box function (with
unit coefficient).

From \eqref{eq:10d}, the 10d correlator can be obtained by uplifting the lowest charge correlator in $D=4$. The only possible issue is that certain linear combination of conformal integrals might vanish at $D=4$ due to Gram identity, which could otherwise contribute to the 10d correlator. However, as verified in \cite{Caron-Huot:2021usw} up to 7 loops, there are no nontrivial Gram identities.

\paragraph{Coulomb-branch amplitudes from 10d lightlike limit}
One of the most important properties of the simplest half-BPS correlator is the amplitude/correlator duality~\cite{Eden:2010ce}. To be more explicit, the amplitude/correlator duality is realized in two steps. First, the light-like limit of the correlator is proportional to a light-like polygon Wilson loop in the adjoint representation~\cite{Alday:2010zy,Adamo:2011dq} where the latter equals the square of the Wilson loop in the fundamental representation in the planar limit. Secondly, in the planar limit, the scattering amplitude is dual to the polygonal Wilson loop in the fundamental representation~\cite{Alday:2007hr, Drummond:2007aua, Brandhuber:2007yx, Bern:2008ap, Drummond:2008aq, Eden:2010zz, Arkani-Hamed:2010zjl, Mason:2010yk, Caron-Huot:2010ryg, Eden:2011yp, Eden:2011ku}. At $4$-point
\begin{equation}
     \lim_{x_{i,i+1}^2\to 0} \frac{\mathcal{G}_{2222}(x_i)}{\mathcal{G}_{2222}^{\rm free}} = A(x_i)^2,
\end{equation}
where $A(x_i)$ is the planar four-point amplitude, divided by the tree-level
amplitude. Equivalently, for the $\ell$-loop correction
\begin{equation}
    \lim_{x_{i,i+1}^2\to0}2\xi_4 \mathcal{H}_{2222}^{(\ell)}=(A(x_i)^2)^{(\ell)},
\end{equation}
where $\xi_4$ is defined earlier and $(A(x_i)^2)^{(\ell)}=\sum_{\ell^\prime=0}^\ell \binom{\ell}{\ell^\prime}A(x_i)^{(\ell^\prime)}A(x_i)^{(\ell-\ell^\prime)}$. Graphically, the $\ell$-loop amplitudes $A(x_i)^{(\ell)}=A(x_i)^{(\ell)}\times A(x_i)^{(0)}$ term can be obtained from the $f$-graph by taking inequivalent four-face, {\it i.e.} four-cycles which encompass no internal points.  

A similar question can be asked about the 10d
light-like limit of the master correlator \eqref{eq:co10}, {\it i.e.} $X^2_{i,i+1}\to 0\Leftrightarrow d_{i,i+1}\to 1 \Leftrightarrow y_{ij}^{2} \to -x_{ij}^{2}$. 
In fact, as shown in \cite{Caron-Huot:2021usw}, the light-like limit in $D=10$ corresponds to the large R charge limit of the master correlator.  Under the large charge limit, the master correlator becomes the square of a certain octagon form
factor $\mathbb{O}$ \cite{Coronado:2018ypq} obtained by gluing two hexagons \cite{Basso:2015zoa} together
\begin{equation}\label{eq:GO}
\mathrm{Res}_{d_{i,i+1}=1} G\big{|}_{\rm integrand}=\mathbb{O}\times \mathbb{O}\big|_{\rm integrand},
\end{equation}
which is similar to the correlator/Wilson loop duality. From the amplitude side, a natural extension to $D=10$ is the scattering amplitude on the Coulomb branch since the nonzero $y_i,i\leq 4$ coordinates can be interpreted as vacuum expectation values for the scalar fields of the $\mathcal{N} = 4$ SYM theory
\begin{equation}\label{eq:OM}
    \frac{\mathbb{O}(x,y)}{\mathbb{O}^{\text{free}}(x,y)} =  M(x,y),
\end{equation}
where $M$ is a massive four-particle amplitude in the Coulomb branch~\footnote{this is a special case of Coulomb branch amplitude where $y_i^2=0$.} normalized so that $M^{\text{free}}=1$,
with external momenta and masses are $p_{i}^\mu \equiv x^\mu_{i,i+1}, m_{i}^2 \equiv y_{i,i+1}^2$. Thus, the light-like condition $X_{i,i+1}^2=0$ becomes the on-shell condition $p_i^2+m_i^2=0$. In addition, the generic external mass $m_i^2\neq0$ also keeps the integrals off-shell, meaning there are no divergences of the integrals. Therefore, \eqref{eq:GO} and \eqref{eq:OM} hold both at the integrand and integrated levels. 

Note that the octagon is known for arbitrary values of the coupling from integrability \cite{Belitsky:2020qrm, Belitsky:2020qir}, and the Coulomb branch amplitude can be computed from the master correlator perturbatively. \eqref{eq:OM} produces some all-loop predictions for the correlators at the integrand level as well as some new results for integrals. Before that, we first discuss the DCI integrals from the light-like limit on the right-hand side of \eqref{eq:OM} at weak coupling. 

\begin{figure}
    \hspace{-5cm}
    \includegraphics[scale=1.5]{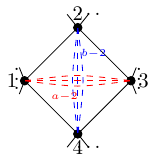}
    \caption{Any $f$-graph with $(a,b)$-structure (on the $4$-face labeled by $1,2,3,4$) contributes to $M_{a,b}$.}
    \label{fig:ab}
\end{figure}

\paragraph{DCI integrals from light-like limit and ladder-type magic identity}
The Coulomb branch amplitude can be obtained by taking the $D=10$ light-like limit of the master correlator.  For example, at $\ell\leq 3$. 
\begin{equation}\label{eq:M3}
    \begin{aligned}
        M&= 1-g^2{X_{13}^2 X_{24}^2}g_{1234} \nonumber \\
&\qquad+g^4\left[\frac{{X_{13}^2 (X_{24}^2)^2}}{x_{24}^2}h_{13;24} +  \frac{{(X_{13}^2)^2 X_{24}^2}}{x_{13}^2}h_{24;13}\right] \nonumber \\
&\qquad-g^{6}\left[\frac{{X_{13}^2 (X_{24}^2)^3}}{(x_{24}^2)^2}L_{13;24} + \frac{{(X_{13}^2)^3 X_{24}^2}}{(x_{13}^2)^2} L_{24;13}+\frac{{X_{13}^2 (X_{24}^2)^2}}{x_{24}^2} (T_{13;24}+T_{13;42})\right. \nonumber\\
&\qquad\qquad\quad\left.+  \frac{{(X_{13}^2)^2 X_{24}^2}}{x_{13}^2}(T_{24;13}+T_{24;31})\right]+\mathcal{O}(g^8)
    \end{aligned}
\end{equation}
We can see that $M_{a,b}$ can be naturally grouped by the power of $X_{13}^2\sim(1-d_{13})$ and $X_{24}^2\sim(1-d_{24})$
\begin{equation}
    M=1+ \sum_{a,b\geq 1} (1-d_{13})^a(1-d_{24})^b M_{a,b},
\end{equation}
where $M_{a,b}=M_{b,a}$ at the integrated level. From the $f$-graph perspective, the $\ell$-loop integrand $M_{a,b}^{(\ell)}$ can be obtained by taking the four-faces with specific $(a,b)$-structure in Fig.~\ref{fig:ab}.
When $a=1$ or $b=1$, the dashed line becomes the solid line. By taking the light-like limit of certain $f$-graph, we focus on the rational functions corresponding to Fig.~\ref{fig:ab} and denote the remaining part as $\tilde{f}$
\begin{equation}\label{eq:deriveab}
    f\rightarrow x_{12}^2x_{23}^2x_{34}^2x_{14}^2x_{13}^4x_{24}^4 \frac{(x_{13}^2)^{a-2}(x_{24}^2)^{b-2}}{x_{12}^2x_{23}^2x_{34}^2x_{14}^2}\tilde{f}\rightarrow (1-d_{13})^a(1-d_{24})^b M_{a,b}^{(\ell)}.
\end{equation}
Therefore, an $f$-graph contributes to $M_{a,b}$ if and only if it has the $(a,b)$-structure.

Notice that when $a=b=1$, the $(1,1)$-structure is just the $k=4$ divergence structure in \cite{Bourjaily:2015bpz}, which means that no $f$-graphs can contribute to $M_{1,1}$ except for the non-planar $f$-graph at $\ell=1$, which gives the box integral. When $a=1,b=2$, this is the double triangle structure discussed in \cite{He:2024cej}, which means that almost all $f$ -graphs contribute to $M_{1,2}$. In fact, one can check that except for the 4 $f$-graphs without double-triangle structure that occur at $\ell=8,11,12$, all $f$-graphs contribute to $M_{1,2}$. The other extremal case is where $a,b$ is large. We will discuss this case in the next subsection. 

At the integrated level, some of these DCI integrals are related by ladder-type magic identity~\cite{Drummond:2006rz}. This kind of magic identity stems from the fact that the $s$-channel double box at $\ell=2$ is equal to the $t$-channel one.
\begin{equation}
\includegraphics[scale=0.5,align=c]{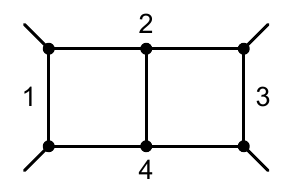}=\includegraphics[scale=0.5,align=c]{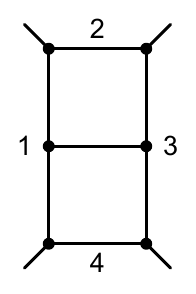}
\end{equation}
This equality has an all-loop generalization by adding boxes to the momentum space Feynman diagram (or the so-called sling-shot rule in position space). For example, at $\ell=3$, it tells us that the ladder integral is equal to the tennis-court integral.
\begin{equation}
\includegraphics[scale=0.5,align=c]{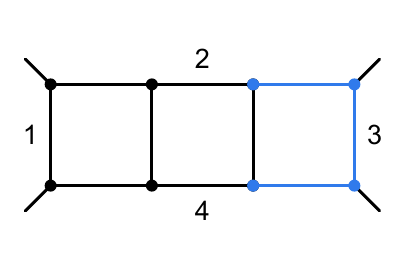}=\includegraphics[scale=0.5,align=c]{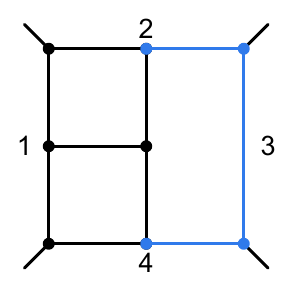}
\end{equation}
We can also translate the ``adding box'' procedure to position space, which turns out to be the ``slingshot'' rule~\cite{Drummond:2006rz} \footnote{Slingshot rule can only generate part of the ladder-type identities. We need to consider relation like $\tilde{\mathcal{I}}^{(\ell)}=\tilde{\mathcal{I}}^{(\ell')}\tilde{\mathcal{I}}^{(\ell-\ell')}$, which has the slingshot rule as its special case when $\ell'=1$.}. To be explicit, we start from the magic identity ${\mathcal{I}}^{(\ell)}_1={\mathcal{I}}^{(\ell)}_2=\ldots={\mathcal{I}}_{m_{\ell}}^{(\ell)}$ at $\ell$ loop whose external points are $1,2,3,4$ in order. We pick up one of the external points, for example, the point 1, then multiply ${\mathcal{I}}^{(\ell)}_i$ by the slingshot factor $\frac{x_{24}^2}{x_{12}^2x_{14}^2x_{\ell+5,1}^2}$
\begin{equation}\label{eq:shot}
{\mathcal{I}}^{(\ell+1)}_j=\includegraphics[scale=1.2,align=c]{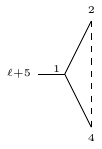}\times {\mathcal{I}}^{(\ell)}_i,\quad i=1,2,\ldots,m_\ell,
\end{equation}
where the vertex $1$ becomes an internal point of ${\mathcal{I}}_j^{(\ell+1)}$ and the external points are now $\ell+5,2,3,4$. All ${\mathcal{I}}_j^{(\ell+1)}$ are equal at the integrated level. At the $f$-graph level, we can see that the ``slingshot'' rule can be translated to the inverse square rule we discuss later.

\subsection{Fishnet and next-to-fishnet from $f$-graphs}
Although all $f$-graphs with $(a,b)$-structures may contribute to $M_{a,b}$ as we discussed in the last section, the physical results are only relevant to those $f$-graphs with non-zero coefficients, which we call non-vanishing $f$-graphs. Practically, we first study all potentially contributing $f$-graphs from a pure graphical view and then distinguish them by their coefficients. Some general observations and rules are brought up for these vanishing and non-vanishing $f$-graph contributions in the leading order (fishnet) and next-to-leading order (next-to-fishnet).

\begin{figure}
    \hspace{-6cm}
    \includegraphics[scale=0.5]{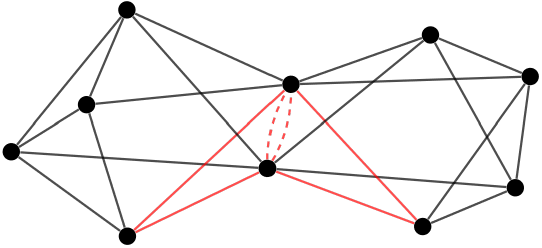}
    \caption{A $6$-loop $f$-graph with $(2,4)$ structure; its coefficient vanishes due to the $k=4$ divergence structure.}
    \label{fig:f624}
\end{figure}

We first discuss some all-loop structures from \eqref{eq:OM} at the integrand level. From integrability, \eqref{eq:OM} tells us that the leading order of $M_{a,b}$ starts from $(g^2)^{ab}$. However, it is not natural from the $f$-graph perspective. A simple combinatorial argument only shows that an upper bound for the $(a,b)$-structure in a $(4+\ell)$-point $f$-graph is $a+b+\mathrm{min}(a,b)\leq\ell+2$. 
For example, the $6$-loop $f$-graph in Fig. \ref{fig:f624} has a $(2,4)$-structure, and its coefficient vanishes from the square rule \cite{Bourjaily:2016evz}. 
Here we conjecture that the coefficient of a $(4+\ell)$-point $f$-graph with $(a,b),ab>\ell$ structure must vanish~\footnote{In fact, $M_{a,b}^{(\ell>ab)}$ vanishes exactly at the integrated level, and we found that they vanish at the integrand level up to $\ell = 11$.}. 
We list the number of $f$-graphs that can potentially contribute to $M_{a,b}, ab>\ell$ up to $\ell=11$ in Table \ref{tab:abgl}. Up to $\ell=10$, all zeros can be understood by the square rule; however, at $\ell=11$, there are $8$ vanishing $c_i^{(11)}$'s that the square rule cannot explain.
\begin{table}[htbp]
  \centering
  \setlength{\tabcolsep}{0.8mm}
    \begin{tabular}{|c|c|c|c|c|c|c|c|c|c|c|c|}
    \hline
    \diagbox{$\ell$}{$(a,b)$}
    & (2,4) & (2,5) & (2,6) & (2,7) & (2,8) & (2,9) & (3,3) & (3,4) & (3,5) & (3,6) & (3,7) \\
    \hline
    6     & 1     &       &       &       &       &       &       &       &       &       &  \\
    \hline
    7     & 4     & 1     &       &       &       &       &       &       &       &       &  \\
    \hline
    8     &   \diagbox[innerleftsep=.3cm]{\;}{\,}    & 9     & 1     &       &       &       & 4     & 2     &       &       &  \\
    \hline
    9     &  \diagbox[innerleftsep=.3cm]{\;}{\,}     & 120   & 9     & 1     &       &       &   \diagbox[innerleftsep=.3cm]{\;}{\,}    & 1     & 2     &       &  \\
    \hline
    10    &   \diagbox[innerleftsep=.3cm]{\;}{\,}    &   \diagbox[innerleftsep=.3cm]{\;}{\,}    & 174   & 12    & 1     &       &   \diagbox[innerleftsep=.3cm]{\;}{\,}    & 738   & 59    & 2     &  \\
    \hline
    11    &    \diagbox[innerleftsep=.3cm]{\;}{\,}   &   \diagbox[innerleftsep=.3cm]{\;}{\,}    & 4049  & 212   & 12    & 1     &   \diagbox[innerleftsep=.3cm]{\;}{\,}    & 21764 & 1845  & 58    & 2 \\
    \hline
    \end{tabular}%
    \caption{Numbers of $\ell$-loop $f$-graphs potentially /contributing to $M_{a,b}$ where $ab>\ell$.}
  \label{tab:abgl}%
\end{table}%

\begin{figure}
    \hspace{-5cm}
    \includegraphics[scale=0.4,align=c]{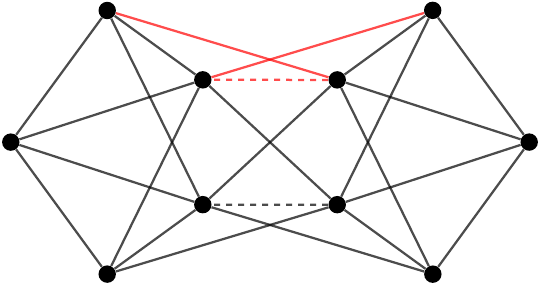}
    \includegraphics[scale=0.4,align=c]{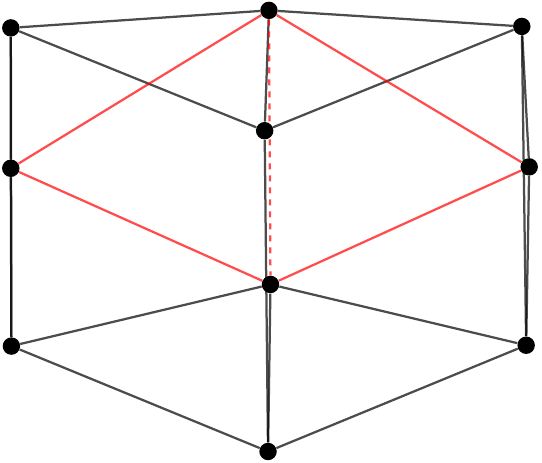}
    \includegraphics[scale=0.4,align=c]{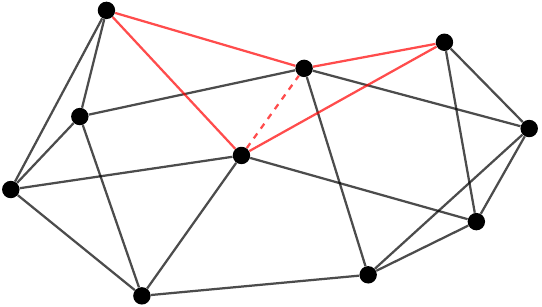}
    \caption{At $\ell=6$, there are 3 $f$-graphs with $(2,3)$ structures: only the middle one has nonzero coefficient, while the other two vanish due to the square rule.}
    \label{fig:M236}
\end{figure}

In the leading order $M_{a,b}^{(\ell=ab)}$, the integrability tells us that $M_{a,b}^{(ab)}$ only has the contribution from the fishnet diagram~\cite{Basso:2017jwq}.
\begin{equation}
    M_{ab}^{(ab)}=(-1)^{(1+a)(1+b-a)}\times\includegraphics[align=c]{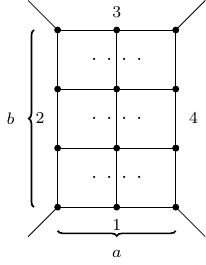}
\end{equation}
However, from the $f$-graph perspective, there are more than one $f$-graph that can potentially contribute to $M_{a,b}^{(\ell=ab)}$, while only one of them does not vanish. For example, at $\ell=6$, there are 3 $f$-graphs with $(2,3)$-structure, but only the middle one in Fig. \ref{fig:M236} has a nonzero coefficient.
For $M_{2,4}^{(8)}, M_{3,3}^{(9)}, M_{2,5}^{(10)}$, there are $60,135,2420$ $f$-graphs contributing to them respectively, but only the one producing the fishnet diagram is non-vanishing. Although up to $\ell=10$, the square rule accounts for all the zero coefficients from this fishnet constraint, there might be some zeros that the square rule cannot explain at higher loops. And the coefficient of the unique $f$-graph contributing to the fishnet is $(-1)^{a(b-a)}$.

In the subleading order $M_{a,b}^{(\ell=ab+1)}$, as indicated in \cite{Caron-Huot:2021usw}, the DCI integrand contributing to $M_{a,b}^{(\ell=ab+1)}$ can be obtained by deforming one of the $2\times2$ sublattice of the regular fishnet. 
The non-vanishing $f$-graphs contributing to the subleading order can be obtained by two rules, {\it i.e.} the inverse square rule and the squeeze rule. The coefficient of other $(ab+1)$-loop $f$-graphs with $(a,b)$ structures vanish.

\paragraph{The inverse square rule} The inverse square rule is related to the square rule; it maps a $\ell$-loop $f$-graph to a $(\ell+1)$-loop one. To be explicit, starting from the $\ell=ab$-loop $f$-graph contributing to the fishnet diagram, we find a $4$-cycle labeled as $1,2,3,4$ in it which does not correspond to the $(a,b)$-structure, then multiply the one-loop box factor $\frac{x_{13}^2x_{24}^2}{x_{1,\ell+5}^2x_{2,\ell+5}^2x_{3,\ell+5}^2x_{4,\ell+5}^2}$ to obtain the new $(5+\ell)$-point $f$-graph.
\begin{equation}\label{eq:is}
\includegraphics[align=c,scale=2]{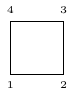}\times\includegraphics[align=c,scale=2]{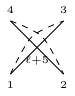}\longrightarrow\includegraphics[align=c,scale=2]{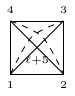}.
\end{equation}
The coefficient of the new graph is equal to the original one. For example, from the $4$-loop $f$-graph contributing to the $2\times2$ fishnet, the inverse square rule gives the following two $f$-graphs contributing to $M_{2,2}^{(5)}$.
\begin{equation}
    \begin{aligned}
&\includegraphics[scale=0.4,align=c]{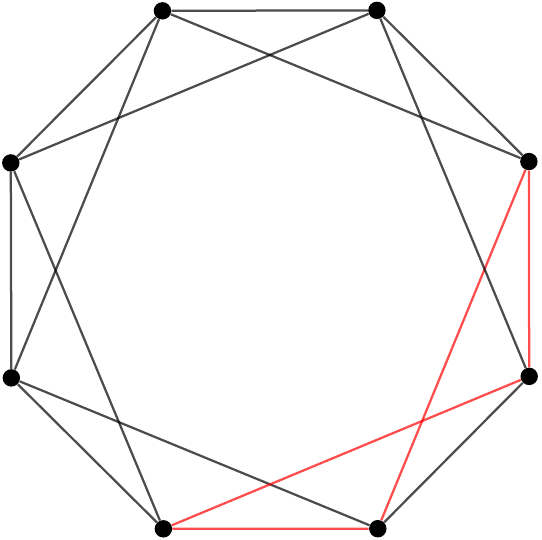}\Rightarrow\includegraphics[scale=0.4,align=c]{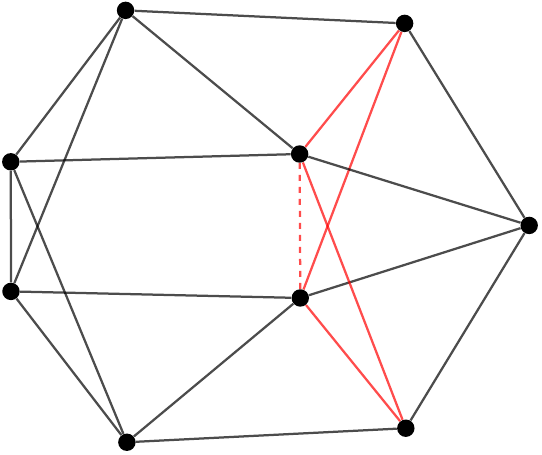}\\
&\includegraphics[scale=0.4,align=c]{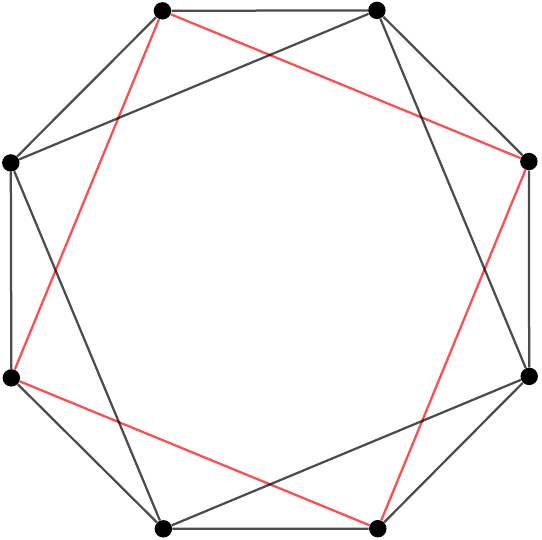}\Rightarrow\includegraphics[scale=0.4,align=c]{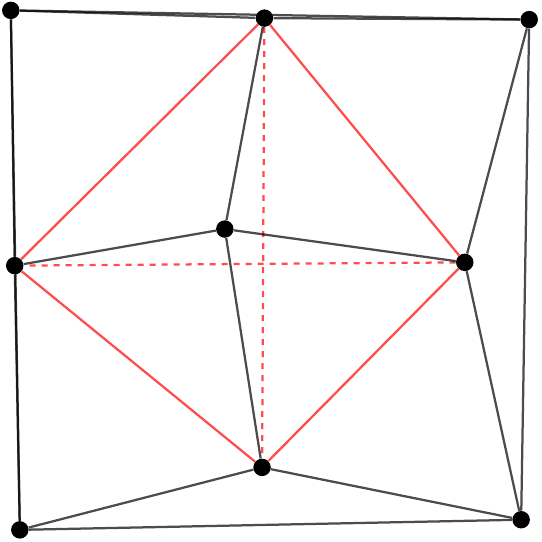}
    \end{aligned}
\end{equation}

\paragraph{The squeeze rule} Meanwhile, the squeeze rule also starts from a $4$-cycle which does not correspond to the $(a,b)$-structure of the $f$-graph, but multiplies a different factor $\frac{x_{12}^2x_{34}^2}{x_{1,\ell+5}^2x_{2,\ell+5}^2x_{3,\ell+5}^2x_{4,\ell+5}^2}$. 
\begin{equation}
\includegraphics[align=c,scale=2]{graph/box.pdf}\times\includegraphics[align=c,scale=2]{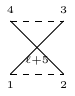}\longrightarrow\includegraphics[align=c,scale=2]{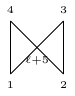}.
\end{equation}
However, the squeeze rule can result in non-planar $f$-graphs, and we only pick up the planar ones. From the $4$-loop $f$-graph contributing to the $2\times2$ fishnet, the squeeze rule gives the following $f$-graph which also contributes to $M_{2,2}^{(5)}$.
\begin{equation}
\includegraphics[scale=0.4,align=c]{graph/M224h2.pdf}\Rightarrow\includegraphics[scale=0.4,align=c]{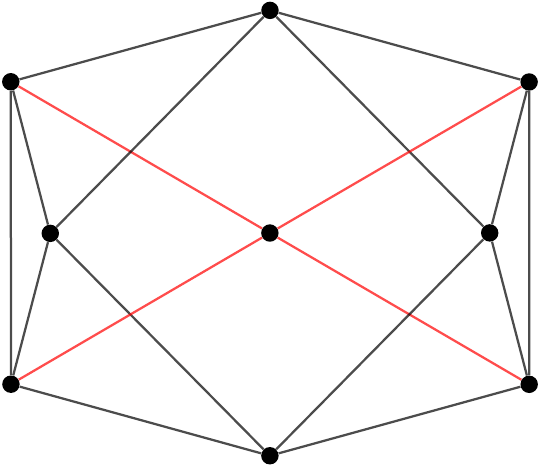},
\end{equation}
One can check that when applied to the $f$-graph contributing to the fishnet integral, the coefficient of the $f$-graph obtained by squeezing equals the minus of the original one, which can be checked up to $\ell=11$. 

\section{Magic identities for DCI integrals}\label{sec:magic}

In this section, we study magic identities for DCI integrals by equating their combinations ($M_{a,b}$) with octagon results as determinants of ladders. We will see that once we obtain a magic identity at a given loop, it implies an infinite family of higher-loop identities, which can in turn be used to simplify new magic identities. We enumerate such simplified magic identities through six loops. 

\subsection{Review: from determinants of ladders to magic identities}
Now we move onto the Octagons at week coupling, which appears on the left-hand side of \eqref{eq:OM}. The hexagonalization~\cite{Fleury:2016ykk, Eden:2016xvg} procedure allows us to uplift the free correlator to the correlator at finite coupling. In this procedure, the octagon can be decomposed into two hexagons by inserting a complete set of states along a mirror cut. This cut can be considered as stretching between the hexagons connected by the bridge of length $l$. 
\begin{equation}
    \mathbb{O} = \mathbb{O}_{0} + \sum_{l=1}^{\infty}(d_{13})^{l}\mathbb{O}_{l} +(d_{24})^{l}\mathbb{O}_{l},
\end{equation}
where $\mathbb{O}_l$ can be expressed as a sum over the number of
particles $n$~\cite{Coronado:2018ypq}
\begin{equation}\label{eq:ObI}
    \mathbb{O}_{l}(z,\bar{z},d_{13},d_{24})= 1+\sum_{n=1}^{\infty}(1-d_{13}d_{24})^{n}I_{n,l}(z,\bar{z}).
\end{equation}
where $z,\bar{z}$ are defined via two cross ratios $u,v$
\begin{equation}
    u\equiv \frac{x_{12}^{2}x_{34}^{2}}{x_{13}^{2}x_{24}^2}\equiv z\bar{z}, \, v\equiv \frac{x_{14}^{2}x_{23}^{2}}{x_{13}^{2}x_{24}^2}\equiv (1-z)(1-\bar{z}).
\end{equation}

At weak coupling,  the $n$-particle contribution $I_{n,l}$ is related to the determinant of ladder integrals as a generalization of the Basso-Dixon fishnet \cite{Basso:2017jwq}. Recall that the integrated result of $p$-loop ladder integral is~\cite{Usyukina:1993ch}
\begin{equation}
    F_{p}(z,\bar{z}) =
(-1)\sum_{j=p}^{2p}\frac{j!\left[-\log(\frac{z}{z-1}\frac{\bar{z}}{\bar{z}-1})\right]^{2p-j}}{p!(j-p)!(2p-j)!}\left[\frac{\text{Li}_{j}(\frac{z}{z-1})-\text{Li}_{j}(\frac{\bar{z}}{\bar{z}-1})}{z-\bar{z}}\right].
\end{equation}
The determinant of ladder integrals is defined as 
\begin{equation}
    F_{i_{1},i_{2},\cdots, i_{n}} =\prod_{m=1}^{n}\frac{1}{i_{m}!(i_{m}-1)!}  \begin{vmatrix}
 f_{i_{1}} & f_{i_{2}-1}  & \cdots  & f_{i_{n}-n+1} \\
f_{i_{1}+1} &  f_{i_{2}} & \cdots  & f_{i_{n}-n+2}\\
\vdots & \vdots & \ddots & \vdots\\
f_{i_{1}+n-1} & f_{i_{2}+n-2} & \dots &  f_{i_{n}}\\
\end{vmatrix}\qquad\text{with }f_{p} =p!(p-1)! F_{p}.
\end{equation}
$I_{n,l}$ in \eqref{eq:ObI} can be written as a linear combination of fishnet integrals perturbatively
\begin{equation}\label{eq:lslI}
    I_{n,l}=\sum_{\ell=l(n+l)}^\infty (-g^{2})^{\ell}I_{n,l}^{(\ell)}
    =\sum_{\ell=l(n+l)}^\infty(-g^{2})^{\ell}(-1)^{nl}\times\sum_{\substack{i_{1}+\cdots + i_{n}=\ell \\ \text{with }(i_{p+1}-i_{p})\geq 2\text{ and }i_{1}> l}} c_{l;\{i\}_{n}}F_{i_{1},i_{2},\cdots,i_{n}}
\end{equation}
where $c_{l;\{i\}}$ are some coefficients. Although these coefficients do not have a closed form generally, the leading and subleading contribution of $I_{n,l}$ is conjectured in \cite{Caron-Huot:2021usw}
\begin{equation}
\begin{aligned}
    &I^{(n(n+{l}))}_{n,{l}} =(-1)^{nl}F_{1+{l},3+{l},\cdots,2n-1+{l}},\\
    &I^{(n(n+{l})+{1})}_{n,{l}} = (-1)^{nl}2(2n-1+l)F_{1+{l},3+{l},\cdots,2n-3+l,2n-1+l+1},
\end{aligned}
\end{equation}
where $F_{1+{l},3+{l},\cdots,2n-1+{l}}$ is just the $(n+l)\times n$ fishnet integral. And for later use,  $I_{2,0}^{(6)}=28F_{1,5}+\frac{18}{5}F_{2,4}$.

Notice that $\mathbb{O}^{\text{free}}=1+\sum_{l=1}^\infty (d_{13})^l+\sum_{l=1}^\infty (d_{24})^l$, we can collect the dependence of $d_{13},d_{24}$ of $\frac{\mathbb{O}}{\mathbb{O}^{\text{free}}}$ as
\begin{equation}\label{eq:OOI}
     \frac{\mathbb{O}}{\mathbb{O}^{\text{free}}}
     =1+\sum_{a,b\geq 1}(1-d_{13})^a(1-d_{24})^b\sum_{n=\min(a,b)}^{a+b-1}\sum_{\substack{l\geq0\\ l+n\geq a,b}}^{\infty}C^{n,l}_{a,b}I_{n,l},
\end{equation}
where
\begin{equation}
    C^{n,l}_{a,b} =\frac{(-1)^{a+b-n-1}}{1+\delta_{l,0}}\left[\binom{n-1}{a-1}\binom{a+l-1}{a+b-n-1}+(a\leftrightarrow b)\right].
\end{equation}
We list the result of independent $M_{a,b}$'s up to $\ell=6$ in Table \ref{tab:mab}.

\begin{table}[htbp]
  \hspace{-15cm}
  \includegraphics[scale=0.35,align=l]{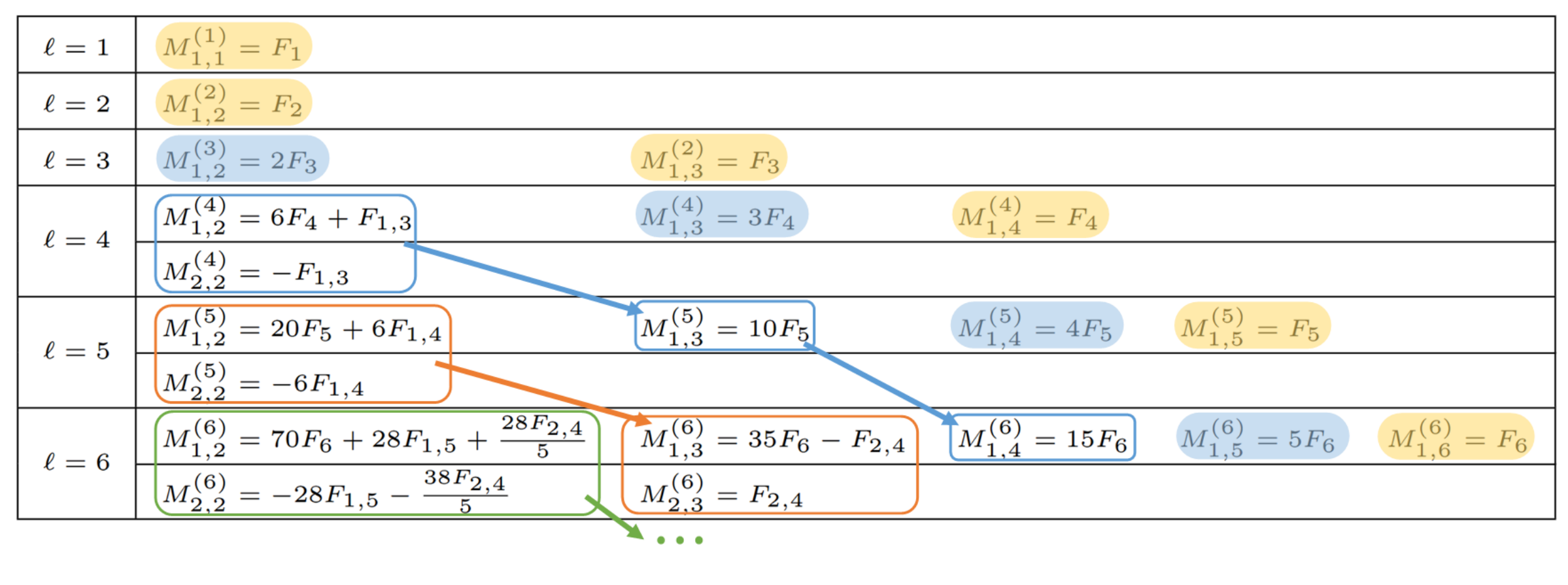}
  \vspace{-1.5em}
    \caption{Result of $M_{a,b}^{(\ell)}$ up to $\ell=6$. The colored highlights and arrows indicate a diagonal structure, which is explained in sec.\,\ref{sec:mab}. The yellow and blue highlight means the results can be obtained by ladder-type magic identities. The arrows mean the ``inverse-boxing" of lower loop magic identity can be used to simplify the higher loop ones.}
  \label{tab:mab}%
\end{table}%

According to \eqref{eq:deriveab} and the discussion between \eqref{eq:M3} and \eqref{eq:deriveab}, $M_{a,b}^{(\ell)}$ can be written formally as 
\begin{equation}\label{eq:PM}
    M_{a,b}^{(\ell)}=\sum_{i=1}^{\mathcal N_{\ell}}c_{i}^{(\ell)}\mathbb{P}_{a,b}f_{i}^{(\ell)},
\end{equation}
where the operator $\mathbb{P}_{a,b}$ is defined as finding all $(a,b)$-structures (which are defined in Fig.~\ref{fig:ab}) in a given $f$-graph, that is,
\begin{equation}
    \mathbb{P}_{a,b}f_{i}^{(\ell)}=\left\{\begin{array}{cl}
        \displaystyle \sum_{\text{all } (a,b)\text{-structure}}\lim_{x_{i,i+1}^2\to 0}\xi_{4}f_{i}^{(\ell)} &  \text{if $f_{i}^{(\ell)}$ contains at least one $(a,b)$-structure},\\
        0 &  \text{if $f_{i}^{(\ell)}$ does not contain any $(a,b)$-structure}.
    \end{array}\right.
\end{equation}
For example, in $\ell=6$, only $f$-graphs which contain at least one square with a dashed diagonal (marked red in Fig.~\ref{fig:M236}) contribute to $M_{2,3}^{(6)}$. The first one in Fig.~\ref{fig:M236} contains two such structures. Other $f$-graphs in $\ell=6$ do not contribute ({\it i.e.} they project to 0 for $M_{2,3}^{(6)}$). $\lim_{x_{i,i+1}^2\to 0}\xi_{4}$ is used to transform $f$-graphs to DCI integrals.
Further comparing \eqref{eq:OOI} to \eqref{eq:OM} and \eqref{eq:M3}, we finally obtain the following identity 
\begin{equation}\label{eq:gma}
   \sum_{i=1}^{\mathcal N_{\ell}}c_{i}^{(\ell)}\mathbb{P}_{a,b}f_{i}^{(\ell)}=\sum_{n=\min(a,b)}^{a+b-1}\sum_{\substack{l\geq0\\ l+n\geq a,b}}^{\infty}C^{n,l}_{a,b}I_{n,l}^{(\ell)}.
\end{equation}
The above equation not only gives us some all-loop predictions about coefficients of $f$-graphs, but also teaches us new magic identities among multi-loop conformal integrals (listed through six loops).

\subsection{Magic identities through six loops}\label{sec:mab}
In this subsection, we derive new relations between the DCI integrals from \eqref{eq:gma} and give some results to all loops.

For $M_{1,\ell}^{(\ell)}$, it is not hard to see that only one $f$-graph contributes, since it is the only graph that contains a $(1,\ell)$-structure. Thus, $M_{1,\ell}^{(\ell)}=F_{\ell}$ for any $\ell$. (The yellow highlight in Table.\,\ref{tab:mab}.)
\begin{equation}
M_{1,\ell}^{(\ell)}=\includegraphics[scale=0.4,align=c]{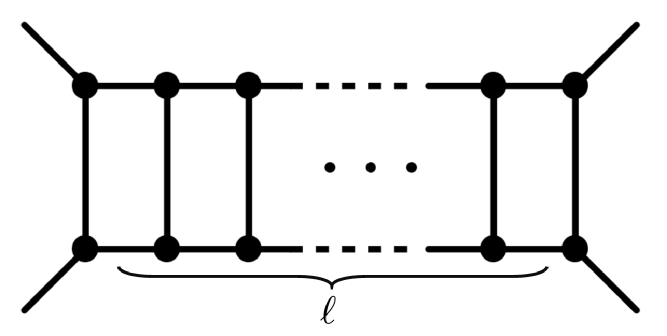}=F_{\ell}.
\end{equation}

Then for $M_{1,\ell-1}^{(\ell)}$, the contributing integrals are obtained by adding a ``rung" to an $(l-1)$-loop ladder, which are all equal to the $l$-loop ladder integral through ladder-type magic identity. This means $M_{1,\ell-1}^{(\ell)}=(\ell-1)F_{\ell}$. (The blue highlight in Table.\,\ref{tab:mab}.) Take $\ell=3$ and $\ell=4$ for example:
\begin{equation}
    M_{1,2}^{(3)}=\includegraphics[scale=0.45,align=c]{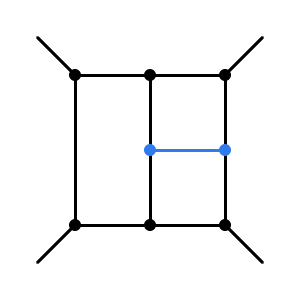}+\text{Dih.}=2 F_3,
\end{equation}
\begin{equation}
    M_{1,3}^{(4)}=\includegraphics[scale=0.45,align=c]{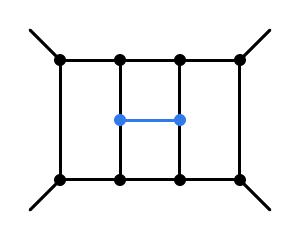}+\includegraphics[scale=0.45,align=c]{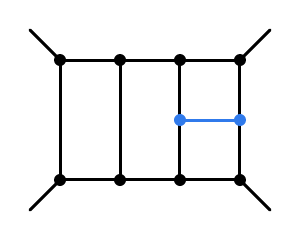}+\text{Dih.} = \mathcal{I}_3^{(4)} + \mathfrak{p}_{2a} \, \mathcal{I}_4^{(4)} = 3 F_4,
\end{equation}
where ``Dih." means all in-equivalent dihedral permutation images of 4 external legs. The definition of the integral is present in Appendix \ref{app:definition} and the permutation operator is defined as $\mathfrak{p}_{2a}X\equiv X+(1{\leftrightarrow}3)$. For later convenience, we also define $\mathfrak{p}_{2b}X\equiv X+(2{\leftrightarrow}4)$ and $\mathfrak{p}_{4}\equiv \mathfrak{p}_{2a}\mathfrak{p}_{2b}$. 

Starting from $M_{1,\ell-2}^{(\ell)}$ (blue arrows in Table.\,\ref{tab:mab}), we have contributions beyond ladder-equivalent integrals, which result in the non-trivial magic identities we are interested in. However, we can still ``count" and ``enumerate" the ladder-equivalent contributions for $M_{1,\ell-2}^{(\ell)}$. In this case, we should add two rungs to an $(l-2)$-loop ladder while maintaining the $(1,\ell-2)$-structure. It is easy to see that the ladder contribution to $M_{1,\ell-2}^{(\ell)}$ is $C_\ell^2 F_\ell$. Explicitly, for $\ell=4$ and $\ell=5$, we have:
\begin{equation}\label{eq:M12(4)ladders}
    M_{1,2}^{(4)} \supset \includegraphics[scale=0.45,align=c]{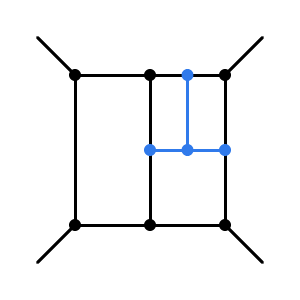}+\includegraphics[scale=0.45,align=c]{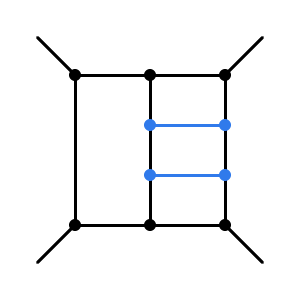}+\text{Dih.} = \mathfrak{p}_{4} \, \mathcal{I}_1^{(4)} +\mathfrak{p}_{2a} \, \mathcal{I}_7^{(4)} = 6 F_4,
\end{equation}
\begin{align}
   M_{1,3}^{5} & \supset \hspace{-0.5em} \includegraphics[scale=0.45,align=c]{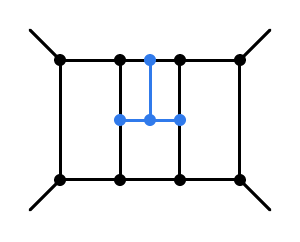} \hspace{-0.5em}+\hspace{-0.5em}\includegraphics[scale=0.45,align=c]{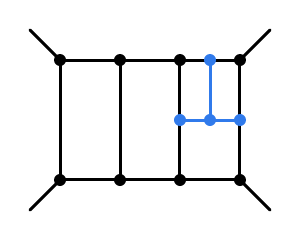} \hspace{-0.5em}+\hspace{-0.5em}\includegraphics[scale=0.45,align=c]{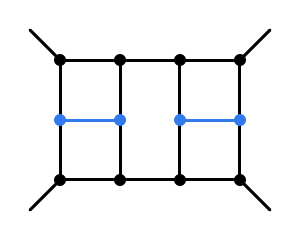}
    \hspace{-0.5em}+\hspace{-0.5em} \includegraphics[scale=0.45,align=c]{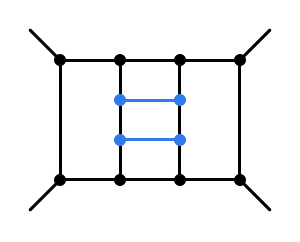} \hspace{-0.5em}+ \hspace{-0.5em}\includegraphics[scale=0.45,align=c]{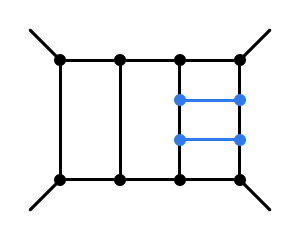} \hspace{-0.5em} + \text{Dih.} \nonumber \\
    & = \mathfrak{p}_4 \, \mathcal{I}_{11}^{(5)} + \mathfrak{p}_{2b} \, \mathcal{I}_{13}^{(5)} + \mathcal{I}_{15}^{(5)} + \mathcal{I}_{17}^{(5)} + \mathfrak{p}_{2a} \, \mathcal{I}_{19}^{(5)} = 10 F_5.
\end{align}
The first non-trivial magic identity starts from $\ell=4$ and has been given in \cite{Caron-Huot:2021usw}. Note that $M_{1,2}^{(2)}+M_{2,2}^{(2)}=6F_4$, and the ladder-equivalent contribution in \eqref{eq:M12(4)ladders} corresponds exactly to $6F_4$. Thus, we are left with
\begin{equation}\label{eq:ma4}
 0=\includegraphics[scale=0.45,align=c]{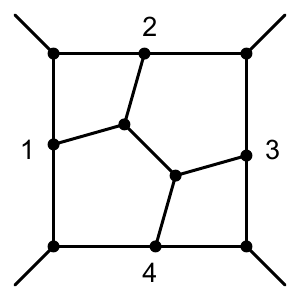} - \includegraphics[scale=0.45,align=c]{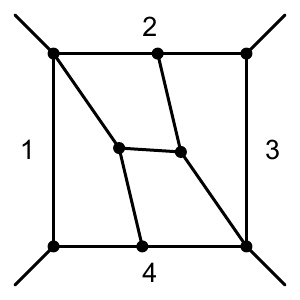} - \includegraphics[scale=0.45,align=c]{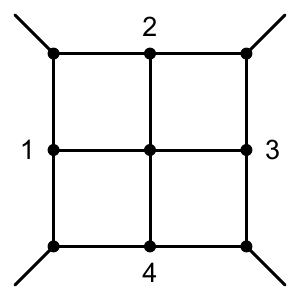} + \text{Dih.} = 
 \mathfrak{p}_{2a} \, \mathcal{I}^{(4)}_{2}-\mathfrak{p}_{2b} \, \mathcal{I}^{(4)}_5-\mathcal{I}^{(4)}_6.
\end{equation}
By attaching a box (we call this operation ``inverse-boxing") to this magic identity, we get higher loop magic identities for $M_{1,\ell-2}^{(\ell)}$. And they are exactly magic identities stemming from $M_{1,\ell-2}^{(\ell\geq 5)}$ after subtracting all ladder-equivalent integrals. For $\ell=5$, the new magic identity reads
\begin{equation}\label{eq:ma45}
    0= \includegraphics[scale=0.45,align=c]{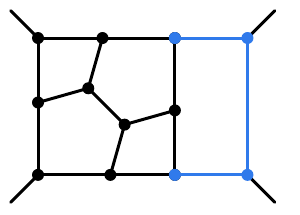}\hspace{-0.5em}-\includegraphics[scale=0.45,align=c]{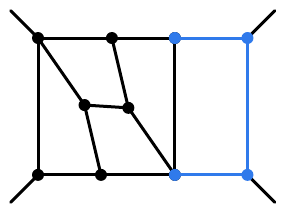}\hspace{-0.5em}-\includegraphics[scale=0.45,align=c]{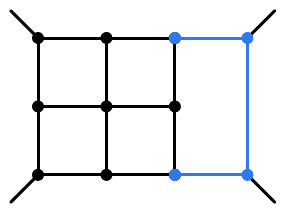} \hspace{-0.5em}+ \text{Dih.} = \mathfrak{p}_{4}\left( \mathcal{I}_{2}^{(5)} - \mathcal{I}_{30}^{(5)} \right) - \mathfrak{p}_{2a} \, \mathcal{I}_{8}^{(5)}.
\end{equation}
For $\ell=6$, we can group the integrals by the places where the boxes are attached.
\begin{equation}
   \begin{aligned}
    0=&\left(\includegraphics[scale=0.45,align=c]{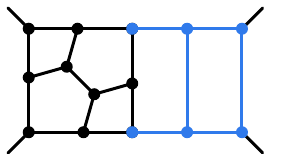}-\includegraphics[scale=0.45,align=c]{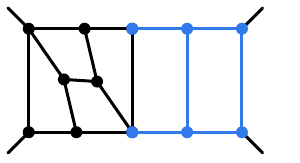}-\includegraphics[scale=0.45,align=c]{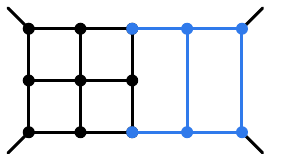} + \text{Dih.} \right)\\
    + &\left(\includegraphics[scale=0.45,align=c]{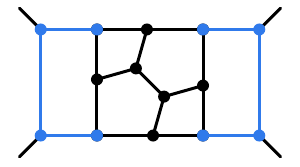}-\includegraphics[scale=0.45,align=c]{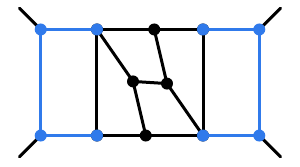}-\includegraphics[scale=0.45,align=c]{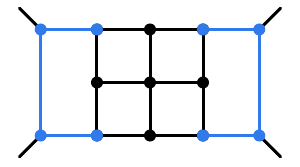} + \text{Dih.} \right)
   \end{aligned}
\end{equation}
And it is easy to generalize this kind of magic identity to higher loops by inverse-boxing.

Now we arrive at $M_{1,\ell-3}^{(\ell)}$ (orange arrows in Table.\,\ref{tab:mab}). Let us first look at $\ell=5$. In Table.\,\ref{tab:mab}, notice that $M_{1,2}^{(5)} +M_{2,2}^{(5)} = 20 F_5$, where all ladder-equivalent integrals contribute $20 F_5$. Subtracting them gives the $5$-loop magic identity already mentioned in \cite{Caron-Huot:2021usw}. We rewrite it in our notation as
\begin{align} \label{eq:ma5}
   0 & = \includegraphics[scale=0.45,align=c]{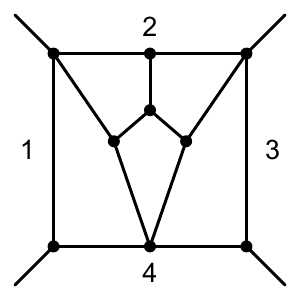} - \includegraphics[scale=0.45,align=c]{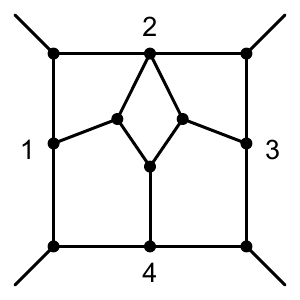} + \includegraphics[scale=0.45,align=c]{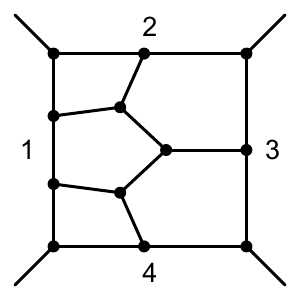} + \includegraphics[scale=0.45,align=c]{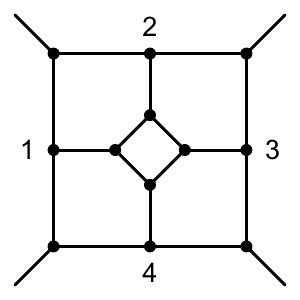} - \includegraphics[scale=0.45,align=c]{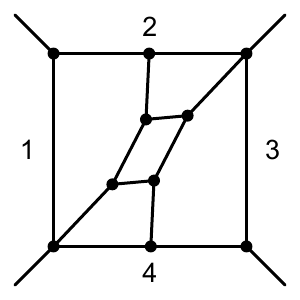} \nonumber \\ 
   & + \includegraphics[scale=0.45,align=c]{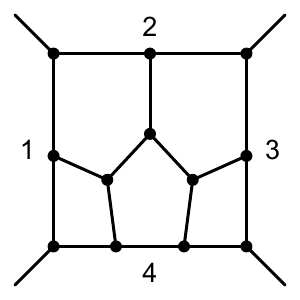} + \includegraphics[scale=0.45,align=c]{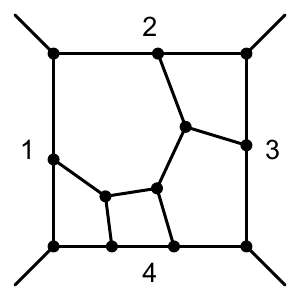} - \includegraphics[scale=0.45,align=c]{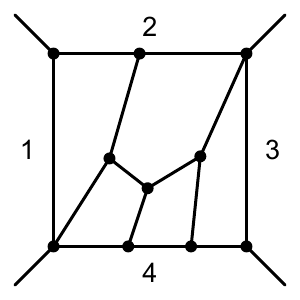} - \includegraphics[scale=0.45,align=c]{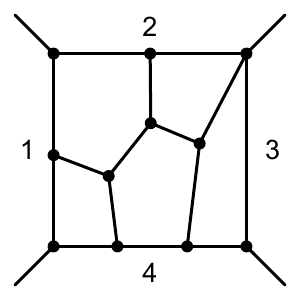}  + \includegraphics[scale=0.45,align=c]{graph/DCI5l14.pdf}\nonumber \\
   & + \includegraphics[scale=0.45,align=c]{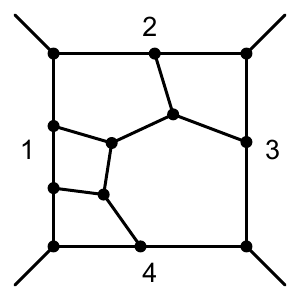} - \includegraphics[scale=0.45,align=c]{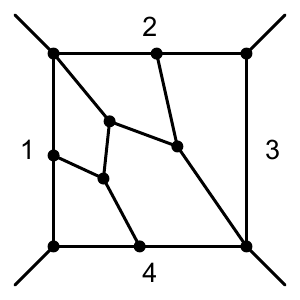} - \includegraphics[scale=0.45,align=c]{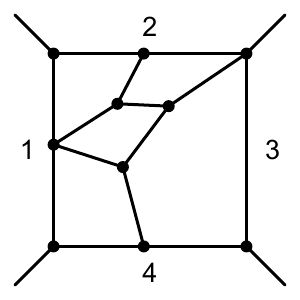} - \includegraphics[scale=0.45,align=c]{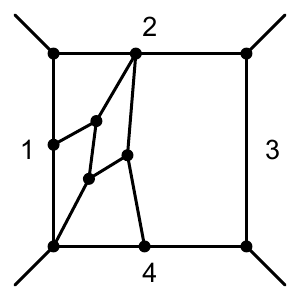} \nonumber \\
    & + \includegraphics[scale=0.45,align=c]{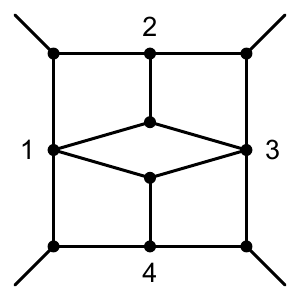} - \includegraphics[scale=0.45,align=c]{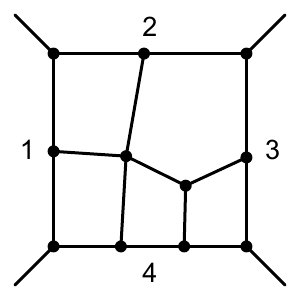} - \includegraphics[scale=0.45,align=c]{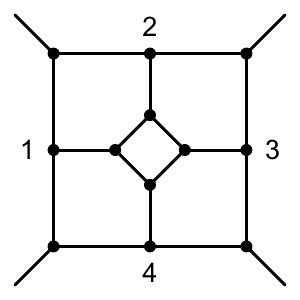} + \text{Dih.} \nonumber \\
    & = \, \mathcal{I}^{(5)}_{33} - \mathcal{I}^{(5)}_{7} + \mathcal{I}^{(5)}_{14} + \mathcal{I}^{(5)}_{1} - \mathcal{I}^{(5)}_{25} + \mathcal{I}^{(5)}_{13} + \mathcal{I}^{(5)}_{12} - \mathcal{I}^{(5)}_{24} - \mathcal{I}^{(5)}_{6} + \mathcal{I}^{(5)}_{4} \nonumber \\
     & \hspace{1em} + \mathcal{I}^{(5)}_{18} - \mathcal{I}^{(5)}_{26} - \mathcal{I}^{(5)}_{28} - \mathcal{I}^{(5)}_{29} + \mathcal{I}^{(5)}_{34} - \mathcal{I}^{(5)}_{27} - \mathcal{I}^{(5)}_{5} + \text{Dih.},
\end{align}
where the diagrams only show the denominators, and the corresponding numerators are defined explicitly in Appendix\,\ref{app:definition}. We will use the inverse-boxing of the above equation to simplify a magic identity at $\ell=6$. Explicitly, after subtracting the ladder-equivalent integrals from $M_{1,3}^{(6)}+M_{2,3}^{(6)}=35 F_6$ \footnote{The ladder-equivalent integrals contribute $36F_6$, that's why the LHS. of \eqref{eq:ma6} is nonzero.}, we get an identity involving 125 integrals (37 dihedral seeds). Applying inverse-boxing from different direction to \eqref{eq:ma5}, \textit{i.e.} attaching box as in \eqref{eq:ma45}, we will have a 6-loop identity. After subtracting it from the identity involving 125 integrals, we have the following nice identity:
\begin{equation}\label{eq:ma6}
\begin{aligned}
     -F_6 = & - \includegraphics[scale=0.45,align=c]{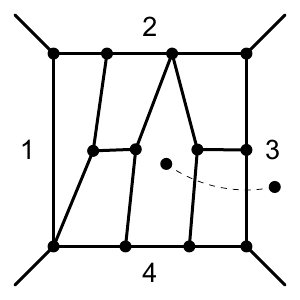} - \includegraphics[scale=0.45,align=c]{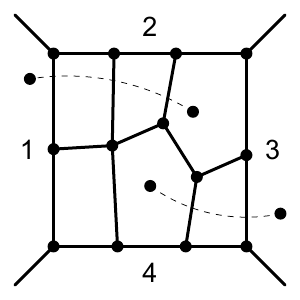} + \includegraphics[scale=0.45,align=c]{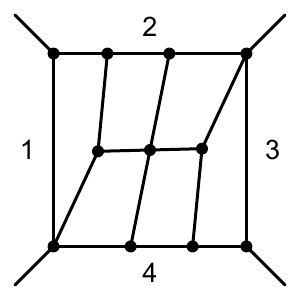} + \includegraphics[scale=0.45,align=c]{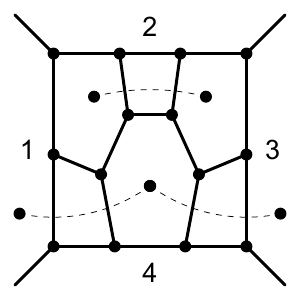}\\
    & + \includegraphics[scale=0.45,align=c]{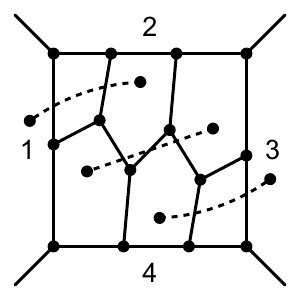} + \includegraphics[scale=0.45,align=c]{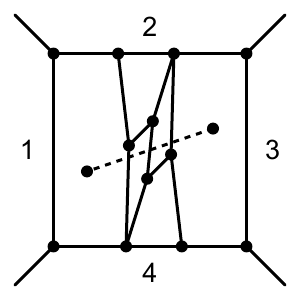} - \includegraphics[scale=0.45,align=c]{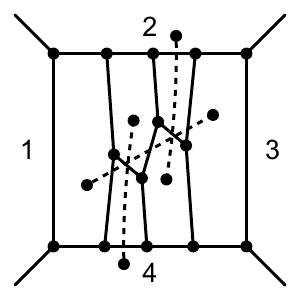} + \includegraphics[scale=0.45,align=c]{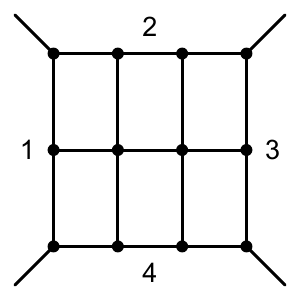} + \text{Dih.}
\end{aligned}
\end{equation}
which only involves 8 integral dihedral seeds, where the dashed lines correspond to numerators. From $70F_5-2F_{2,4} = M_{1,2}^{(6)} +M_{2,2}^{(6)}$, we will have another magic identity that is too long to be explicitly shown. The last comment is that, from the above analysis, we can detect a diagonal structure shown in Table.\,\ref{tab:mab}, which indicates the inverse-boxing of magic identity stemming from lower loop can be used to simplify the higher loop ones when they lie on the same diagonal.

\subsection{Comments on periods and integrated correlators}
Before proceeding, we briefly comment on the periods of conformal integrals, which have been studied quite extensively in the literature~\cite{Broadhurst:1995km, Schnetz:2008mp, Brown:2009ta, Brown:2013gia, Schnetz:2013hqa, Panzer:2016snt, Panzer:2014caa, Borinsky:2020rqs, Borinsky:2022lds} partly due to the connection to integrated correlator~\cite{Binder:2019jwn,Chester:2019pvm,Chester:2020vyz,Dorigoni:2021guq,Dorigoni:2022zcr,Paul:2022piq,Wen:2022oky,Brown:2023zbr}. Recall that for an integrand $I^{(\ell)}$ with $(4+\ell)$ vertex\footnote{Here $I^{(\ell)}$ denotes the integrand of DCI integrals defined in the appendix. Note that we always suppress a factor $1/(x_{13}^2x_{24}^2)$ in the definition. It should be distinguished from $I_{n,l}^{(l)}$ in the octagons, which always has two indices in the subscript.}, its period is defined as integrating over all space-time coordinates with a specific measure~\cite{Wen:2022oky,Brown:2023zbr}
\begin{equation}
    \mathcal{P}(I^{(\ell)})=\frac{1}{\pi^{2(\ell+1)}}\int\frac{\mathrm{d}^4x_1\cdots\mathrm{d}^4x_{\ell+4}}{\mathrm{vol}(SO(2,4))}\frac{I^{(\ell)}}{x_{12}^{2}x_{23}^{2}x_{34}^{2}x_{41}^{2}x_{13}^{2}x_{24}^2}
\end{equation}
where $\mathrm{vol}(SO(2,4))$ allows us to fix three $x_i$. Since we are integrating over all space-time coordinates, periods of all conformal integrals coming from one $f$-graph are equal. So we only need to consider periods of $f$-graphs, which is related to the periods of integrals by
\begin{equation}
    \mathcal{P}(f_{i^\prime}^{(\ell)})=\frac{(4+\ell)!}{\left|\mathrm{aut}(f_{i^\prime}^{(\ell)})\right|}\mathcal{P}(I_i^{(\ell)}),
\end{equation}
where $I_{i}^{(\ell)}$ is one of the conformal integrals from the $f$-graph $f_{i^\prime}^{(\ell)}$.

As perhaps the simplest constraints, the periods of integrals related by the original magic identity are all equal. Now we consider the ``slingshot'' \eqref{eq:shot} rule at the $f$-graph level: recall that integrals ${I}_i^{(\ell)}$ can be obtained from certain $f$-graph $f_{i^\prime}^{(\ell)}$ by multiplying $x_{12}^2x_{23}^2x_{34}^2x_{41}^2x_{13}^2x_{24}^2$, \eqref{eq:shot} can be translated to
\begin{equation}  
\begin{aligned}
f_{j^\prime}^{(\ell+1)}
&=\frac{1}{x_{\ell+5,2}^2x_{23}^2x_{34}^2x_{4,\ell+5}^2x_{\ell+5,3}^2x_{24}^2}\times \frac{x_{24}^2}{x_{12}^2x_{41}^2x_{\ell+5,1}^2}\times \left(x_{12}^2x_{23}^2x_{34}^2x_{41}^2x_{13}^2x_{24}^2\right)f_{i^\prime}^{(\ell)}\\
&=\frac{x_{13}^2x_{24}^2}{x_{\ell+5,1}^2x_{\ell+5,2}^2x_{\ell+5,3}^2x_{\ell+5,4}^2}f_{i^\prime}^{(\ell)},
\end{aligned}
\end{equation}
which is just the inverse square rule \eqref{eq:is}. It should be noted that the four-cycle $1-2-3-4$ in $f_{i^\prime}$ corresponds to the DCI integrals involved in the $\ell$-loop magic identity. The periods of all $f^{(\ell+1)}_{j^\prime}$ are equal up to the symmetry factor.

Moreover, by integrating over all $x_i$ in \eqref{eq:gma}, we obtain some constraints between the periods of $f$-graphs.
\begin{equation}\label{eq:gmap}
   \sum_{i=1}^{\mathcal N_{\ell}}c_{i}^{(\ell)}\left|\mathrm{aut}(f_i^{(\ell)})\right|\left|\mathbb{P}_{a,b}\right|\mathcal{P}(f_{i}^{(\ell)})=\sum_{n=\min(a,b)}^{a+b-1}\sum_{\substack{l\geq0\\ l+n\geq a,b}}^{\infty}C^{n,l}_{a,b}\mathcal{P}(I_{n,l}^{(\ell)}),
\end{equation}
where $\left|\mathbb{P}_{a,b}\right|$ is the number of $(a,b)$-structures in the $f$-graph. For example, at four loops, $\mathcal{I}^{(4)}_2$, $\mathcal{I}^{(4)}_5$, and $\mathcal{I}^{(4)}_6$ in \eqref{eq:ma4} come from $f^{(4)}_1,f^{(4)}_2,f^{(4)}_2$, thus we have
\begin{equation}
    2\times 8\times \mathcal{P}(f_1^{(4)})-2\times 16\times \mathcal{P}(f_2^{(4)})= 16\times \mathcal{P}(f_2^{(4)}),
\end{equation}
which can be easily verified since $\mathcal{P}({f_1^{(4)}})=\frac{8!}{8}\times 252\zeta_9,\mathcal{P}({f_2^{(4)}})=\frac{8!}{16}\times 168\zeta_9$. Similarly, the constraints between periods of five-loop $f$-graphs reduce all seven periods to three independent ones~\cite{Wen:2022oky}. For example, the identity $M_{2,2}^{(5)}=-6F_{1,4}$ in Table.~\ref{tab:mab} gives a linear relation (see also \eqref{eq:fivelooprelation}) among $\mathcal{P}(f_{1}^{(5)})$, $\mathcal{P}(f_{5}^{(5)})$ and $\mathcal{P}(f_{7}^{(5)})$ (the order of five-loop $f$-graphs is defined as in \cite{Eden:2012tu}):
\begin{equation}
    -\frac{8}{9!}\mathcal{P}(f_{5}^{(5)})+2\times\frac{12}{9!}\mathcal{P}(f_{7}^{(5)})-8\times \frac{2}{9!}\mathcal{P}(f_{1}^{(5)})=-5\times 924\zeta_{11}.
\end{equation}
This relation supplements the constraint from localization result of the integrated correlator presented in \cite{Wen:2022oky}. For periods of 26 six-loop $f$-graphs with non-zero coefficients, the relations between Coulomb branch amplitudes and octagons also play a key role in determining some of them. Here we take $M_{1,3}^{(6)}+M_{2,3}^{(6)}=35F_6$ as an example, which is simplified to \eqref{eq:ma6}. Since the integrated result of $F_p$ is given by $C_{2p+2}^{p+1} \zeta_{2p+1}$, after lifting the DCI integrals on the RHS of \eqref{eq:ma6} to $f$-graphs, we obtain the following constraint on the integrated result:
\begin{equation}
\begin{aligned}
    -4\times\frac{1}{10!}\mathcal{P}(f_{25}^{(6)})
    -4\times\frac{2}{10!}\mathcal{P}(f_{17}^{(6)})
    +2\times\frac{2}{10!}\mathcal{P}(f_{8}^{(6)})
    +2\times\frac{2}{10!}\mathcal{P}(f_{2}^{(6)})
    +2\times\frac{4}{10!}\mathcal{P}(f_{1}^{(6)})\\
    +2\times\frac{2}{10!}\mathcal{P}(f_{23}^{(6)})
    -2\times\frac{2}{10!}\mathcal{P}(f_
    {6}^{(6)})
    +\frac{4}{10!}\mathcal{P}(f_{14}^{(6)})= -3432 \zeta_{13},
\end{aligned}
\end{equation}
which provides a non-trivial linear relation between $\mathcal{P}(f_{1}^{(6)})$ and other periods. It supplements the constraints derived from the simplification of correlators by substitution $d_{ij}\to\gamma_{i}\gamma_{j}$, as described in \cite{Brown:2023zbr}. It also serves as a consistency check of our results, as we already know the periods of these $f$-graphs from \cite{Brown:2023zbr}.

\section{Bootstrapping DCI integrals at four loops and beyond }\label{sec:bootstrap}

In this section, we turn our attention to individual DCI integrals (as opposed to those special combinations studied above), which have been studied systematically only up to three loops~\cite{Drummond:2013nda}. Starting at four loops, although some of these DCI integrals are rather trivial, {\it e.g.} equivalent to the ladder integral (or generalized ladders~\cite{Usyukina:1993ch,Broadhurst:2010ds,Drummond:2012bg}), the remaining ones have not been computed with a few exceptions such as fishnet integrals~\cite{Basso:2017jwq,Basso:2021omx,Aprile:2023gnh}, {\it e.g.} the four-loop combination $M_{2,2}$ consists of a single integral denoted as ${\cal I}^{f2}$ in~\cite{Caron-Huot:2021usw}. As we have mentioned, some four-loop conformal integrals involve elliptic cuts which prevent us from computing all of them, but it turns out to be much easier for DCI integrals at four and even higher loops. 

Based on our analysis of leading singularities and the very restrictive function space, we will see that it is straightforward to bootstrap the $3$ four-loop DCI integrals that are not equal to the ladder integral, which also provides an independent check for the four-loop magic identity. Moreover, we will see that the same bootstrap method can be applied to five-loop DCI integrals. We will again enumerate all possible leading singularities and also obtain some results for the pure functions, though the latter become more complicated: there are also lower-weight pieces for some of these integrals, and in general they involve functions beyond harmonic polylogarithms. 

\subsection{All four-loop DCI integrals}
Among the magic identities we have studied in sec.~\ref{sec:magic}, there is a simple enough but nontrivial one
\begin{equation}\label{eq:magicf}
    \mathcal{I}^{d2}(u,v)+\mathcal{I}^{d2}(v,u)+\mathcal{I}^{f2}(u,v)=\mathcal{I}^{f}(u,v)+\mathcal{I}^{f}(v,u).
\end{equation}
Here, to maintain consistency with the notation in \cite{Caron-Huot:2021usw}, we rename these functions as follows.
\begin{equation}
    \mathcal{I}^{d2}\equiv \mathcal{I}^{(4)}_{5}, \, \mathcal{I}^{f2}\equiv \mathcal{I}^{(4)}_{6}, \, \mathcal{I}^{f}\equiv \mathcal{I}^{(4)}_{2}.
\end{equation}
The arguments of $\mathcal{I}^{f}(u,v)$ specify the order of the external legs. For example, $\mathcal{I}^{f}(v,u)$ indicates that the external legs $1$ and $3$ are exchanged in the original definition of $\mathcal{I}^{f}(u,v)$. 
Although the analytic expressions for $\mathcal{I}^{d2}$ and $\mathcal{I}^{f2}$ (the fishnet) have already been calculated and involve only SVHPL functions~\cite{Drummond:2013nda,Eden:2016dir}, $\mathcal{I}^{f}(u,v)$ remains unknown. \eqref{eq:magicf} gives the combination of $\mathcal{I}^{f}(u,v)$ and $\mathcal{I}^{f}(v,u)$, offering constraints on bootstrapping the analytic form of $\mathcal{I}^{f}$. The above magic identity suggests that $\mathcal{I}^{f}$ should also be expressible in terms of SVHPLs with the same prefactors in $\mathcal{I}^{d2}$ and $\mathcal{I}^{f2}$. Furthermore, the asymptotic expansion of $\mathcal{I}^{f}$ in various kinematic regions, as provided in \cite{Chicherin:2018avq}, can be used in the bootstrap process.

Before proceeding with the bootstrap, we review some properties of SVHPL and analyze the leading singularities of $\mathcal{I}^{f}$. The symbol alphabet of SVHPLs consists of $\{z,\bar{z},1-z,1-\bar{z}\}$, and SVHPLs satisfy the property that they have no branch cuts in the whole complex plane. However, this does not imply that they are purely real or imaginary. Nevertheless, we can construct such bases that have definite parity under exchanging $z$ and $\bar{z}$:
\begin{equation}
\begin{aligned}
    \text{parity odd: }&\mathcal{L}^{odd}(z,\bar{z})=-\mathcal{L}^{odd}(\bar{z},z), \\
    \text{parity even: }&\mathcal{L}^{even}(z,\bar{z})=\mathcal{L}^{even}(\bar{z},z)
\end{aligned}
\end{equation}
This basis has been constructed up to weight 8 using \texttt{HyperlogProcedure}. The number of independent SVHPL basis functions for each weight $n$ is summarized in Table.~\ref{tab:SVHPL}.
\begin{table}[htbp]
    \centering
    \begin{tabular}{c|c|c|c|c|c|c|c|c}
    \hline\hline
        weight & 1 & 2 & 3 & 4 & 5 & 6 & 7 & 8 \\
        \hline
        \# of parity odd & 0 & 1 & 2 & 6 & 12 & 28 & 56 & 120 \\
        \# of parity even & 2 & 3 & 6 & 10 & 20 & 36 & 72 & 136 \\
        \# of total & 2 & 4 & 8 & 16 & 32 & 64 & 128 & 256 \\
    \hline\hline
    \end{tabular}
    \caption{The counting of SVHPL up to weight 8.}
    \label{tab:SVHPL}
\end{table}
It has been proven that the total number is $2^{n}$ at weight $n$, while the number of parity even is $2^{n-1}+2^{\lfloor \frac{n-1}{2} \rfloor}$~\cite{Brown:2004ugm,Schnetz:2013hqa,Alday:2024ksp}. This basis can be constructed directly using the method described in \cite{Schnetz:2013hqa} and is implemented in the package \texttt{HyperlogProcedures} up to very high weights. Although the expressions for $\mathcal{I}^{d2}$ and $\mathcal{I}^{f2}$ are available in the literature, as mentioned earlier, they can also be bootstrapped using the asymptotic expansions provided in \cite{Chicherin:2018avq}.
The ansatz for $\mathcal{I}^{d2}$ or $\mathcal{I}^{f2}$ can be set up as
\begin{equation}
    \mathcal{I}^{d2}_{ans}=\sum_{i}\frac{\mathcal{L}^{w=8}_{odd,i}}{\Delta}, \quad \mathcal{I}^{f2}_{ans}=\sum_{i}\frac{\mathcal{L}^{w=8}_{even,i}}{\Delta^2},
\end{equation}
where $\Delta$ is defined as
\begin{equation}
    \Delta\equiv z-\bar{z}=\sqrt{1-2u-2v+u^2-2uv+v^2}.
\end{equation}
Note that $\mathcal{L}^{w=8}_{\bullet}$ indicates all SVHPL functions with total weight 8 which contain functions like $\zeta_{3}\mathcal{L}^{w=5}$ where $\mathcal{L}^{w=5}$ consists of weight-5 SVHPL\footnote{This counting of $\mathcal{L}^{w=n}$ in Tab.~\ref{tab:SVHPL} at weight $n$ does not include terms such as $\zeta_{m}\mathcal{L}^{w=n-m},m>0$.}. Using the asymptotic expansions around $(u,v)\sim (0,1)$, $(1/u,v/u)\sim (0,1)$ and $(1/v,u/v)\sim (0,1)$, the results are determined to be\footnote{In the definition of $\mathcal{I}^{d2}$ and $\mathcal{I}^{f2}$ of \cite{Chicherin:2018avq}, the external legs are ordered as $1,3,2,4$. Therefore, in the expressions for $\mathcal{I}^{d2}$ and $\mathcal{I}^{f}$, $z$ and $\bar{z}$ should be replaced by $1/z$ and $1/\bar{z}$, and an overall factor $z\bar{z}$ should be factored out to get the expressions of order $1,2,3,4$. }
\begin{equation}\label{eq:d2f2}
    \begin{aligned}
        &\mathcal{I}^{d2}(\frac{v}{u},\frac{1}{u})=\frac{1}{\Delta}\left(\mathrm{I}_{z,1,0,1,0,1,1,0,1,0}-\mathrm{I}_{z,1,0,1,1,0,1,0,1,0}-15\zeta_{5}\mathrm{I}_{z,1,0,1,0}-\frac{441}{8}\zeta_{7}\mathrm{I}_{z,1,0}\right), \\
        &\mathcal{I}^{f2}(\frac{1}{u},\frac{v}{u})=\frac{4z\bar{z}}{\Delta^2}\Big(\mathrm{I}_{z,0,1,1,0,1,1,1,1,0}-\mathrm{I}_{z,0,1,1,1,1,0,1,1,0}-\mathrm{I}_{z,1,0,1,1,0,1,1,1,0}+\mathrm{I}_{z,1,0,1,1,1,1,0,1,0}\\
        &+\mathrm{I}_{z,1,1,0,1,1,0,1,1,0}-\mathrm{I}_{z,1,1,0,1,1,1,1,0,0}-\mathrm{I}_{z,1,1,1,0,1,1,0,1,0}+\mathrm{I}_{z,1,1,1,1,0,1,1,0,0}+2\zeta_{3}\mathrm{I}_{z,0,1,1,1,1,0}\\
        &+4\zeta_{3}\mathrm{I}_{z,1,0,1,1,1,0}\!-\!4\zeta_{3}\mathrm{I}_{z,1,1,1,0,1,0}\!-\!2\zeta_{3}\mathrm{I}_{z,1,1,1,1,0,0}\!+\!10\zeta_{5}\mathrm{I}_{z,0,1,1,0}\!-\!10\zeta_{5}\mathrm{I}_{z,1,1,0,0}\!-\!6\zeta_{3}^{2}\mathrm{I}_{z,1,1,0}\Big),
    \end{aligned}
\end{equation}
We apply the function $\mathrm{I}_{z,...,0}$ defined within \texttt{HyperlogProcedures} as a basis for SVHPLs. $\mathrm{I}_{z,...,0}$ is defined recursively as follows \cite{Schnetz:2013hqa}:
\begin{equation}
    \begin{aligned}
    &\mathrm{I}_{z,a_{n},a_{n-1},\ldots,a_{1},0}=\int_{sv}\frac{\mathrm{d}z}{z-a_{n}}\mathrm{I}_{z,a_{n-1},\ldots,a_{1},0}, \, a_{n}=0,1; \\
    &\mathrm{I}_{z,0}=1; \mathrm{I}_{z,0,0}=\log z\bar{z}, \, \mathrm{I}_{z,1,0}=\log(1-z)(1-\bar{z}).
    \end{aligned}
\end{equation}
$\int_{sv}$ is the single-valued integration defined by Eq.~(2.53) in \cite{Schnetz:2013hqa}, which not only is an integration of $z$, but also accounts for the antiholomorphic part in $\bar{z}$ and the boundary condition that requires the integration vanishes when $z$ approaches 0 (regularized around $z=0$). It keeps the result single-valued. This definition can be taken as the single-valued version for HPL functions (here, we adopt the definition within \texttt{HyperlogProcedures} for consistency, which is slightly different from the standard definition in \cite{Remiddi:1999ew})
\begin{equation}
    \begin{aligned}
    &\mathrm{i}_{z,a_{n},a_{n-1},\ldots,a_{1},0}=\int_{0}^{z}\frac{\mathrm{d}z^{\prime}}{z^{\prime}-a_{n}}\mathrm{i}_{z,a_{n-1},\ldots,a_{1},0}, \, a_{n}=0,\pm 1; \\
        &\mathrm{i}_{z,0}=1.
    \end{aligned}
\end{equation}
where $\mathrm{i}_{z,...,0}$ is taken to be already regularized around $z=0$.
SVHPLs can be generally expressed as products of HPLs:
\begin{equation}
    \mathrm{I}_{z,\vec{a},0}=\sum_{i,j}c_{i,j}\mathrm{i}_{z,\vec{a}_{i},0}\mathrm{i}_{\bar{z},\vec{a}_{j},0}.
\end{equation}
For instance, $\mathrm{I}_{z,1,0,0}$ is a weight-2 function given by $\mathrm{i}_{z,1,0}\mathrm{i}_{\bar{z},0,0}+\mathrm{i}_{z,1,0,0}+\mathrm{i}_{\bar{z},0,1,0}$. $\mathrm{i}_{z,1,0,0}$ is a HPL function with the symbol $z\otimes (1-z)$. \eqref{eq:d2f2} agrees with direct computations using \texttt{HyperlogProcedures}. For example, $\mathcal{I}^{d2}$ can be computed in \texttt{HyperlogProcedures} using the following syntax:
\begin{alltt}
    int:=\{[\{1,5\},1],[\{2,6\},1],[\{3,5\},1],[\{3,7\},1],[\{5,7\},1],[\{5,8\},1],
    [\{7,8\},1],[\{6,7\},1],[\{6,8\},1]\}:
    Id2:=GraphicalFunction(int,[1,3,2],4,0,[z,zz]);
\end{alltt}
where the first line defines the integrand using edges of the corresponding graph. The basic element is \texttt{\{edge,weight\}}. For example, \texttt{[\{1,5\},1]} represents $x_{15}^{2}$ in the denominator. A weight of \texttt{-1} would indicate a numerator term. The second line computes the integral using the graph function method described in \cite{Schnetz:2013hqa}. We have set $x_4$ to infinity and fixed $x_1=0,x_3=1,x_2=z$\footnote{For more details on the function's usage and the meaning of every argument, refer to the documentation of \texttt{HyperlogProcedures}.}. \texttt{zz} is a shorthand for $\bar{z}$. $\mathcal{I}^{f2}$ is a fishnet integral whose expression can be obtained by the determinant of ladders $F_{n}$~\cite{Basso:2017jwq} which has a rather compact form expressed using $\mathrm{I}_{z,...,0}$:
\begin{equation}
    F_{n}(\frac{u}{v},\frac{1}{v})=\frac{(-1)^{n}}{\Delta}\times \big(\mathrm{I}_{z,{\scriptscriptstyle \underbrace{0,...,0}_{n}},1,{\scriptscriptstyle \underbrace{0,...,0}_{n-1}},0}-\mathrm{I}_{z,{\scriptscriptstyle \underbrace{0,...,0}_{n-1}},1,{\scriptscriptstyle \underbrace{0,...,0}_{n}},0}\big).
\end{equation}
The expression of all fishnets have been implemented into \texttt{HyperlogProcedures}.

Although $\mathcal{I}^{d2}$ and $\mathcal{I}^{f2}$ can be computed directly in this manner, $\mathcal{I}^{f}$ cannot be evaluated using \texttt{HyperlogProcedures}~\footnote{$\mathcal{I}^{f2}$ is a fishnet, and its result is pre-implemented in \texttt{HyperlogProcedures} \cite{Borinsky:2022lds}. Thus, it is not explicitly calculated. In this sense, our result for $\mathcal{I}^{f}$ can also serve as a seed for \texttt{HyperlogProcedures}. }. This necessitates the usage of the bootstrap method.
Now we bootstrap $\mathcal{I}^{f}$ in the same way as above. To construct an ansatz for $\mathcal{I}^{f}$, we first analyze its leading singularities (or equivalently, construct $d\log$ forms for the integrand).
\begin{equation}
    \mathcal{I}^{f}= \int\mathrm{d}^{4}x_8\mathrm{d}^{4}x_7\mathrm{d}^{4}x_6\mathrm{d}^{4}x_5\frac{x_{18}^{2}x_{37}^{2}x_{24}^{2}}{x_{52}^{2}x_{53}^{2}x_{57}^{2}x_{58}^{2}x_{61}^{2}x_{64}^{2}x_{67}^{2}x_{68}^{2}x_{71}^{2}x_{72}^{2}x_{83}^{2}x_{84}^{2}x_{78}^{2}}
\end{equation}
It ($\mathcal{I}_{2}^{(4)}$) is depicted in Appendix.~\ref{app:definition}.
After first integrating out $x_{5}$ and $x_6$, the remaining integrand takes the form
\begin{equation}
    I^{f}_{78}=\frac{x_{18}^{2}x_{37}^{2}x_{24}^{2}}{x_{17}^{2}x_{27}^{2}x_{38}^{2}x_{48}^{2}x_{78}^{2}\lambda_{2378}\lambda_{1478}}.
\end{equation}
where $\lambda_{ijkl}=\sqrt{\det(x_{mn}^{2})_{m,n=i,j,k,l}}$ represents the leading singularity of a four-mass box. Now the question is how many leading singularities can be derived from $I^{f}_{78}$. Since the integrand is symmetric of $x_7$ and $x_8$, we can first cut $x_7$ and then $x_8$. The cut for $x_7$ and the corresponding remaining terms are summarized in Table.~\ref{tab:fcut}.
\begin{table}[htbp]
    \centering
    \begin{tabular}{c|c|c|c}
    \hline\hline
        cut for $x_7$ &  remaining term & cut for $x_8$ & leading singularity\\
        \hline
        $x_{17}^{2},x_{27}^{2},x_{78}^{2},\lambda_{1478}$  & $\displaystyle \frac{x_{24}^{2}}{x_{28}^{2}x_{38}^{2}x_{48}^{2}\lambda_{1248}}$ &  $x_{28}^{2},x_{38}^{2},x_{48}^{2},\lambda_{1248}$ & $\displaystyle \frac{1}{\Delta}$\\
        $x_{17}^{2},x_{27}^{2},x_{78}^{2},\lambda_{2378}$  & 0\tablefootnote{The cut vanishes due to numerator $x_{37}^{2}$. Same reason for other 0's in the table.} &  0 & 0\\
        $x_{17}^{2},x_{78}^{2},\lambda_{2378},\lambda_{1478}$  & $\displaystyle \frac{x_{24}^{2}}{x_{28}^{2}x_{48}^{2}\lambda_{148\{23\}}}$ &  $x_{18}^{2},x_{28}^{2},x_{38}^{2},x_{48}^{2}$ & $\displaystyle \frac{1}{\Delta}$\\
        $x_{27}^{2},x_{78}^{2},\lambda_{2378},\lambda_{1478}$  & 0 &  0 & 0\\
        $x_{17}^{2},x_{27}^{2},\lambda_{2378},\lambda_{1478}$  & $\displaystyle \frac{x_{18}^{2}x_{23}^{2}x_{24}^{2}}{x_{28}^{2}x_{38}^{2}x_{48}^{2}\lambda_{12\{348\}}} $ & $\displaystyle x_{18}^{2},x_{28}^{2},x_{38}^{2},x_{48}^{2}$  & \thead{$\displaystyle \frac{1}{\Delta}$ \\[6pt] $\displaystyle\frac{u-v-1}{\Delta^2}$}\\
    \hline\hline
    \end{tabular}
    \caption{Cuts for $I_{78}^{f}$. $\lambda^{2}_{148\{23\}}= [x_{18}^{2}(x_{38}^{2}x_{24}^{2}\!-\!x_{28}^{2}x_{34}^{2})\!-\!x_{48}^{2}(x_{38}^{2}x_{12}^{2}\!-\!x_{28}^{2}x_{13}^{2})]^2-4x_{14}^{2}x_{23}^{2}x_{18}^{2}x_{28}^{2}x_{38}^{2}x_{48}^{2}$ and $\lambda_{12\{348\}}^{2}= [x_{12}^{2}(x_{18}^{2}x_{28}^{2}x_{34}^{2}-x_{14}^{2}x_{34}^{2}x_{28}^{2}-x_{18}^{2}x_{48}^{2}x_{23}^{2})-(x_{14}^{2}x_{28}^{2}-x_{18}^{2}x_{24}^2)(x_{23}^{2}x_{18}^{2}-x_{28}^{2}x_{13}^{2})]^2-4x_{12}^{4}x_{14}^{2}x_{23}^{2}x_{18}^{2}x_{28}^{2}x_{38}^{2}x_{48}^{2}$.}
    \label{tab:fcut}
\end{table}
Two $\lambda$s come from Gram determinants:
\begin{equation}
    \begin{aligned}
        &\lambda^{2}_{148\{23\}}\equiv G(x_{71}^{\mu},x_{74}^{\mu},x_{78}^{\mu},x_{38}^{2}x_{72}^{\mu}-x_{28}^{2}x_{73}^{\mu})|_{cut}, \,\\
        &\text{cut: } x_{71}^{2}=x_{74}^{2}=x_{78}^2=x_{38}^{2}x_{72}^{2}-x_{28}^{2}x_{73}^{2}=0. \\
        &\lambda_{12\{348\}}^{2}\equiv G(x_{71}^{\mu},x_{72}^{\mu},x_{14}^{2}x_{78}^{\mu}-x_{18}^{2}x_{74}^{\mu},x_{23}^{2}x_{78}^{\mu}-x_{28}^{2}x_{73}^{\mu})|_{cut}, \, \\
        &\text{cut: } x_{71}^2=x_{72}^2=x_{14}^{2}x_{78}^{2}-x_{18}^{2}x_{74}^{2}=x_{23}^{2}x_{78}^{2}-x_{28}^{2}x_{73}^{2}=0.
    \end{aligned}
\end{equation}
Using the magic identity \eqref{eq:magicf} and the leading singularities analysis in Table.~\ref{tab:fcut}, we propose the following ansatz for $\mathcal{I}^{f}$: 
\begin{equation}
    \mathcal{I}_{ans}^{f}=\sum_{i}\frac{\mathcal{L}^{w=8}_{odd,i}}{\Delta}+(u-v-1)\sum_{j}\frac{\mathcal{L}^{w=8}_{even,j}}{\Delta^2}.
\end{equation}
In this case, we find that the asymptotic expansion alone suffices to determine all parameters. However, we can also directly apply the functional constraints from \eqref{eq:magicf}, which must hold at the function level without asymptotic expansion:
\begin{equation}\label{eq:magicconstraint}
    \begin{aligned}
        &\frac{1}{\Delta}\left(\sum_{i}\mathcal{L}^{w=8}_{odd,i}(z,\bar{z})-\sum_{i}\mathcal{L}^{w=8}_{odd,i}(1-z,1-\bar{z})\right)=\mathcal{I}^{d2}(z,\bar{z})+\mathcal{I}^{d2}(1-z,1-\bar{z}), \\
        &\frac{1}{\Delta^2}\left((u\!-\!v\!-\!1)\sum_{j}\mathcal{L}^{w=8}_{even,j}(z,\bar{z})+(v\!-\!u\!-\!1)\sum_{j}\mathcal{L}^{w=8}_{even,j}(1-z,1-\bar{z})\right)=\mathcal{I}^{f2}(z,\bar{z}).
    \end{aligned}
\end{equation}
Note that to apply the above identity, external points must be consistently labeled to ensure that $z$ and $\bar{z}$ are defined in the same way. Although magic identity alone cannot fix all parameters, it reduces the number of free parameters to 63. This significantly simplifies the bootstrap process when combined with asymptotic expansion results. For example, applying the magic identity constraint first reduces the required number of asymptotic expansions from four to two. 
Although more complicated than $\mathcal{I}^{d2}$ and $\mathcal{I}^{f2}$, the explicit result for $\mathcal{I}^{f}$ can still fit into a few lines (with the usual relabeling of external points as $x_2,x_1,x_3,x_4$, which simplifies the expression):
\begin{equation}
    \begin{aligned}
        &\mathcal{I}^{f}(\frac{u}{v},\frac{1}{v})=\frac{u-v-1}{\Delta^2}\Bigg(2 \mathrm{I}_{z ,1,0,0,0,0,1,0,0,0}-2 \mathrm{I}_{z ,1,0,0,1,0,0,0,0,0}-2 \mathrm{I}_{z ,0,0,0,0,1,0,0,1,0}\\
        &+2 \mathrm{I}_{z ,0,0,1,0,0,0,0,1,0}\!-\!2 \mathrm{I}_{z ,0,1,0,0,0,0,1,0,0}\!+\!2 \mathrm{I}_{z ,0,1,0,0,1,0,0,0,0}\!-\!2 \mathrm{I}_{z ,0,0,1,0,0,1,0,0,0}\!+\!2 \mathrm{I}_{z ,0,0,0,1,0,0,1,0,0} \Bigg) \\
        &-\frac{1}{\Delta}\Bigg(\!40\zeta_{5} \mathrm{I}_{z ,0,0,1,0}\!-\!20 \zeta_{5} \mathrm{I}_{z ,0,1,0,0}\!-\!8\zeta_{3} \mathrm{I}_{z ,0,0,0,0,1,0}+8\zeta_{3} \mathrm{I}_{z ,0,0,0,1,0,0}+4 \zeta_{3} \mathrm{I}_{z ,0,1,0,0,1,0} \\
        &-4 \zeta_{3} \mathrm{I}_{z ,0,1,0,1,0,0}-2 \mathrm{I}_{z ,0,0,0,0,1,0,0,1,0}+2 \mathrm{I}_{z ,0,0,0,0,1,0,1,0,0}+2 \mathrm{I}_{z ,0,0,1,0,0,0,0,1,0}-2 \mathrm{I}_{z ,0,0,1,0,0,0,1,0,0}\\
        &+2 \mathrm{I}_{z ,0,1,0,0,0,1,0,0,0}-\mathrm{I}_{z ,0,1,0,0,1,0,1,0,0}-2 \mathrm{I}_{z ,0,1,0,1,0,0,0,0,0}+\mathrm{I}_{z ,0,1,0,1,0,0,1,0,0}-2 \mathrm{I}_{z ,1,0,0,0,0,1,0,0,0}\\
        &+2 \mathrm{I}_{z ,1,0,0,1,0,0,0,0,0}\Bigg).
    \end{aligned}
\end{equation}
This has also been verified numerically using \texttt{pySecDec}~\cite{Borowka:2017idc} with a precision of $\mathcal{O}(10^{-6})$.

\subsection{Leading singularities and some results for five-loop DCI integrals}
Among the 34 five-loop DCI integrals defined in Appendix~\ref{app:definition}, 23 can be directly computed using \texttt{HyperlogProcedures}. There are several identities among these integrals, and they will be discussed first. Next, we study integrals which are unknown but can be calculated by boxing, {\it i.e.} a second-order differential equation~\cite{Drummond:2010cz}. The second-order differential operator of $z,\bar{z}$ acting on the integrals will relate them to lower-loop DCI integrals. In contrast, inverse boxing is the technique for directly solving the second-order differential equation through single-valued integration of lower-loop integrals~\cite{Schnetz:2013hqa,Borinsky:2022lds}.  Two of the unknown integrals can be related to the four-loop integrals $\mathcal{I}^{f}$ calculated in the previous section. They can no longer be expressed using only SVHPLs and they are not of uniform weight as well, due to the appearance of a leading singularity, $1/\Delta^2$, in the four-loop DCI integrals\footnote{In the inverse boxing of $\mathcal{I}^{f}$, $1/\Delta^2$ will result in the single-valued integration of the following type: $\int\frac{\mathrm{d}z\mathrm{d}\bar{z}}{z\bar{z}(1-z)(1-\bar{z})(z-\bar{z})}\times\text{pure functions}$, the measure of which is not of $\mathrm{d}\log$ type for $z,\bar{z}$. Thus, integration will result in a nonuniform-weight part.}. Although the rest unknown integrals can be studied by bootstrap in principle, we have less control over both the corresponding ansatz and required constraints, due to the possible loss of uniform-weight property and the loss of higher-rank asymptotic expansions.  Here, we only take a first step in studying all leading singularities of the 34 DCI integrals and present them in Table~\ref{tab:fiveloopls}\footnote{For those integrals that are not of uniform weight, the leading singularity analysis only accounts for the coefficients of the weight-10 part.}. Finally, there is a five-loop relation ($M_{2,2}^{(5)}=-6F_{1,4}$ in Table~\ref{tab:mab} ) as presented in \cite{Caron-Huot:2021usw} which is similar to the four-loop one:
\begin{equation}\label{eq:fivelooprelation}
    -\mathcal{I}_{5}^{(5)}(u,v)+2\mathcal{I}_{34}^{(5)}(u,v)-4\left[\mathcal{I}_{27}^{(5)}(u,v)+\mathcal{I}_{27}^{(5)}(v,u)\right]=-6F_{1,4}=-6F_1F_4+F_2F_3,
\end{equation}
where $\mathcal{I}_{34}^{(5)}$ which only involves SVHPLs, is known from \texttt{HyperlogProcedures}. This magic identity will give additional constraints when bootstrapping $\mathcal{I}_{27}^{(5)}$ and $\mathcal{I}_{5}^{(5)}$. And it indicates that $\mathcal{I}_{27}^{(5)}$ and $\mathcal{I}_{5}^{(5)}$ may be expressed using only SVHPLs.

In the 23 integrals that can be computed using \texttt{HyperlogProcedures}, 14 equal to the five-loop ladder $F_{5}$. They satisfy the original magic identity discussed in \cite{Drummond:2006rz} and are listed as follows.
\begin{equation}
\begin{aligned}
    &\mathcal{I}^{(5)}_{9}\!=\!\mathcal{I}^{(5)}_{10}\!=\!\mathcal{I}^{(5)}_{11}\!=\!\mathcal{I}^{(5)}_{13}\!=\!\mathcal{I}^{(5)}_{15}\!=\!\mathcal{I}^{(5)}_{16}\!=\!\mathcal{I}^{(5)}_{17}\!=\!\mathcal{I}^{(5)}_{19}\!=\!\mathcal{I}^{(5)}_{20}\!=\!\mathcal{I}^{(5)}_{21}\!=\!\mathcal{I}^{(5)}_{22}\!=\!\mathcal{I}^{(5)}_{23}\!=\!\mathcal{I}^{(5)}_{31}\!=\!\mathcal{I}^{(5)}_{32}.\\
    &\mathcal{I}_{9}^{(5)}(\frac{u}{v},\frac{1}{v})=\frac{\mathrm{I}_{z,0,0,0,0,1,0,0,0,0,0,0}-\mathrm{I}_{z,0,0,0,0,0,1,0,0,0,0,0}}{\Delta}.
\end{aligned}
\end{equation}
The other two identities are
\begin{equation}
    \begin{aligned}
        &\mathcal{I}_{24}^{(5)}(\frac{1}{v},\frac{u}{v})=\mathcal{I}_{25}^{(5)}(\frac{1}{v},\frac{u}{v})=\mathcal{I}_{26}^{(5)}(\frac{u}{v},\frac{1}{v})\\
        &=\frac{70\zeta_{7}\mathrm{I}_{z,0,1,0,0}\!-\!8\zeta_{5}\mathrm{I}_{z,0,1,0,0,1,0}+20\zeta_{5}\mathrm{I}_{z,0,1,0,1,0,0}-\mathrm{I}_{z,0,1,0,0,0,1,0,0,1,0,0}+\mathrm{I}_{z,0,1,0,0,1,0,0,0,1,0,0}}{\Delta},
    \end{aligned}
\end{equation}
and
\begin{equation}
    \begin{aligned}
        &\mathcal{I}_{28}^{(5)}(\frac{1}{v},\frac{u}{v})=\mathcal{I}_{29}^{(5)}(\frac{1}{v},\frac{u}{v})=\mathcal{I}_{30}^{(5)}(\frac{1}{v},\frac{u}{v})\\
        &=\frac{1}{\Delta}\Big(-42 \zeta_{7} \mathrm{I}_{z ,0,0,1,0}-20 \zeta_{5} \mathrm{I}_{z ,0,0,1,0,0,0}+12 \zeta_{5} \mathrm{I}_{z ,0,0,1,0,1,0}+4 \zeta_{3} \mathrm{I}_{z ,0,0,1,0,0,1,0,0}\\
        &-4 \zeta_{3} \mathrm{I}_{z ,0,0,1,0,1,0,0,0}-\mathrm{I}_{z ,0,0,1,0,0,1,0,1,0,0,0}+\mathrm{I}_{z ,0,0,1,0,1,0,0,1,0,0,0}
\Big).
    \end{aligned}
\end{equation}
They take the form of generalized ladders~\cite{Drummond:2012bg}. The first identity can be proved by directly applying the magic identity to the two-loop subdiagrams. The second identity can be proved by performing the boxing twice for $x_3$ and $x_4$. In this process, we have used the same fact that any DCI integral is invariant under the exchange of legs, $1\leftrightarrow 4,2\leftrightarrow 3$. Once the seeds are provided, they can ``grow'' in various ways by inverse boxing. Therefore, these three identities have the same origin. In the remaining three integrals that can be computed by \texttt{HyperlogProcedures}, $\mathcal{I}_{8}^{(5)}$ can be related to $\mathcal{I}^{f2}$ in the previous section through boxing and $\mathcal{I}_{33}^{(5)}$ can be related to some lower-loop DCI integral as well, but it is not any one that we discussed before. Interestingly, it is a purely lower-weight integral which takes the form:
\begin{equation}\label{eq:I33}
    \begin{aligned}
        \mathcal{I}_{33}^{(5)}(u,v)=\frac{20\zeta_{5}}{\Delta}\Big(&\mathrm{I}_{z ,0,0,1,0,0}-\mathrm{I}_{z ,0,0,1,1,0}-\mathrm{I}_{z ,0,1,0,0,0}+\mathrm{I}_{z ,0,1,0,1,0}-\mathrm{I}_{z ,1,0,1,0,0}+\mathrm{I}_{z ,1,0,1,1,0}\\
        &+\mathrm{I}_{z ,1,1,0,0,0}-\mathrm{I}_{z ,1,1,0,1,0}\Big).
    \end{aligned}
\end{equation}
It is a weigh-9 function. This agrees with the leading singularity analysis for the integrand, since the multivariate residue of the integrand is 0. Finally, $\mathcal{I}_{34}^{(5)}$ is computed with a unique leading singularity $1/\Delta/(u-v)$.

\begin{figure}
    \hspace{-6cm}
    \includegraphics[width=0.25\linewidth]{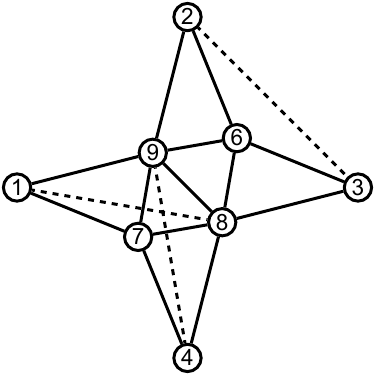}
    \caption{The four-loop DCI integral after acting boxing on $\mathcal{I}_{6}^{(5)}$. It is not one of the four-loop integrals that originate from taking four cycles of $f$-graphs.}
    \label{fig:boxingI6}
\end{figure}

Next, we study three unknown integrals, $\mathcal{I}_{2}^{(5)},\mathcal{I}_{4}^{(5)},\mathcal{I}_{6}^{(5)}$, which can also be related to some lower-loop integrals through boxing. $\mathcal{I}_{2}^{(5)}$ and $\mathcal{I}_{4}^{(5)}$ are the same with each other. Because through the boxing they are related to two identical DCI integrals with legs $1\leftrightarrow 4,2\leftrightarrow 3$ exchanged, which is $\mathcal{I}^{f}$ calculated in the previous section. Then $\mathcal{I}_{4}^{(5)}$($\mathcal{I}_{2}^{(5)}$) is obtained by the single-value integration and its boundary values provided in \cite{Chicherin:2018avq}. $\mathcal{I}_{6}^{(5)}$ is related to a four-loop integral that we have not encountered before. It is depicted in Fig.~\ref{fig:boxingI6}.

\begin{table}[htbp]
    \centering
    \begin{tabular}{c|c}
    \hline\hline
       integrals  &  leading singularities (of weight-10) \\
       \hline
      $\mathcal{I}_{1}^{(5)}$   & $\frac{1-u+v}{\Delta^2}$, $\frac{1}{\Delta}$,$-\frac{1}{(u-v)\Delta}$ \\
      $\mathcal{I}_{3}^{(5)}$   &  $\frac{1-u+v}{\Delta^2}$,$\frac{1}{\Delta}$ \\
      $\mathcal{I}_{5}^{(5)}$   &  $\frac{1}{\Delta^2}$,$-\frac{1}{(u-v)\Delta}$ \\
       $\mathcal{I}_{7}^{(5)}$   &  $\frac{1}{\Delta}$,$-\frac{1}{(u-v)\Delta}$ \\
       $\mathcal{I}_{12}^{(5)}=\mathcal{I}_{18}^{(5)}$   &  $\frac{1-u+v}{\Delta^2}$,$\frac{1}{\Delta}$ \\
       $\mathcal{I}_{14}^{(5)}$   &  $\frac{1-u+v}{\Delta^2}$,$\frac{1+u-v}{\Delta^2}$,$\frac{1}{\Delta}$ \\
       $\mathcal{I}_{27}^{(5)}$   &  $\frac{1}{\Delta^2}$ \\
       $\mathcal{I}_{33}^{(5)}$   &  - \\
       $\mathcal{I}_{34}^{(5)}$   &  $-\frac{1}{(u-v)\Delta}$ \\
       the rest   &  $\frac{1}{\Delta}$ \\
    \hline\hline
    \end{tabular}
    \caption{The leading analysis for five-loop DCI integrals. For those integrals that are not of uniform weight, the leading singularities correspond to the coefficients of weight-10 part of the full results. It should also be noted that $\mathcal{I}_{33}^{(5)}$ is a purely lower-weight integral whose form is given in \eqref{eq:I33}.}
    \label{tab:fiveloopls}
\end{table}

The results of $\mathcal{I}_{2}^{(5)}$ and $\mathcal{I}_{4}^{(5)}$ are provided as follows:
\begin{equation}
    \begin{aligned}
        &\mathcal{I}_{2}^{(5)}(\frac{1}{v},\frac{u}{v})=\mathcal{I}_{4}^{(5)}(\frac{1}{v},\frac{u}{v})=\frac{1}{\Delta}\Big(-4 \mathrm{I}_{z ,0,\bar{z} ,0,0,0,0,1,0,0,1,0}-4 \mathrm{I}_{z ,0,1,0,0,1,0,0,0,0,1,0}\\
        &+4 \mathrm{I}_{z ,0,\bar{z} ,1,0,0,0,0,1,0,0,0}-4 \mathrm{I}_{z ,0,\bar{z} ,1,0,0,1,0,0,0,0,0}-4 \mathrm{I}_{z ,0,\bar{z} ,0,1,0,0,0,0,1,0,0}+4 \mathrm{I}_{z ,0,\bar{z} ,0,0,0,1,0,0,1,0,0}\\
        &+4 \mathrm{I}_{z ,0,\bar{z} ,0,1,0,0,1,0,0,0,0}+\mathrm{I}_{z ,0,0,0,0,1,0,0,1,0,0,0}+3 \mathrm{I}_{z ,0,0,1,0,0,1,0,0,0,0,0}+4 \mathrm{I}_{z ,0,\bar{z} ,0,0,1,0,0,0,0,1,0}\\
        &+4 \mathrm{I}_{z ,0,1,0,0,0,0,1,0,0,1,0}-4 \mathrm{I}_{z ,0,\bar{z} ,0,0,1,0,0,1,0,0,0}-2 \mathrm{I}_{z ,0,0,1,0,0,0,0,1,0,0,0}-2 \mathrm{I}_{z ,0,0,0,1,0,0,1,0,0,0,0}+\\
        &56 \zeta_{7} \mathrm{I}_{z ,0,0,1,0}-\mathrm{I}_{z ,0,0,0,0,0,1,0,1,0,0,0}-2 \mathrm{I}_{z ,0,0,1,0,0,0,1,0,0,0,0}+4 \zeta_{3}\left(\mathrm{I}_{z ,0,0,0,0,0,1,0,0}-\mathrm{I}_{z ,0,0,0,0,1,0,0,0}\right.\\
        &\left.+\!\mathrm{I}_{z ,0,0,1,0,0,1,0,0}\!-\!\mathrm{I}_{z ,0,0,1,0,1,0,0,0}\right)\!+\!4 \zeta_5\! \left(3 \mathrm{I}_{z ,0,0,0,0,1,0}\!-\!10 \mathrm{I}_{z ,0,0,0,1,0,0}\!+\!3 \mathrm{I}_{z ,0,0,1,0,1,0}\!\right. \\
        &\left.-12 \mathrm{I}_{z ,0,1,0,0,1,0}\right)-\mathrm{I}_{z ,0,0,1,0,0,1,0,1,0,0,0}+2 \mathrm{I}_{z ,0,0,0,1,0,0,0,1,0,0,0}+\mathrm{I}_{z ,0,0,0,0,1,0,0,0,0,0,0}\\
        &-\mathrm{I}_{z ,0,0,0,0,0,1,0,0,0,0,0}+\mathrm{I}_{z ,0,0,1,0,1,0,0,0,0,0,0}+\mathrm{I}_{z ,0,0,1,0,1,0,0,1,0,0,0}\Big)\\
        &+\frac{2}{\Delta}\Bigg(2\mathrm{I}_{z ,1,0,0,0,0,1,0,0,1,0}-2\mathrm{I}_{z ,1,0,0,1,0,0,0,0,1,0}-24\zeta_{5}\mathrm{I}_{z,1,0,0,1,0}+(z+\bar{z}) \Big(\mathrm{I}_{z ,0,0,0,0,1,0,0,1,0,0}\\
        &-\mathrm{I}_{z ,0,0,0,1,0,0,1,0,0,0}-\mathrm{I}_{z ,0,0,1,0,0,0,0,1,0,0}+\mathrm{I}_{z ,0,0,1,0,0,1,0,0,0,0}+\mathrm{I}_{z ,0,1,0,0,0,0,1,0,0,0} \\
        &-\mathrm{I}_{z ,0,1,0,0,1,0,0,0,0,0}-\mathrm{I}_{z ,1,0,0,0,0,1,0,0,1,0}+\mathrm{I}_{z ,1,0,0,1,0,0,0,0,1,0}+12\zeta_{5}(\mathrm{I}_{z,1,0,0,1,0}-\mathrm{I}_{z,0,0,0,1,0})\Big)\Bigg)\\
        &+2\left(\mathrm{I}_{z ,0,0,0,0,1,0,0,1,0,0}-\mathrm{I}_{z ,0,0,0,1,0,0,1,0,0,0}-\mathrm{I}_{z ,0,0,1,0,0,0,0,1,0,0}+\mathrm{I}_{z ,0,0,1,0,0,1,0,0,0,0}\right. \\
        &+\mathrm{I}_{z ,0,1,0,0,0,0,1,0,0,0}-\mathrm{I}_{z ,0,1,0,0,1,0,0,0,0,0}-\mathrm{I}_{z ,1,0,0,0,0,1,0,0,1,0}+\mathrm{I}_{z ,1,0,0,1,0,0,0,0,1,0}\\
        &+2\mathrm{I}_{z ,\bar{z} ,0,0,0,0,1,0,0,1,0}-2\mathrm{I}_{z ,\bar{z} ,0,0,0,1,0,0,1,0,0}-2\mathrm{I}_{z ,\bar{z} ,0,0,1,0,0,0,0,1,0}+2\mathrm{I}_{z ,\bar{z} ,0,0,1,0,0,1,0,0,0}\\
        &\left.+2\mathrm{I}_{z ,\bar{z} ,0,1,0,0,0,0,1,0,0}-2\mathrm{I}_{z ,\bar{z} ,0,1,0,0,1,0,0,0,0}-2\mathrm{I}_{z ,\bar{z} ,1,0,0,0,0,1,0,0,0}+2\mathrm{I}_{z ,\bar{z} ,1,0,0,1,0,0,0,0,0}\right)\\
        &+24\zeta_{5}(\mathrm{I}_{z,1,0,0,1,0}-\mathrm{I}_{z,0,0,0,1,0})+168\zeta_{9}.
    \end{aligned}
\end{equation}
The first part is the weight-10 part of the result, and the remaining part is lower-weight (weight-9) terms.

Then we perform the leading singularity analysis for all 34 five-loop DCI integrals and list the results in Table~\ref{tab:fiveloopls}. It should be noted that, for those integrals that are not of uniform weight, the leading singularities given only correspond to the weight-10 part of the full results. Among all the unknown integrals, $\mathcal{I}_{12}^{(5)}=\mathcal{I}_{18}^{(5)}$ due to the magic identity existing in the two-loop subdiagrams. 

For completeness, we record analytic results for all integrals we computed/bootstrapped (including the most interesting ones like $\mathcal{I}_{2}^{(5)}$, $\mathcal{I}_{4}^{(5)}$, and ${\cal I}_2^{(4)}={\cal I}^f$) in an ancillary file \texttt{intresults.nb} along with this paper.

\section{Discussions and Outlook}
\label{sec:discussions}
In this paper, we have continued explorations for multi-loop integrands and integrated results for four-point correlators and amplitudes, and especially those DCI integrals contributing to Coulomb-branch amplitudes in the ten-dimensional light-like limit~\cite{Caron-Huot:2021usw}. At the integrand level, we identified graphical structures in $f$-graphs which contribute to component $M_{a,b}$, which in turn provides all-loop predictions for coefficients of some $(4+\ell)$-point $f$-graphs containing $(a,b)$-structure with $ab\geq \ell-1$. By comparing the Coulomb-branch amplitudes with octagons at the integrated level, we obtain three infinite families of magic identities, which originate from those prototypes for $M_{1,2}^{(4)},M_{1,2}^{(5)}$ and $M_{1,3}^{(6)}$. For individual DCI integrals, we have bootstrapped the $3$ non-trivial ones at four loops, in terms of nice weight-$8$ SVHPL functions; at five loops, $25$ of $34$ DCI integrals can be computed either by using \texttt{HyperlogProcedures} or by solving ``boxing" differential equations, and we have analyzed all leading singularities and obtained functions beyond SVHPL at both weight $10$ and $9$. Our preliminary studies have opened up several new avenues for future investigations. 

We have obtained some constraints on the coefficients of $f$-graphs from integrability, and it would be interesting to understand them by using the pinching limit~\cite{He:2024cej} (and other graphical constraints~\cite{Bourjaily:2016evz}), and even improve the efficiency of graphical bootstrap from these inputs. In general, it would be highly desirable to see whether the octagon could tell us more about higher-loop $f$-graphs. Perhaps we could first understand the rule for coefficients of next-to-fishnet, which seems to be the tip of an iceberg of such constructive rules for certain $f$-graphs. Besides, all the discussions here are restricted to four-point Coulomb branch amplitudes, but it would be very interesting if one could produce integrands of higher-point amplitudes with more general 10d lightlike limits. 

For integrated results, it would be highly desirable if one could systematically classify these (infinite families of) ``irreducible" magic identities provided by octagons at higher loops: is there some all-loop structure to be discovered here, and can one understand them from the point of view of conformal/Feynman integrals {\it e.g.} via some representations such as Mellin-Barnes? Any progress in this direction may shed new light into mathematical structures and relations for multi-loop DCI integrals. Note that these Coulomb branch amplitudes from 10d lightlike limit differ from those considered in {\it e.g.}~\cite{Arkani-Hamed:2023epq, Flieger:2025ekn}, which have been partly motivated by geometric structures, and it is tempting to ask if they are related. Also, it would be interesting to explore all-loop infrared divergences when we take the massless limit as discussed in~\cite{Caron-Huot:2021usw}. The infrared divergence extracted in this way is governed by the octagon anomalous dimension, $\Gamma_{\text{oct}}$, rather than the cusp anomalous dimension, $\Gamma_{\text{cusp}}$. And this intriguing phenomenon appears to extend to other off-shell observables, such as the off-shell form factors~\cite{Belitsky:2022itf,Belitsky:2023ssv,Belitsky:2024agy,Belitsky:2024dcf}. Another important direction concerns periods of $f$-graphs, which are strongly constrained by magic identities as well as exact results from integrated correlators {\it etc.}; in particular, an interesting open question is if there exists closed-formula for periods associated with general determinants of ladders from octagons.

We have illustrated the power of the bootstrap method for both four- and five-loop DCI integrals, based on their leading singularities and function space of SVHPL (or its extensions). It would be interesting to reorganize the integrands so that individual integrals all have a uniform, maximal transcendentality (weight $2\ell$), which would greatly simplify such a bootstrap program for higher loop integrals. Knowing the analytic results for DCI integrals at four and higher loops would be very useful for computing other amplitudes or correlators that can be expressed in terms of these integrals, and we leave these potential applications of our results to future investigations. 

Recall that our original motivation was to study all four-loop conformal integrals, extending the classic results in~\cite{Drummond:2013nda} at $\ell=3$. Indeed, by looking at the $30$ conformal integrals at $\ell=4$ (excluding the two containing elliptic pieces), we have bootstrapped most of them in terms of SVHPL or SVMPL functions, but there are still a few integrals that we cannot fix due to lack of boundary data. It would be highly desirable to determine these $30$ integrals, which can be applied to the computation of other physical quantities. Last but not least, although we have focused on the general goal of bootstrapping individual conformal integrals, a similar bootstrap program can also be set up for the four-point correlator without solving all individual integrals, where {\it e.g.} through four loops we do know all possible leading singularities and accompanying function spaces, but more physical constraints/boundary data are needed (perhaps from integrability or other inputs). It would be extremely interesting to study this problem in the future.

\begin{CJK*}{UTF8}{}
\CJKfamily{gbsn}
\begin{acknowledgments}
    It is our pleasure to thank Jacob Bourjaily, Yu-tin Huang, Chia-Kai Kuo, Canxin Shi, Yichao Tang, Qinglin Yang for collaborations on related topics and/or helpful discussions. The work of SH is supported by the National Natural Science Foundation of China under Grant No. 12225510, 12447101, 12247103, and by the New Cornerstone Science Foundation through the XPLORER PRIZE.
\end{acknowledgments}
\end{CJK*}

\appendix
\section{The definition of four and five-loop DCI integrals}\label{app:definition}
Here we present the definition of DCI integrals at $\ell=4,5$ used in the main text 
. They have all been normalized by $x_{13}^{2}x_{24}^2$.
\begin{align*}
 \mathcal{I}^{(4)}_{1}&=\includegraphics[scale=0.4,align=c]{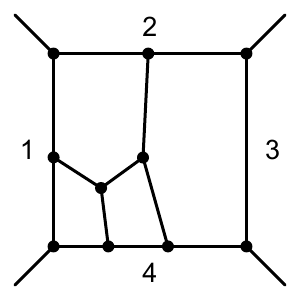}&=\includegraphics[scale=0.4,align=c]{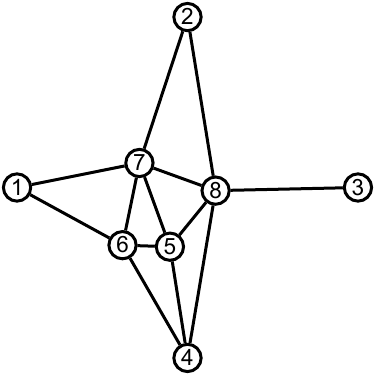}&=\int \prod_{a=5}^8 d^4 x_a\frac{x_{13}^2 x_{18}^2 x_{24}^4 x_{47}^2}{x_{16}^2 x_{17}^2 x_{27}^2 x_{28}^2 x_{38}^2 x_{45}^2 x_{46}^2 x_{48}^2 x_{56}^2 x_{57}^2 x_{58}^2 x_{67}^2 x_{78}^2}\\
 \mathcal{I}^{(4)}_{2}&=\includegraphics[scale=0.4,align=c]{graph/DCI4l2.pdf}&=\includegraphics[scale=0.4,align=c]{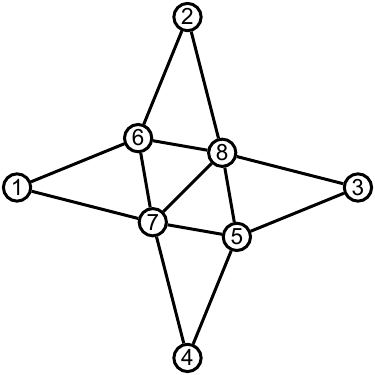}&=\int \prod_{a=5}^8 d^4 x_a \frac{x_{13}^2 x_{18}^2 x_{24}^4 x_{37}^2}{x_{16}^2 x_{17}^2 x_{26}^2 x_{28}^2 x_{35}^2 x_{38}^2 x_{45}^2 x_{47}^2 x_{57}^2 x_{58}^2 x_{67}^2 x_{68}^2 x_{78}^2}\\
 \mathcal{I}^{(4)}_{3}&=\includegraphics[scale=0.4,align=c]{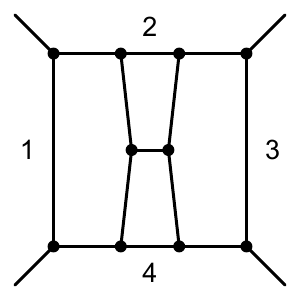}&=\includegraphics[scale=0.4,align=c]{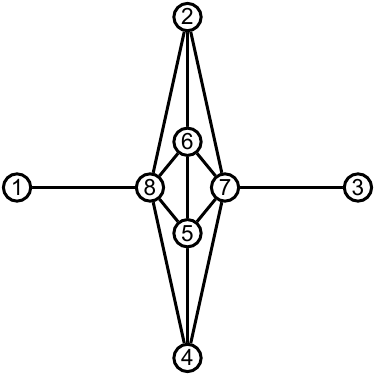}&=\int \prod_{a=5}^8 d^4 x_a\frac{x_{13}^2 x_{24}^6 x_{78}^2}{x_{18}^2 x_{26}^2 x_{27}^2 x_{28}^2 x_{37}^2 x_{45}^2 x_{47}^2 x_{48}^2 x_{56}^2 x_{57}^2 x_{58}^2 x_{67}^2 x_{68}^2}\\
 \mathcal{I}^{(4)}_{4}&=\includegraphics[scale=0.4,align=c]{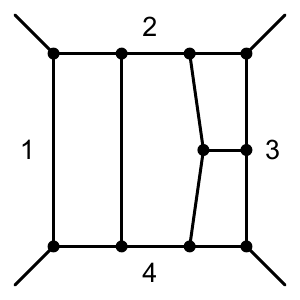}&=\includegraphics[scale=0.4,align=c]{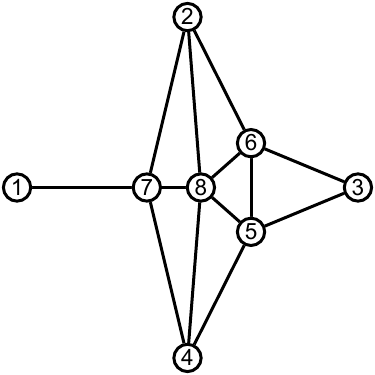}&=\int \prod_{a=5}^8 d^4 x_a\frac{x_{13}^2 x_{24}^6 x_{38}^2}{x_{17}^2 x_{26}^2 x_{27}^2 x_{28}^2 x_{35}^2 x_{36}^2 x_{45}^2 x_{47}^2 x_{48}^2 x_{56}^2 x_{58}^2 x_{68}^2 x_{78}^2}\\
 \mathcal{I}^{(4)}_{5}&=\includegraphics[scale=0.4,align=c]{graph/DCI4l5.pdf}&=\includegraphics[scale=0.4,align=c]{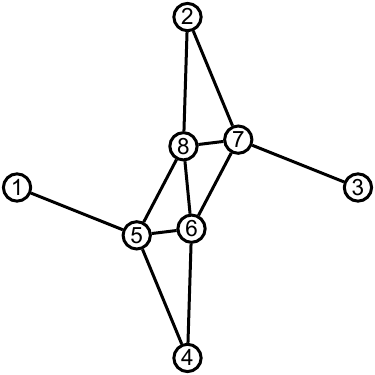}&=\int \prod_{a=5}^8 d^4 x_a\frac{x_{13}^2 x_{24}^4}{x_{15}^2 x_{27}^2 x_{28}^2 x_{37}^2 x_{45}^2 x_{46}^2 x_{56}^2 x_{58}^2 x_{67}^2 x_{68}^2 x_{78}^2}\\
 \mathcal{I}^{(4)}_{6}&=\includegraphics[scale=0.4,align=c]{graph/DCI4l6.pdf}&=\includegraphics[scale=0.4,align=c]{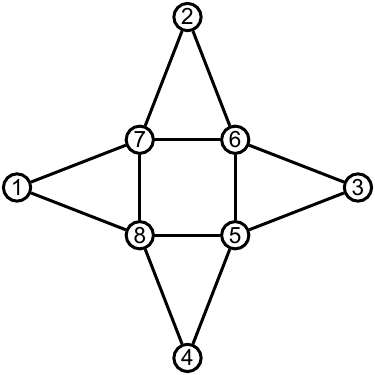}&=\int \prod_{a=5}^8 d^4 x_a\frac{x_{13}^4 x_{24}^4}{x_{17}^2 x_{18}^2 x_{26}^2 x_{27}^2 x_{35}^2 x_{36}^2 x_{45}^2 x_{48}^2 x_{56}^2 x_{58}^2 x_{67}^2 x_{78}^2}\\
 \mathcal{I}^{(4)}_{7}&=\includegraphics[scale=0.4,align=c]{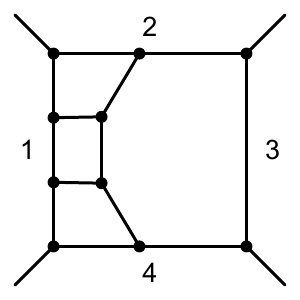}&=\includegraphics[scale=0.4,align=c]{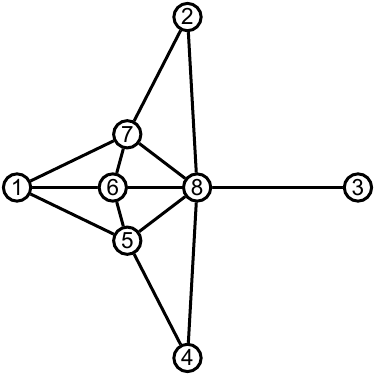}&=\int \prod_{a=5}^8 d^4 x_a\frac{x_{13}^2 x_{18}^4 x_{24}^4}{x_{15}^2 x_{16}^2 x_{17}^2 x_{27}^2 x_{28}^2 x_{38}^2 x_{45}^2 x_{48}^2 x_{56}^2 x_{58}^2 x_{67}^2 x_{68}^2 x_{78}^2}\\
 \mathcal{I}^{(4)}_{8}&=\includegraphics[scale=0.4,align=c]{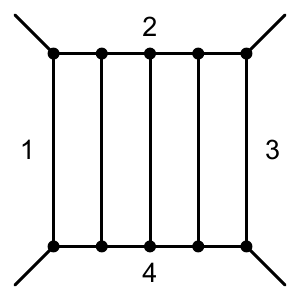}&=\includegraphics[scale=0.4,align=c]{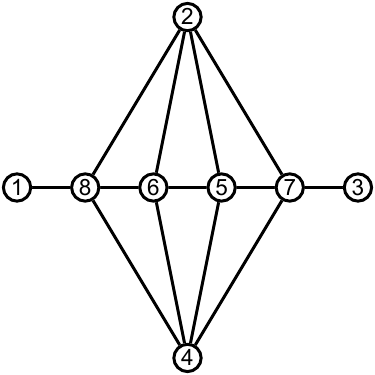}&=\int \prod_{a=5}^8 d^4 x_a\frac{x_{13}^2 x_{24}^8}{x_{18}^2 x_{25}^2 x_{26}^2 x_{27}^2 x_{28}^2 x_{37}^2 x_{45}^2 x_{46}^2 x_{47}^2 x_{48}^2 x_{56}^2 x_{57}^2 x_{68}^2}
\end{align*}

These $8$ DCI integrals at $\ell=4$ were denoted as $(e), (d2), (d), (c), (f), (f2), (b), (a)$ in Figure 14 of~\cite{Caron-Huot:2021usw}. Note that the original magic identity implies $\mathcal{I}^{(4)}_{1}=\mathcal{I}^{(4)}_3=\mathcal{I}^{(4)}_4=\mathcal{I}^{(4)}_7=\mathcal{I}^{(4)}_8$, while the new magic identity involve ${\cal I}_2, {\cal I}_5, {\cal I}_6$ as we discussed in eq~\eqref{eq:magicf}.

\begin{footnotesize}
\begin{align*}
\mathcal{I}^{(5)}_{1}&=\includegraphics[scale=0.4,align=c]{graph/DCI5l1.pdf}&=\includegraphics[scale=0.4,align=c]{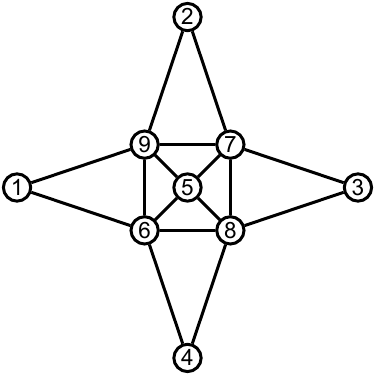}&=\int \prod_{a=5}^9 d^4 x_a\frac{x_{13}^2 x_{18}^2 x_{24}^4 x_{39}^2 x_{67}^2}{x_{16}^2 x_{19}^2 x_{27}^2 x_{29}^2 x_{37}^2 x_{38}^2 x_{46}^2 x_{48}^2 x_{56}^2 x_{57}^2 x_{58}^2 x_{59}^2 x_{68}^2 x_{69}^2 x_{78}^2 x_{79}^2}\\
\mathcal{I}^{(5)}_{2}&=\includegraphics[scale=0.4,align=c]{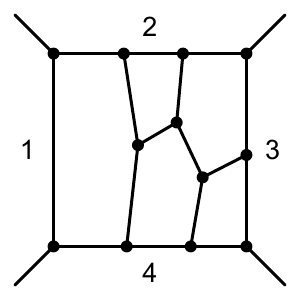}&=\includegraphics[scale=0.4,align=c]{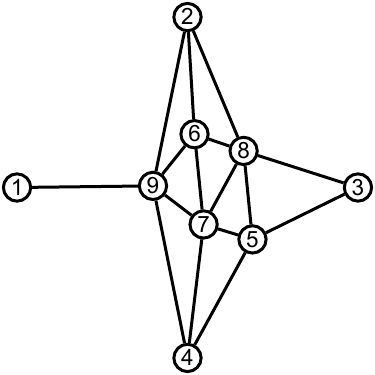}&= \int \prod_{a=5}^9 d^4 x_a \frac{x_{13}^2 x_{24}^6 x_{37}^2 x_{89}^2}{x_{19}^2 x_{26}^2 x_{28}^2 x_{29}^2 x_{35}^2 x_{38}^2 x_{45}^2 x_{47}^2 x_{49}^2 x_{57}^2 x_{58}^2 x_{67}^2 x_{68}^2 x_{69}^2 x_{78}^2 x_{79}^2}\\
\mathcal{I}^{(5)}_{3}&=\includegraphics[scale=0.4,align=c]{graph/DCI5l3.pdf}&=\includegraphics[scale=0.4,align=c]{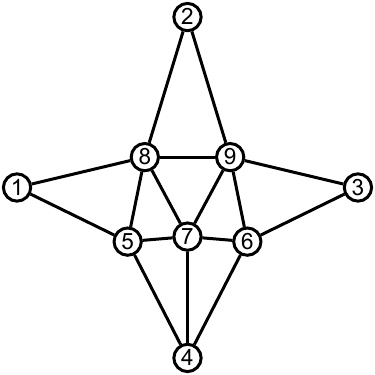}&= \int \prod_{a=5}^9 d^4 x_a \frac{x_{13}^2 x_{17}^2 x_{24}^4 x_{38}^2 x_{49}^2}{x_{15}^2 x_{18}^2 x_{28}^2 x_{29}^2 x_{36}^2 x_{39}^2 x_{45}^2 x_{46}^2 x_{47}^2 x_{57}^2 x_{58}^2 x_{67}^2 x_{69}^2 x_{78}^2 x_{79}^2 x_{89}^2}\\
\mathcal{I}^{(5)}_{4}&=\includegraphics[scale=0.4,align=c]{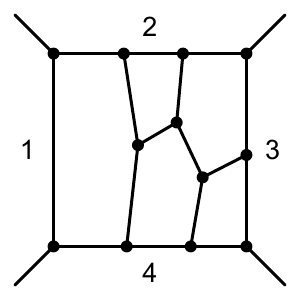}&=\includegraphics[scale=0.4,align=c]{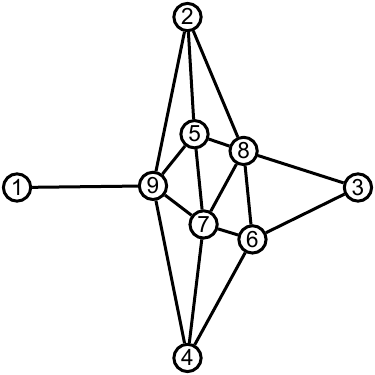}&= \int \prod_{a=5}^9 d^4 x_a \frac{x_{13}^2 x_{24}^4 x_{27}^2 x_{39}^2 x_{48}^2}{x_{19}^2 x_{25}^2 x_{28}^2 x_{29}^2 x_{36}^2 x_{38}^2 x_{46}^2 x_{47}^2 x_{49}^2 x_{57}^2 x_{58}^2 x_{59}^2 x_{67}^2 x_{68}^2 x_{78}^2 x_{79}^2}\\
\mathcal{I}^{(5)}_{5}&=\includegraphics[scale=0.4,align=c]{graph/DCI5l5.pdf}&=\includegraphics[scale=0.4,align=c]{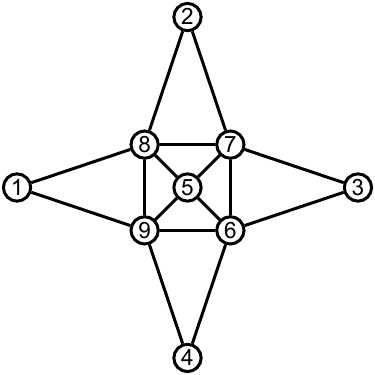}&= \int \prod_{a=5}^9 d^4 x_a \frac{x_{13}^4 x_{24}^4 x_{68}^2 x_{79}^2}{x_{18}^2 x_{19}^2 x_{27}^2 x_{28}^2 x_{36}^2 x_{37}^2 x_{46}^2 x_{49}^2 x_{56}^2 x_{57}^2 x_{58}^2 x_{59}^2 x_{67}^2 x_{69}^2 x_{78}^2 x_{89}^2}\\
\mathcal{I}^{(5)}_{6}&=\includegraphics[scale=0.4,align=c]{graph/DCI5l6.pdf}&=\includegraphics[scale=0.4,align=c]{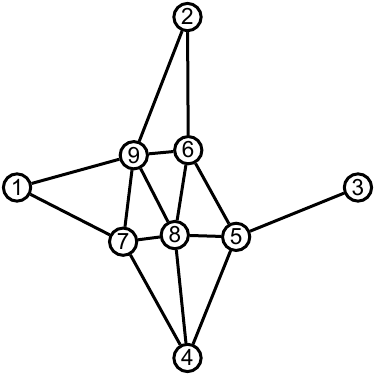}&= \int \prod_{a=5}^9 d^4 x_a \frac{x_{13}^2 x_{18}^2 x_{24}^4 x_{49}^2}{x_{17}^2 x_{19}^2 x_{26}^2 x_{29}^2 x_{35}^2 x_{45}^2 x_{47}^2 x_{48}^2 x_{56}^2 x_{58}^2 x_{68}^2 x_{69}^2 x_{78}^2 x_{79}^2 x_{89}^2}\\
\mathcal{I}^{(5)}_{7}&=\includegraphics[scale=0.4,align=c]{graph/DCI5l7.pdf}&=\includegraphics[scale=0.4,align=c]{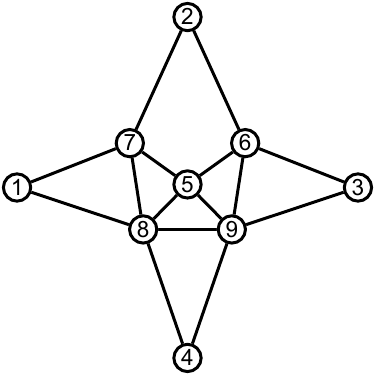}&= \int \prod_{a=5}^9 d^4 x_a \frac{x_{13}^2 x_{19}^2 x_{24}^4 x_{38}^2}{x_{17}^2 x_{18}^2 x_{26}^2 x_{27}^2 x_{36}^2 x_{39}^2 x_{48}^2 x_{49}^2 x_{56}^2 x_{57}^2 x_{58}^2 x_{59}^2 x_{69}^2 x_{78}^2 x_{89}^2}\\
\mathcal{I}^{(5)}_{8}&=\includegraphics[scale=0.4,align=c]{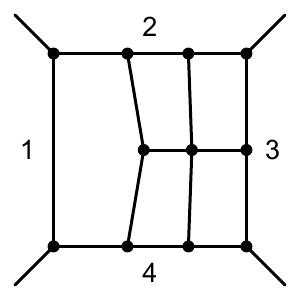}&=\includegraphics[scale=0.4,align=c]{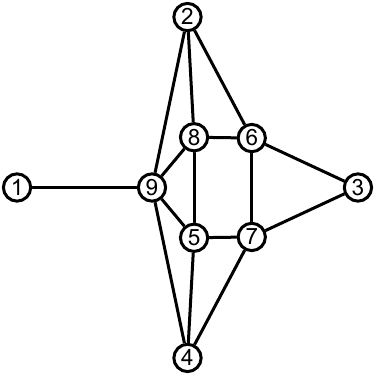}&= \int \prod_{a=5}^9 d^4 x_a \frac{x_{13}^2 x_{24}^6 x_{39}^2}{x_{19}^2 x_{26}^2 x_{28}^2 x_{29}^2 x_{36}^2 x_{37}^2 x_{45}^2 x_{47}^2 x_{49}^2 x_{57}^2 x_{58}^2 x_{59}^2 x_{67}^2 x_{68}^2 x_{89}^2}\\
\mathcal{I}^{(5)}_{9}&=\includegraphics[scale=0.4,align=c]{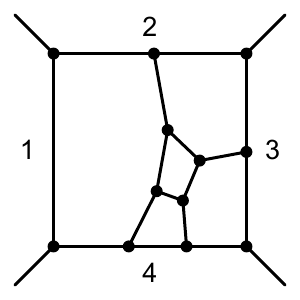}&=\includegraphics[scale=0.4,align=c]{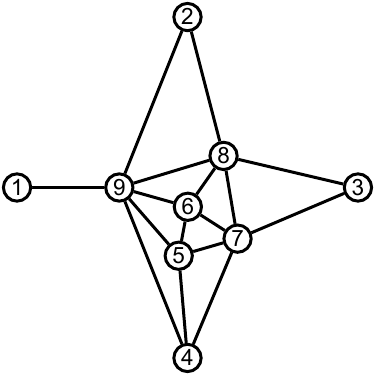}&= \int \prod_{a=5}^9 d^4 x_a \frac{x_{13}^2 x_{24}^4 x_{39}^2 x_{48}^2 x_{79}^2}{x_{19}^2 x_{28}^2 x_{29}^2 x_{37}^2 x_{38}^2 x_{45}^2 x_{47}^2 x_{49}^2 x_{56}^2 x_{57}^2 x_{59}^2 x_{67}^2 x_{68}^2 x_{69}^2 x_{78}^2 x_{89}^2}\\
\mathcal{I}^{(5)}_{10}&=\includegraphics[scale=0.4,align=c]{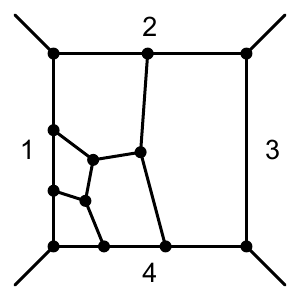}&=\includegraphics[scale=0.4,align=c]{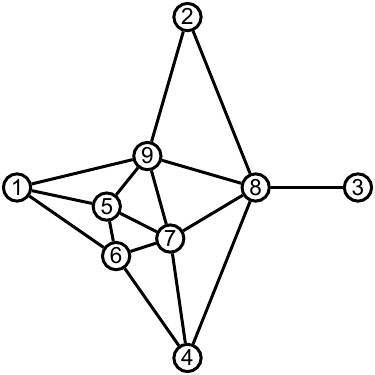}&= \int \prod_{a=5}^9 d^4 x_a \frac{x_{13}^2 x_{17}^2 x_{18}^2 x_{24}^4 x_{49}^2}{x_{15}^2 x_{16}^2 x_{19}^2 x_{28}^2 x_{29}^2 x_{38}^2 x_{46}^2 x_{47}^2 x_{48}^2 x_{56}^2 x_{57}^2 x_{59}^2 x_{67}^2 x_{78}^2 x_{79}^2 x_{89}^2}\\
\mathcal{I}^{(5)}_{11}&=\includegraphics[scale=0.4,align=c]{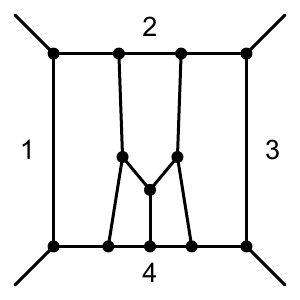}&=\includegraphics[scale=0.4,align=c]{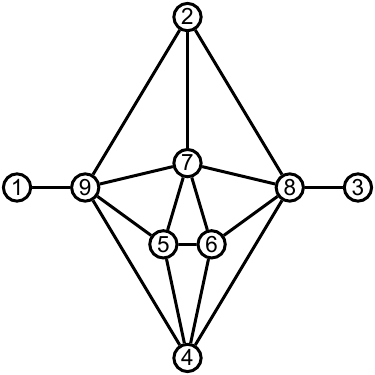}&= \int \prod_{a=5}^9 d^4 x_a \frac{x_{13}^2 x_{24}^6 x_{47}^2 x_{89}^2}{x_{19}^2 x_{27}^2 x_{28}^2 x_{29}^2 x_{38}^2 x_{45}^2 x_{46}^2 x_{48}^2 x_{49}^2 x_{56}^2 x_{57}^2 x_{59}^2 x_{67}^2 x_{68}^2 x_{78}^2 x_{79}^2}\\
\mathcal{I}^{(5)}_{12}&=\includegraphics[scale=0.4,align=c]{graph/DCI5l12.pdf}&=\includegraphics[scale=0.4,align=c]{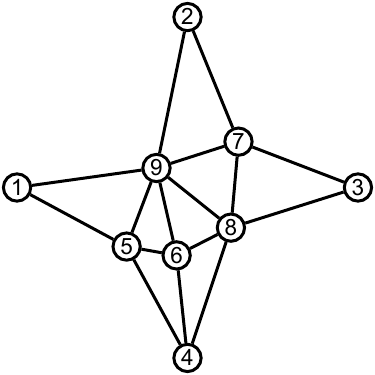}&= \int \prod_{a=5}^9 d^4 x_a \frac{x_{13}^2 x_{18}^2 x_{24}^4 x_{39}^2 x_{49}^2}{x_{15}^2 x_{19}^2 x_{27}^2 x_{29}^2 x_{37}^2 x_{38}^2 x_{45}^2 x_{46}^2 x_{48}^2 x_{56}^2 x_{59}^2 x_{68}^2 x_{69}^2 x_{78}^2 x_{79}^2 x_{89}^2}\\
\mathcal{I}^{(5)}_{13}&=\includegraphics[scale=0.4,align=c]{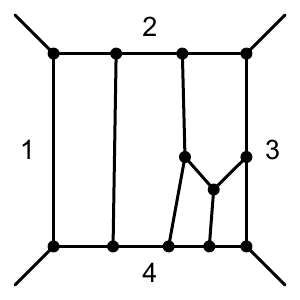}&=\includegraphics[scale=0.4,align=c]{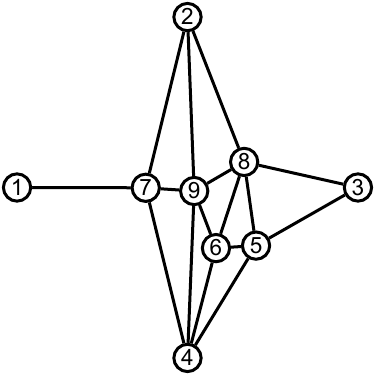}&= \int \prod_{a=5}^9 d^4 x_a \frac{x_{13}^2 x_{24}^6 x_{39}^2 x_{48}^2}{x_{17}^2 x_{27}^2 x_{28}^2 x_{29}^2 x_{35}^2 x_{38}^2 x_{45}^2 x_{46}^2 x_{47}^2 x_{49}^2 x_{56}^2 x_{58}^2 x_{68}^2 x_{69}^2 x_{79}^2 x_{89}^2}\\
\mathcal{I}^{(5)}_{14}&=\includegraphics[scale=0.4,align=c]{graph/DCI5l14.pdf}&=\includegraphics[scale=0.4,align=c]{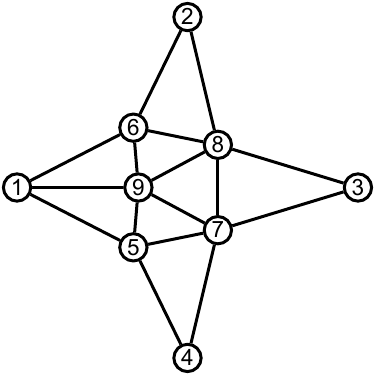}&=\int \prod_{a=5}^9 d^4 x_a \frac{x_{13}^2 x_{17}^2 x_{18}^2 x_{24}^4 x_{39}^2}{x_{15}^2 x_{16}^2 x_{19}^2 x_{26}^2 x_{28}^2 x_{37}^2 x_{38}^2 x_{45}^2 x_{47}^2 x_{57}^2 x_{59}^2 x_{68}^2 x_{69}^2 x_{78}^2 x_{79}^2 x_{89}^2}\\
\mathcal{I}^{(5)}_{15}&=\includegraphics[scale=0.4,align=c]{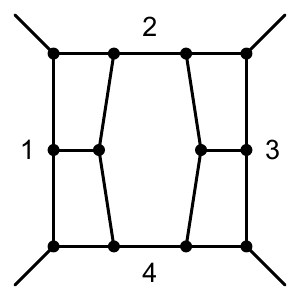}&=\includegraphics[scale=0.4,align=c]{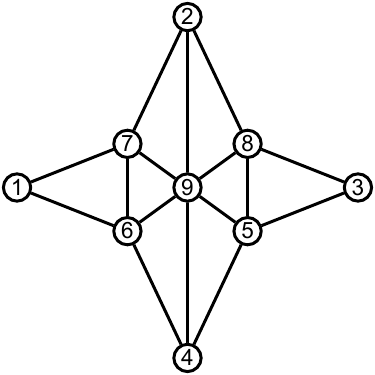}&=\int \prod_{a=5}^9 d^4 x_a \frac{x_{13}^2 x_{19}^2 x_{24}^6 x_{39}^2}{x_{16}^2 x_{17}^2 x_{27}^2 x_{28}^2 x_{29}^2 x_{35}^2 x_{38}^2 x_{45}^2 x_{46}^2 x_{49}^2 x_{58}^2 x_{59}^2 x_{67}^2 x_{69}^2 x_{79}^2 x_{89}^2}\\
\mathcal{I}^{(5)}_{16}&=\includegraphics[scale=0.4,align=c]{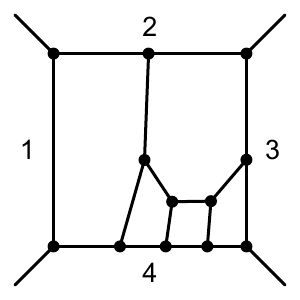}&=\includegraphics[scale=0.4,align=c]{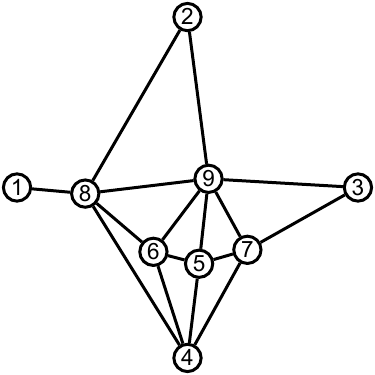}&= \int \prod_{a=5}^9 d^4 x_a \frac{x_{13}^2 x_{24}^4 x_{38}^2 x_{49}^4}{x_{18}^2 x_{28}^2 x_{29}^2 x_{37}^2 x_{39}^2 x_{45}^2 x_{46}^2 x_{47}^2 x_{48}^2 x_{56}^2 x_{57}^2 x_{59}^2 x_{68}^2 x_{69}^2 x_{79}^2 x_{89}^2}\\
\mathcal{I}^{(5)}_{17}&=\includegraphics[scale=0.4,align=c]{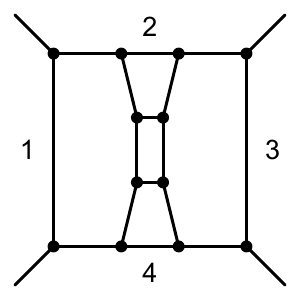}&=\includegraphics[scale=0.4,align=c]{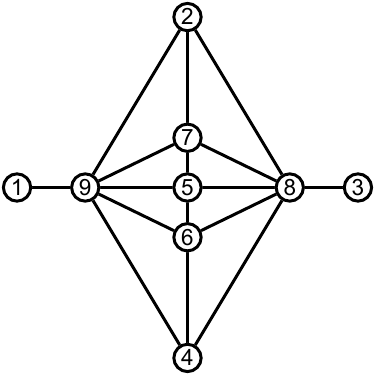}&= \int \prod_{a=5}^9 d^4 x_a \frac{x_{13}^2 x_{24}^6 x_{89}^4}{x_{19}^2 x_{27}^2 x_{28}^2 x_{29}^2 x_{38}^2 x_{46}^2 x_{48}^2 x_{49}^2 x_{56}^2 x_{57}^2 x_{58}^2 x_{59}^2 x_{68}^2 x_{69}^2 x_{78}^2 x_{79}^2}\\
\mathcal{I}^{(5)}_{18}&=\includegraphics[scale=0.4,align=c]{graph/DCI5l18.pdf}&=\includegraphics[scale=0.4,align=c]{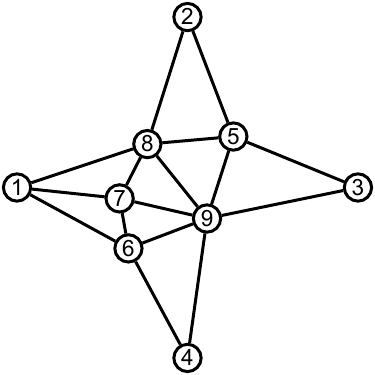}&=\int \prod_{a=5}^9 d^4 x_a \frac{x_{13}^2 x_{19}^4 x_{24}^4 x_{38}^2}{x_{16}^2 x_{17}^2 x_{18}^2 x_{25}^2 x_{28}^2 x_{35}^2 x_{39}^2 x_{46}^2 x_{49}^2 x_{58}^2 x_{59}^2 x_{67}^2 x_{69}^2 x_{78}^2 x_{79}^2 x_{89}^2}\\
\mathcal{I}^{(5)}_{19}&=\includegraphics[scale=0.4,align=c]{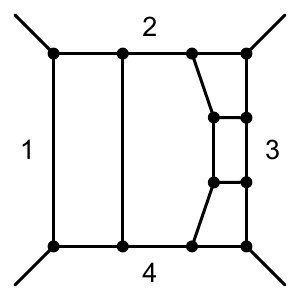}&=\includegraphics[scale=0.4,align=c]{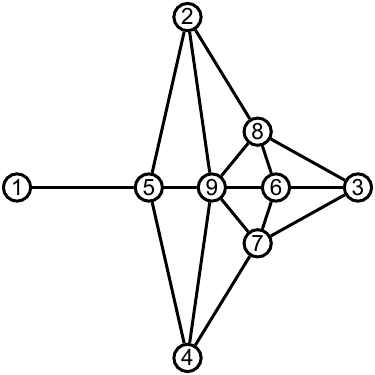}&= \int \prod_{a=5}^9 d^4 x_a \frac{x_{13}^2 x_{24}^6 x_{39}^4}{x_{15}^2 x_{25}^2 x_{28}^2 x_{29}^2 x_{36}^2 x_{37}^2 x_{38}^2 x_{45}^2 x_{47}^2 x_{49}^2 x_{59}^2 x_{67}^2 x_{68}^2 x_{69}^2 x_{79}^2 x_{89}^2}\\
\mathcal{I}^{(5)}_{20}&=\includegraphics[scale=0.4,align=c]{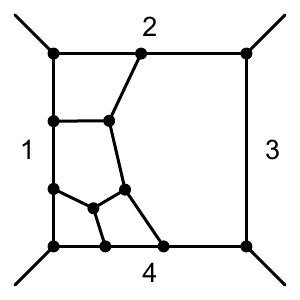}&=\includegraphics[scale=0.4,align=c]{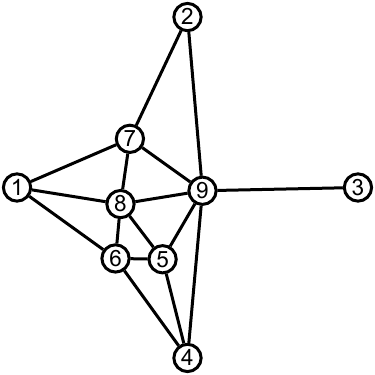}&= \int \prod_{a=5}^9 d^4 x_a \frac{x_{13}^2 x_{19}^4 x_{24}^4 x_{48}^2}{x_{16}^2 x_{17}^2 x_{18}^2 x_{27}^2 x_{29}^2 x_{39}^2 x_{45}^2 x_{46}^2 x_{49}^2 x_{56}^2 x_{58}^2 x_{59}^2 x_{68}^2 x_{78}^2 x_{79}^2 x_{89}^2}\\
\mathcal{I}^{(5)}_{21}&=\includegraphics[scale=0.4,align=c]{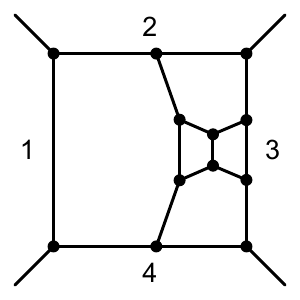}&=\includegraphics[scale=0.4,align=c]{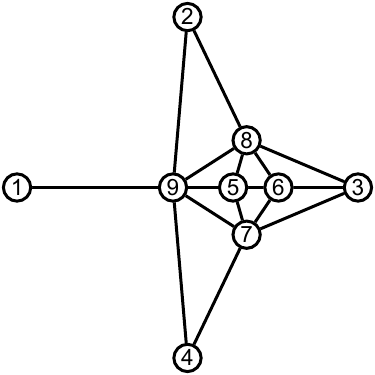}&= \int \prod_{a=5}^9 d^4 x_a \frac{x_{13}^2 x_{24}^4 x_{39}^4 x_{78}^2}{x_{19}^2 x_{28}^2 x_{29}^2 x_{36}^2 x_{37}^2 x_{38}^2 x_{47}^2 x_{49}^2 x_{56}^2 x_{57}^2 x_{58}^2 x_{59}^2 x_{67}^2 x_{68}^2 x_{79}^2 x_{89}^2}\\
\mathcal{I}^{(5)}_{22}&=\includegraphics[scale=0.4,align=c]{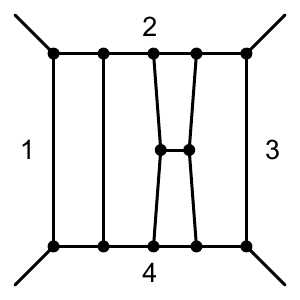}&=\includegraphics[scale=0.4,align=c]{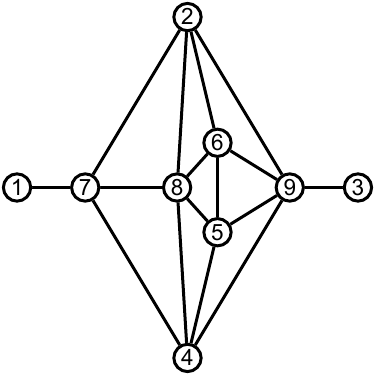}&= \int \prod_{a=5}^9 d^4 x_a \frac{x_{13}^2 x_{24}^8 x_{89}^2}{x_{17}^2 x_{26}^2 x_{27}^2 x_{28}^2 x_{29}^2 x_{39}^2 x_{45}^2 x_{47}^2 x_{48}^2 x_{49}^2 x_{56}^2 x_{58}^2 x_{59}^2 x_{68}^2 x_{69}^2 x_{78}^2}\\
\mathcal{I}^{(5)}_{23}&=\includegraphics[scale=0.4,align=c]{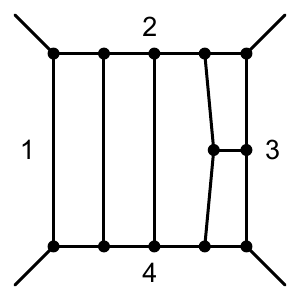}&=\includegraphics[scale=0.4,align=c]{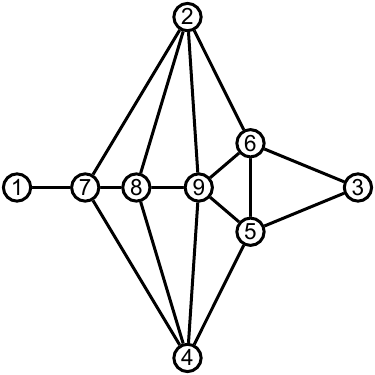}&= \int \prod_{a=5}^9 d^4 x_a \frac{x_{13}^2 x_{24}^8 x_{39}^2}{x_{17}^2 x_{26}^2 x_{27}^2 x_{28}^2 x_{29}^2 x_{35}^2 x_{36}^2 x_{45}^2 x_{47}^2 x_{48}^2 x_{49}^2 x_{56}^2 x_{59}^2 x_{69}^2 x_{78}^2 x_{89}^2}\\
\mathcal{I}^{(5)}_{24}&=\includegraphics[scale=0.4,align=c]{graph/DCI5l24.pdf}&=\includegraphics[scale=0.4,align=c]{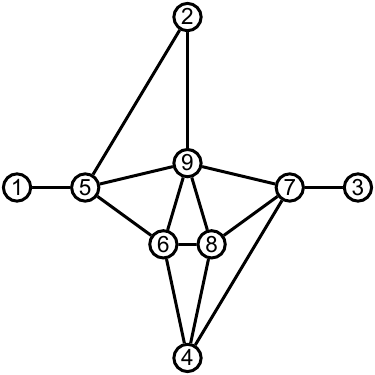}&= \int \prod_{a=5}^9 d^4 x_a \frac{x_{13}^2 x_{24}^4 x_{49}^2}{x_{15}^2 x_{25}^2 x_{29}^2 x_{37}^2 x_{46}^2 x_{47}^2 x_{48}^2 x_{56}^2 x_{59}^2 x_{68}^2 x_{69}^2 x_{78}^2 x_{79}^2 x_{89}^2}\\
\mathcal{I}^{(5)}_{25}&=\includegraphics[scale=0.4,align=c]{graph/DCI5l25.pdf}&=\includegraphics[scale=0.4,align=c]{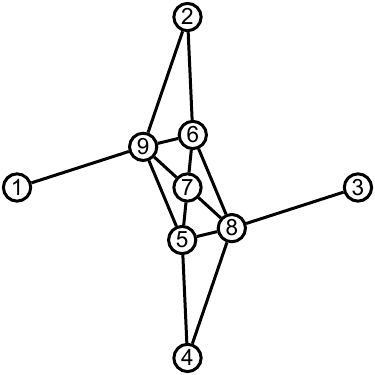}&= \int \prod_{a=5}^9 d^4 x_a \frac{x_{13}^2 x_{24}^4 x_{89}^2}{x_{19}^2 x_{26}^2 x_{29}^2 x_{38}^2 x_{45}^2 x_{48}^2 x_{57}^2 x_{58}^2 x_{59}^2 x_{67}^2 x_{68}^2 x_{69}^2 x_{78}^2 x_{79}^2}\\
\mathcal{I}^{(5)}_{26}&=\includegraphics[scale=0.4,align=c]{graph/DCI5l26.pdf}&=\includegraphics[scale=0.4,align=c]{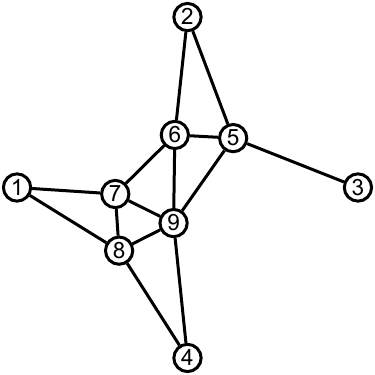}&= \int \prod_{a=5}^9 d^4 x_a \frac{x_{13}^2 x_{19}^2 x_{24}^4}{x_{17}^2 x_{18}^2 x_{25}^2 x_{26}^2 x_{35}^2 x_{48}^2 x_{49}^2 x_{56}^2 x_{59}^2 x_{67}^2 x_{69}^2 x_{78}^2 x_{79}^2 x_{89}^2}\\
\mathcal{I}^{(5)}_{27}&=\includegraphics[scale=0.4,align=c]{graph/DCI5l27.pdf}&=\includegraphics[scale=0.4,align=c]{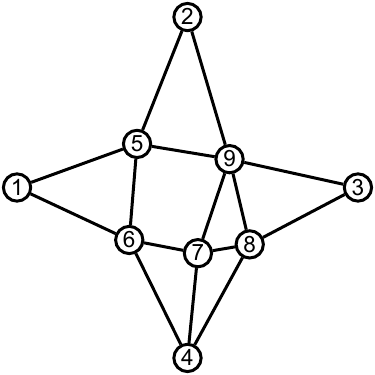}&= \int \prod_{a=5}^9 d^4 x_a \frac{x_{13}^4 x_{24}^4 x_{49}^2}{x_{15}^2 x_{16}^2 x_{25}^2 x_{29}^2 x_{38}^2 x_{39}^2 x_{46}^2 x_{47}^2 x_{48}^2 x_{56}^2 x_{59}^2 x_{67}^2 x_{78}^2 x_{79}^2 x_{89}^2}\\
\mathcal{I}^{(5)}_{28}&=\includegraphics[scale=0.4,align=c]{graph/DCI5l28.pdf}&=\includegraphics[scale=0.4,align=c]{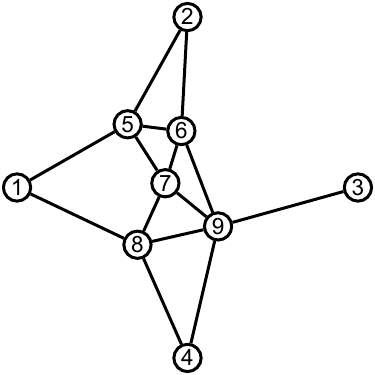}&= \int \prod_{a=5}^9 d^4 x_a \frac{x_{13}^2 x_{19}^2 x_{24}^4}{x_{15}^2 x_{18}^2 x_{25}^2 x_{26}^2 x_{39}^2 x_{48}^2 x_{49}^2 x_{56}^2 x_{57}^2 x_{67}^2 x_{69}^2 x_{78}^2 x_{79}^2 x_{89}^2}\\
\mathcal{I}^{(5)}_{29}&=\includegraphics[scale=0.4,align=c]{graph/DCI5l29.pdf}&=\includegraphics[scale=0.4,align=c]{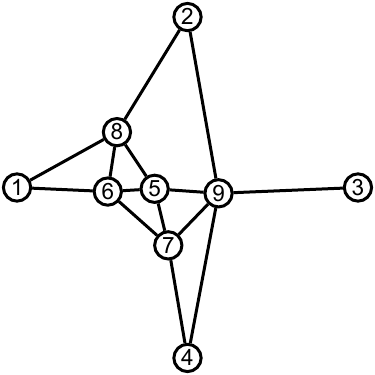}&= \int \prod_{a=5}^9 d^4 x_a \frac{x_{13}^2 x_{19}^2 x_{24}^4}{x_{16}^2 x_{18}^2 x_{28}^2 x_{29}^2 x_{39}^2 x_{47}^2 x_{49}^2 x_{56}^2 x_{57}^2 x_{58}^2 x_{59}^2 x_{67}^2 x_{68}^2 x_{79}^2}\\
\mathcal{I}^{(5)}_{30}&=\includegraphics[scale=0.4,align=c]{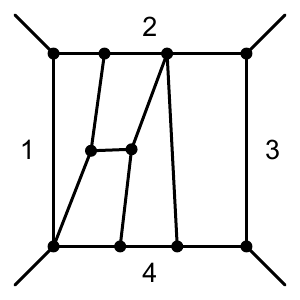}&=\includegraphics[scale=0.4,align=c]{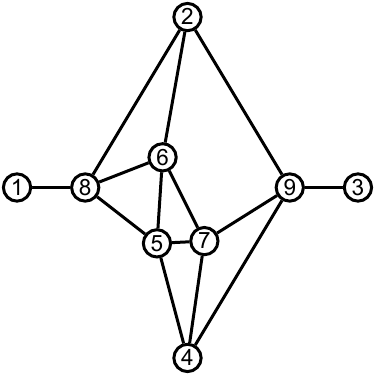}&= \int \prod_{a=5}^9 d^4 x_a \frac{x_{13}^2 x_{24}^6}{x_{18}^2 x_{26}^2 x_{28}^2 x_{29}^2 x_{39}^2 x_{45}^2 x_{47}^2 x_{49}^2 x_{56}^2 x_{57}^2 x_{58}^2 x_{67}^2 x_{68}^2 x_{79}^2}\\
\mathcal{I}^{(5)}_{31}&=\includegraphics[scale=0.4,align=c]{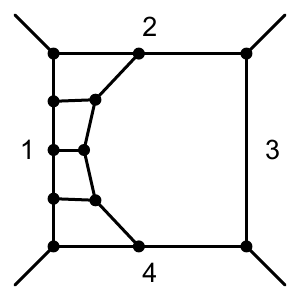}&=\includegraphics[scale=0.4,align=c]{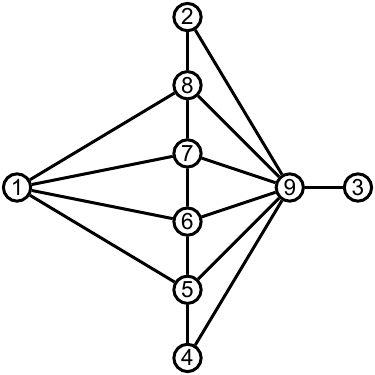}&= \int \prod_{a=5}^9 d^4 x_a \frac{x_{13}^2 x_{19}^6 x_{24}^4}{x_{15}^2 x_{16}^2 x_{17}^2 x_{18}^2 x_{28}^2 x_{29}^2 x_{39}^2 x_{45}^2 x_{49}^2 x_{56}^2 x_{59}^2 x_{67}^2 x_{69}^2 x_{78}^2 x_{79}^2 x_{89}^2}\\
\mathcal{I}^{(5)}_{32}&=\includegraphics[scale=0.4,align=c]{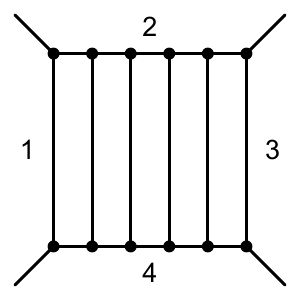}&=\includegraphics[scale=0.4,align=c]{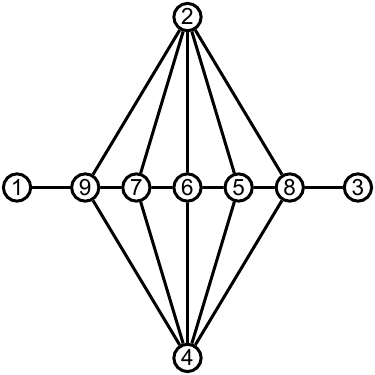}&= \int \prod_{a=5}^9 d^4 x_a \frac{x_{13}^2 x_{24}^{10}}{x_{19}^2 x_{25}^2 x_{26}^2 x_{27}^2 x_{28}^2 x_{29}^2 x_{38}^2 x_{45}^2 x_{46}^2 x_{47}^2 x_{48}^2 x_{49}^2 x_{56}^2 x_{58}^2 x_{67}^2 x_{79}^2}\\
\mathcal{I}^{(5)}_{33}&=\includegraphics[scale=0.4,align=c]{graph/DCI5l33.pdf}&=\includegraphics[scale=0.4,align=c]{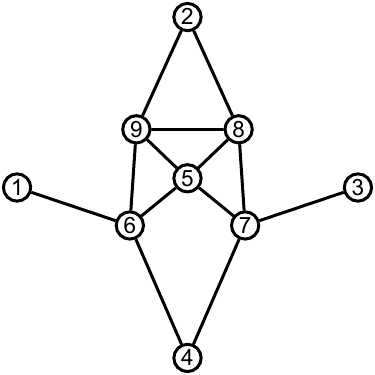}&= \int \prod_{a=5}^9 d^4 x_a \frac{x_{13}^2 x_{24}^4}{x_{16}^2 x_{28}^2 x_{29}^2 x_{37}^2 x_{46}^2 x_{47}^2 x_{56}^2 x_{57}^2 x_{58}^2 x_{59}^2 x_{69}^2 x_{78}^2 x_{89}^2}\\
\mathcal{I}^{(5)}_{34}&=\includegraphics[scale=0.4,align=c]{graph/DCI5l34.pdf}&=\includegraphics[scale=0.4,align=c]{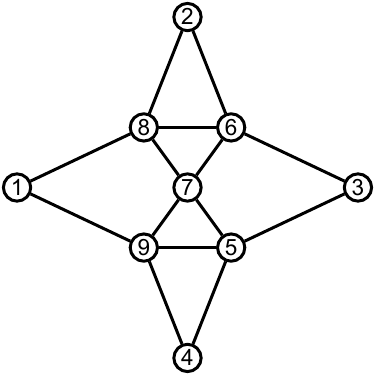}&= \int \prod_{a=5}^9 d^4 x_a \frac{x_{13}^4 x_{24}^4}{x_{18}^2 x_{19}^2 x_{26}^2 x_{28}^2 x_{35}^2 x_{36}^2 x_{45}^2 x_{49}^2 x_{57}^2 x_{59}^2 x_{67}^2 x_{68}^2 x_{78}^2 x_{79}^2}
\end{align*}
\end{footnotesize}
At $\ell=5$, the original magic identity implies $\mathcal{I}^{(5)}_{9}=\mathcal{I}^{(5)}_{10}=\mathcal{I}^{(5)}_{11}=\mathcal{I}^{(5)}_{13}=\mathcal{I}^{(5)}_{15}=\mathcal{I}^{(5)}_{16}=\mathcal{I}^{(5)}_{17}=\mathcal{I}^{(5)}_{19}=\mathcal{I}^{(5)}_{20}=\mathcal{I}^{(5)}_{21}=\mathcal{I}^{(5)}_{22}=\mathcal{I}^{(5)}_{23}=\mathcal{I}^{(5)}_{31}=\mathcal{I}^{(5)}_{32}$.

\bibliography{inspire}


\end{document}